\documentclass[pdflatex,sn-mathphys-num]{sn-jnl}

\usepackage{graphicx}%
\usepackage{multirow}%
\usepackage{amsmath,amssymb,amsfonts}%
\usepackage{amsthm}%
\usepackage{mathrsfs}%
\usepackage[title]{appendix}%
\usepackage{xcolor}%
\usepackage{textcomp}%
\usepackage{manyfoot}%
\usepackage{booktabs}%
\usepackage{algorithm}%
\usepackage{algorithmicx}%
\usepackage{algpseudocode}%
\usepackage{listings}%

\theoremstyle{thmstyleone}%
\newtheorem{theorem}{Theorem}

\newtheorem{lemma}[theorem]{Lemma}

\newtheorem{proposition}[theorem]{Proposition}%

\theoremstyle{thmstyletwo}%
\newtheorem{remark}{Remark}%

\theoremstyle{thmstylethree}%

\begin{document}

\renewcommand\vec[1]{\mathbf{#1}}

\title{A unified cross-scale framework for the dynamic Waddington landscape}

\author[1]{\fnm{Yuling} \sur{Xue}}\email{xyl981274902@163.com}
\author*[1,2]{\fnm{Jinzhi} \sur{Lei}}\email{jzlei@tiangong.edu.cn} 

\affil[1]{\orgdiv{School of Mathematical Sciences}, \orgname{Tiangong University}, \orgaddress{\city{Tianjin}, \postcode{300387}, \country{China}}}
\affil[2]{\orgdiv{Center for Applied Mathematics},\orgname{Tiangong University}, \orgaddress{\city{Tianjin}, \postcode{300387}, \country{China}}}

\abstract{
The Waddington epigenetic landscape is a foundational concept in developmental biology, yet its quantitative formulations remain largely disconnected from the population dynamics that originally motivated the metaphor. Existing approaches typically describe either intracellular gene-regulatory dynamics or population-level cell kinetics, but rarely both within a unified framework. Here, we develop a unified PDE-integral framework that couples stochastic gene-regulatory dynamics with probability-conserving population renewal, yielding a dynamic Waddington landscape $U(\vec{x}, t) = -\log f(\vec{x}, t)$ defined directly from the evolving epigenetic probability density $f(\vec{x}, t)$. Rigorous mathematical analysis of this framework establishes its well-posedness and provides conditions for the existence and uniqueness of stationary solutions. Application to canonical gene regulatory circuits shows that proliferation is not merely a determinant of population size, but an active regulator of landscape topology. Population renewal reshapes attractor stability and transition barriers in a circuit-dependent manner, generating transient plasticity in some regulatory architectures while stabilizing progenitor states in others. Under highly proliferative conditions, these interactions further produce oscillatory ``breathing'' landscapes. Together, these results establish population renewal as an integral component of cell-fate regulation and extend the Waddington landscape from a static description of intracellular dynamics to a predictive multiscale framework linking gene regulation, cell-state dynamics, and tissue homeostasis.  
}


\maketitle

\section{Introduction}
\label{sec:1}

The Waddington epigenetic landscape remains a foundational concept in developmental biology, providing a theoretical paradigm for understanding cell fate determination \cite{Waddington1957,Waddington:1940}. Originally envisioned in the 1940s and 1950s, Waddington's classic heuristic of a ball traversing branching valleys was intended to conceptually describe the phenotypic evolution of a proliferating cell population, such as a developing tissue, rather than the isolated state transitions of a single cell. From this perspective, the topography reflects the collective properties of the population, where valley depths represent the relative abundance of specific cell types under the combined forces of stochastic gene regulation and proliferative selection. Today, translating this conceptual paradigm into a fully quantitative model (essential for insights into tumorigenesis, cellular reprogramming, and regenerative medicine) remains an open challenge, requiring mathematical frameworks that can explicitly capture its inherently multiscale nature \cite{Shakiba2022Waddington,Vittadello:2025aa}. 

Formulating the landscape through the lens of single-cell probability has yielded fundamental insights into how molecular networks dictate stable cell states. Pioneering theoretical frameworks, grounded in foundational advances in non-equilibrium statistical physics \cite{AO200963,wang2011landscape, Wang02012015,Zhou:2016gd}, have rigorously defined the landscape utilizing stochastic gene regulatory networks (GRNs). While these approaches successfully capture the intracellular non-equilibrium dynamics that maintain stable phenotypic attractors, they typically derive the landscape from the steady-state probability of an isolated cell. This formulation implicitly assumes a static equilibrium, which cannot fully capture scenarios such as morphogenesis or developmental progression, in which the cell population itself is a macroscopic non-equilibrium entity. In such systems, continuous cell population fluxes driven by asymmetric division, apoptosis, and proliferative competition actively shape the relative prevalence of cell types \cite{Lander:2009aa,Klein:2011aa,Huang:2012aa}. Consequently, a molecularly stable state might be virtually absent in a tissue if it undergoes rapid apoptosis, while a metastable state might become a dominant macroscopic valley through clonal expansion; this presents a cross-scale reality that fundamentally extends beyond the scope of single-cell steady-state models.

Recent theoretical and computational efforts, particularly driven by advances in single-cell transcriptomics, have sought to address this limitation by incorporating proliferation effects into trajectory inference and landscape definitions \cite{weinreb2018,shi2022energy,La-Manno:2018aa,Sha:2024aa,Peng:2026aa}. These models typically rely on phenomenological birth-death rates, which have laid important groundwork for coupling population growth with phenotypic states. However, extending these formulations to time-varying, non-equilibrium systems remains a mathematical challenge, as ad hoc source terms do not inherently guarantee probability conservation during transient continuous dynamics \cite{shi2022energy}. Furthermore, a mechanistic representation of the cellular life cycle has yet to be systematically integrated, particularly regarding the mitotic inheritance of epigenetic states \cite{Probst:2009aa,Anreiter:2017aa,Allis:2016aa}. During cell division, the stochastic redistribution of molecular marks, such as DNA methylation and histone modifications, introduces subtle variations in daughter cells. While these individual fluctuations may not immediately alter the cellular phenotype, their accumulation over successive divisions drives a stochastic drift in gene expression, ultimately facilitating macroscopic phenotypic transitions. Without a physically consistent and mechanistically grounded mathematical treatment of this cross-scale coupling, our understanding of how population-level expansion feeds back to actively reshape the phenotypic landscape remains incomplete.

In this work, we establish a unified PDE-integral framework that provides a mathematically rigorous, cross-scale formulation of the dynamic Waddington landscape. Rather than treating the landscape as a static property of isolated cells, our framework defines it as $U(\vec{x}, t) = -\log f(\vec{x}, t)$, where $f(\vec{x}, t)$ is the epigenetic probability density over the continuous epigenetic state space. By coupling stochastic GRN dynamics with a probability-conserving population renewal operator $\mathcal{R}[f; Q_{\mathrm{tot}}]$ (also written as $\mathcal{R}[f]$ for simplicity), the framework integrates intracellular regulation with cell division, differentiation, apoptosis, and stochastic epigenetic inheritance into a single mathematical description while preserving total probability throughout development. Furthermore, we validate this framework through rigorous mathematical analysis, establishing the well-posedness of the coupled system and delineating the conditions for the existence and uniqueness of its stationary solutions.

Application to canonical gene circuits shows that proliferation acts not merely as a passive scaling factor for cell abundance but as a topological driver of the dynamic landscape. Depending on the underlying circuit architecture, population renewal reshapes transition barriers, reorganizes source-sink relationships, and either promotes transient proliferation-induced plasticity or preserves stable progenitor reservoirs through autoregulatory memory. Under hyper-proliferative conditions, the framework further predicts a previously unrecognized breathing landscape, in which landscape topology oscillates in synchrony with tissue-level population dynamics. Together, these results establish a quantitative framework for understanding how population renewal actively governs cell-fate landscapes during development and regeneration.

\section{Mathematical formulation: A unified cross-scale framework}
\label{sec:model}

\begin{figure}[htbp]
\centering
\includegraphics[width=0.6\textwidth]{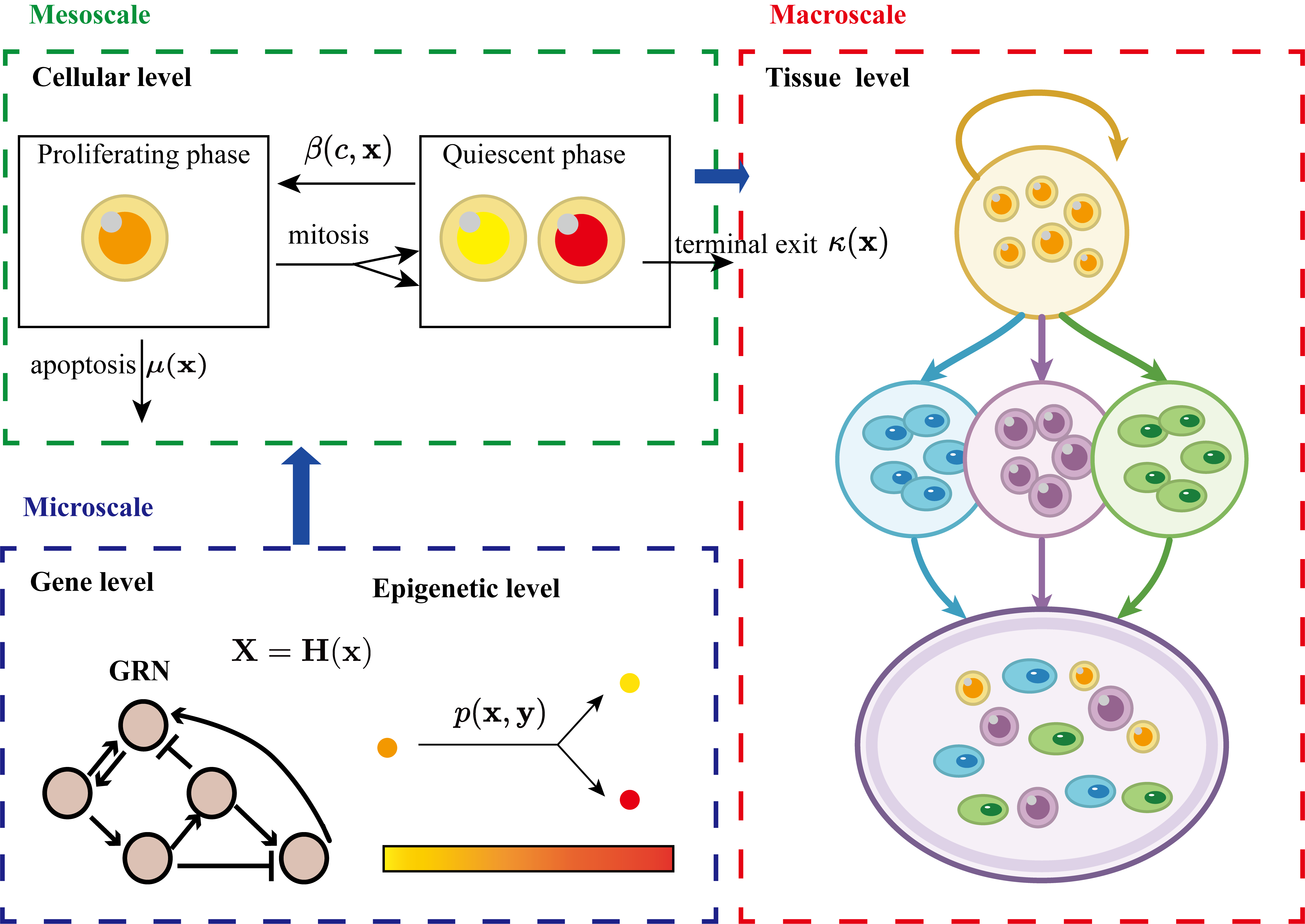}
\caption{\textbf{Unified multiscale modeling framework for the dynamic Waddington landscape.} \textit{Microscale GRN and epigenetic encoding:} Intracellular GRN dynamics govern the molecular state $\vec{X}$, described by a stochastic differential equation (SDE) with deterministic drift $\vec{F}(\vec{X})$ and noise perturbation $\boldsymbol{\eta}(t)$. The nonlinear mapping $\vec{X} = \vec{H}(\vec{x})$ defines the connection from the epigenetic state $\vec{x}$ to the gene expression profile. \textit{Mesoscale and macroscale population dynamics:} At the cellular and tissue levels, the $G_0$ cell cycle dynamics encode the core events of heterogeneous tissue regeneration: quiescent cells re-enter the proliferating phase at rate $\beta(c, \vec{x})$ or undergo terminal exit from the regenerative pool at rate $\kappa(\vec{x})$, while proliferating cells undergo apoptosis at rate $\mu(\vec{x})$ or complete mitosis after a duration $\tau(\vec{x})$. \textit{Cross-scale integration:} The inheritance kernel $p(\vec{x},\vec{y})$, combined with the cell cycle kinetics, serves as the mathematical coupler. It mechanistically translates cellular epigenetic inheritance during division into the macroscopic population renewal operator $\mathcal{R}[f; Q_{\mathrm{tot}}]$. The integrated formulation determines the time-evolving epigenetic probability density $f(\vec{x}, t)$, which explicitly shapes the macroscopic dynamic Waddington landscape, mathematically characterized by the phenotypic potential $U(\vec{x}, t) = -\log f(\vec{x}, t)$.}
	\label{fig1}
\end{figure}

To establish a physically consistent formulation of the dynamic Waddington landscape, we develop a closed PDE-integral framework that directly connects intracellular molecular stochasticity with macroscopic population dynamics (Fig.~\ref{fig1}). This multiscale approach bridges two distinct regimes: the continuous microscale fluctuations of gene regulatory networks and the discrete macroscale events of tissue regeneration. We formally link these scales via a mechanistic population renewal operator, $\mathcal{R}[f; Q_{\mathrm{tot}}]$, which encapsulates macroscale cellular fluxes and the stochastic redistribution of epigenetic states during mitosis. By integrating this non-equilibrium operator into the single-cell-scale Fokker--Planck equations, we determined the time-evolving epigenetic probability density $f(\vec{x}, t)$. This integration enables us to generalize the classical non-equilibrium potential \cite{Wang:2008aa,Zhou:2016gd} and explicitly construct the dynamic phenotypic landscape as $U(\vec{x}, t) = -\log f(\vec{x}, t)$. The governing equations for each scale are developed in the subsections below.

\subsection{Microscale GRN dynamics and epigenetic encoding}
\label{sec:epigenetic_dynamics}

At the molecular level, biological phenotypes are characterized by the expression profiles of key marker genes, governed by underlying gene regulatory networks (GRNs) \cite{Graf:2009aa,Davidson:2010aa,Wagner:2016aa}. The state of a single cell is denoted by $\vec{X} = (X_1, X_2, \dots, X_n) \in \mathbb{R}^n_{\geq 0}$, where $X_i$ represents the expression level of the $i$-th gene. In the regime of moderate molecular copy numbers, this intracellular dynamics is conventionally modeled by a Langevin-type stochastic differential equation (SDE) \cite{Elowitz:2002aa, Rao:2002aa}
\begin{equation}
    \frac{d\vec{X}}{dt} = \vec{F}(\vec{X}) + \boldsymbol{\eta}(t),
    \label{eq:SDE}
\end{equation}
where $\vec{F}(\vec{X})$ captures deterministic interactions and $\boldsymbol{\eta}(t)$ represents Gaussian white noise with cross-correlation $\langle \boldsymbol{\eta}(t)\boldsymbol{\eta}(t')^\top\rangle = 2\vec{D}\delta(t-t')$. Stable phenotypic states correspond to the deterministic attractors $\vec{F}(\vec{X}) = \vec{0}$. In this context, the physical landscape theory pioneered by Wang et al.~\cite{Wang:2008aa, wang2011landscape} defines a potential $U(\vec{X}) = - \log P_{\mathrm{ss}}(\vec{X})$ via the steady-state probability density of \eqref{eq:SDE} (see \emph{SI Appendix} A.1), enabling quantitative analysis of attractor stability.

However, continuous GRN dynamics alone fail to capture the discrete events of the cell cycle. During mitosis, molecular states $\vec{X}$ undergo drastic partitioning fluctuations, whereas epigenetic states (e.g., histone modifications and chromatin accessibility) are systematically conserved and reconstituted, providing the physical basis for phenotypic inheritance and cell fate plasticity \cite{Probst:2009aa,Allis:2016aa}. To mathematically bridge continuous molecular kinetics with discrete mitotic division, we introduce a latent vector $\vec{x}\in \Omega \subset \mathbb{R}^m$ to represent the heritable epigenetic state. Physically, the high-dimensional expression profile $\vec{X}\in \mathbb{R}_{\geq 0}^n$ acts as a rapidly fluctuating observable, while the epigenetic state $\vec{x}$ serves as the slowly varying latent variable, effectively constraining the system's dynamics onto a lower-dimensional epigenetic manifold ($m \leq n$). 

This conceptual link is formalized by a smooth mapping $\vec{X} = \vec{H}(\vec{x})$, enabling the stochastic dynamics of the epigenetic state to be derived directly from the underlying GRN logic. Under this coordinate transformation, the original SDE \eqref{eq:SDE} rigorously maps to a multiplicative-noise SDE in the latent $\vec{x}$-space. The temporal evolution of the autonomous state PDF (in the absence of population renewal), $f_0(\vec{x}, t)$, is thus governed by the Fokker--Planck equation (see \emph{SI Appendix} A.2):
\begin{equation}
    \frac{\partial f_0}{\partial t} =  \nabla \cdot \bigl(\tilde{\vec{D}}(\vec{x})\,\nabla f_0\bigr) - \nabla \cdot \bigl(\tilde{\vec{F}}(\vec{x})\,f_0\bigr).
    \label{eq:FP_epi}
\end{equation}
Here, $\tilde{\vec{D}}(\vec{x}) = \vec{A}(\vec{x}) \vec{D} \vec{A}(\vec{x})^\top$ is the state-dependent diffusion tensor, and the flux drift in the conservative Fickian form is defined as $\tilde{\vec{F}}(\vec{x}) = \vec{b}(\vec{x}) - \nabla \cdot \tilde{\vec{D}}(\vec{x})$. The local projection matrix $\vec{A}(\vec{x})$ is determined as the Moore--Penrose left pseudoinverse of the Jacobian $\vec{B}(\vec{x}) = \partial \vec{H}(\vec{x})/\partial \vec{x}\in \mathbb{R}^{n\times m}$. Furthermore, $\vec{b}(\vec{x}) = \vec{A}(\vec{x})\vec{F}(\vec{H}(\vec{x})) + \vec{c}_{\mathrm{cg}}^{\,I}(\vec{x})$ is the It\^{o} drift of the latent SDE, where $\vec{c}_{\mathrm{cg}}^{\,I}$ denotes the It\^{o} coarse-graining drift. Under a physically consistent closure that sets the Stratonovich coarse-graining drift to zero ($\vec{c}_{\mathrm{cg}}^{\,S} = \vec{0}$), $\vec{c}_{\mathrm{cg}}^{\,I}(\vec{x})$ emerges naturally as the noise-induced spurious drift dictated by the standard Stratonovich-to-It\^{o} drift conversion formula (detailed in \emph{SI Appendix} A.2).

For the illustrative models throughout this work, we adopt the component-wise exponential map $H_i(x_i) = e^{x_i} - 1$, equivalent to the logarithmic transformation $x_i = \log(1 + X_i)$. Mathematically, this invertible mapping ensures that our unified perspective recovers the coordinate transformation prescribed by It\^{o}'s formula (see \emph{SI Appendix} A.3). Beyond this mathematical consistency, the choice satisfies both empirical and mechanistic requirements. Empirically, it aligns with standard scale-normalization practices in single-cell transcriptomics (e.g., scRNA-seq) \cite{Stuart:2019aa,Hafemeister:2019aa}. Mechanistically, it reflects the physical reality that epigenetic modifications dictate chromatin accessibility by directly modulating the energy barriers for transcription factor binding \cite{Jaenisch:2003aa}. By mathematically amplifying regulatory variations at low molecular concentrations while compressing stochastic fluctuations at high expression levels, this scaling behavior directly mirrors the nonlinear saturation of the transcriptional machinery, establishing a physically and biologically consistent formulation.

While \eqref{eq:FP_epi} effectively captures the state transitions of an isolated cell or a non-dividing ensemble, it is inherently inadequate for describing how macroscopic population fluxes dynamically reshape cell-type proportions within a proliferating population.

\subsection{Mesoscale and macroscale population dynamics: kinetotype-driven cell cycle kinetics}
\label{sec:population_renewal}

While the microscale SDE describes single-cell state transitions, macroscopic tissue dynamics are fundamentally driven by discrete mesoscale cellular behaviors: cell division, apoptosis, and terminal exit from the cell cycle. Building upon our previous discrete-time \cite{Lei:2014aa} and continuous \cite{lei2020general,Lei:2020gd} cell-cycle frameworks, we model the entire proliferatively competent pool (including stem, transit-amplifying, and progenitor cells) as a heterogeneous continuum across the epigenetic state space, conceptually bypassing rigid cell-type boundaries (see \emph{SI Appendix} B.1). 

Utilizing the classical $G_0$ cell cycle model \cite{Burns:1970tm}, we track the cell-number density $Q(\vec{x}, t)$ over the epigenetic state space, i.e., the number of biologically quiescent (yet regeneratively capable) cells per unit epigenetic state. To explicitly link intracellular states to macroscopic tissue growth, we postulate that cellular dynamics are parametrically governed by their underlying epigenetic coordinates $\vec{x}$ through a dependency we previously formalized as the cell's \emph{kinetotype} \cite{lei2020general}. Specifically, the kinetotype encapsulates the population-level kinetic phenotype associated with a given epigenetic state, defined by the state-dependent rates of cell-cycle entry, proliferating-phase survival, mitotic completion, and irreversible exit. The resulting temporal evolution of the population is governed by the following integro-differential equation (see \emph{SI Appendix} B.2):
\begin{equation}
\begin{aligned}
\frac{\partial Q(\vec{x}, t)}{\partial t} = & -Q(\vec{x}, t)\bigl(\beta(c(t),\vec{x}) + \kappa(\vec{x})\bigr)\\
&{} + 2\int_\Omega M(\vec{y}, t)\,p(\vec{x}, \vec{y})\,d\vec{y}.
\end{aligned}
\label{eq:macro_Q}
\end{equation}
Here, $\beta(c(t), \vec{x})$ is the kinetotype-specific cell cycle re-entry rate, which responds to a global cytokine field $c(t)=\int_\Omega \xi(\vec{x})Q(\vec{x}, t) \, d\vec{x}$, where $\xi(\vec{x})$ denotes the effective cytokine secretion rate of cells with state $\vec{x}$. The cycle exit rate $\kappa(\vec{x})$ functionally abstracts irreversible processes that permanently deplete proliferative capacity, such as terminal differentiation, senescence, or resting-phase cell death. 

The integral term in \eqref{eq:macro_Q} captures the influx of newborn cells arising from mitosis, with 
\begin{equation*}
    M(\vec{y}, t) = \beta\bigl(c(t-\tau(\vec{y})), \vec{y}\bigr)\,Q\bigl(\vec{y}, t-\tau(\vec{y})\bigr)\,e^{-\mu(\vec{y})\tau(\vec{y})}.
\end{equation*}
This term represents cells that entered the cell cycle from state $\vec{y}$ at time $t - \tau(\vec{y})$ and completed division after a duration $\tau(\vec{y})$, surviving with probability $e^{-\mu(\vec{y})\tau(\vec{y})}$ against a state-dependent apoptosis rate $\mu(\vec{y})$. Finally, the epigenetic inheritance kernel $p(\vec{x}, \vec{y})$ (normalized such that $\int_\Omega p(\vec{x}, \vec{y})\, d\vec{x} = 1$) defines the conditional probability of state redistribution from a mother cell ($\vec{y}$) to a daughter cell ($\vec{x}$) during mitosis \cite{lei2020general} (see \emph{SI Appendix} B.3). This kernel mechanistically encapsulates the imperfect fidelity of epigenetic inheritance, providing the fundamental biological source of division-coupled cell plasticity.

In our continuous formulation, phenotypic differentiation emerges naturally from the underlying epigenetic drift, whereas terminal differentiation and senescence are mathematically aggregated as an explicit efflux from the regenerative pool.

\subsection{Cross-scale integration: the unified PDE-integral system}
\label{sec:landscape_density}

To rigorously bridge single-cell GRN dynamics with tissue-level population evolution, we defined the normalized probability density of the resting population as 
\begin{equation}
\label{eq:total_population}
f_q(\vec{x}, t) = \frac{Q(\vec{x}, t)}{Q_{\mathrm{tot}}(t)},\qquad Q_{\mathrm{tot}}(t) = \int_\Omega Q(\vec{x}, t)\,d\vec{x}.
\end{equation} 
Here, $Q_{\mathrm{tot}}(t)$ is the total number of quiescent cells, obtained by integrating the cell-number density $Q(\vec{x},t)$ over the epigenetic state space. Hereafter referred to simply as the \emph{total cell number}, it evolves following the macroscopic equation (see \emph{SI Appendix} C.1)
\begin{equation}
\label{eq:Qtot}
\frac{dQ_{\mathrm{tot}}(t)}{dt}    = -\int_\Omega Q(\vec{x},t)    \bigl(\beta(c(t),\vec{x})+\kappa(\vec{x})\bigr)\,d\vec{x}  + 2\int_\Omega M(\vec{y},t)\,d\vec{y}.
\end{equation}
Differentiating $f_q(\vec{x}, t)$ with respect to time and substituting the macroscale flux from \eqref{eq:macro_Q}, we obtain the evolution equation for this state density (see \emph{SI Appendix} C.2):
\begin{equation}
\label{eq:evl_f}
\frac{\partial f_q(\vec{x}, t)}{\partial t}   =  \frac{2}{Q_{\mathrm{tot}}(t)} \int_\Omega M(\vec{y}, t)\bigl[p(\vec{x},\vec{y}) - f_q(\vec{x}, t)\bigr]\,d\vec{y}  - f_q(\vec{x}, t) \bigl[r(\vec{x}, t) - \bar{r}_{f_q}(t)\bigr].
\end{equation}
Here, $r(\vec{x}, t) = \beta(c(t), \vec{x}) + \kappa(\vec{x})$ denotes the state-dependent resting-phase removal rate, and 
$$\bar{r}_{f_q}(t) = \int_\Omega r(\vec{y}, t) f_q(\vec{y}, t)\,d\vec{y}$$ 
represents its population average over $f_q$.

The right-hand side of \eqref{eq:evl_f} explicitly captures the discrete population-level redistribution induced by cellular turnover. In contrast, the Fokker--Planck equation \eqref{eq:FP_epi}  describes continuous epigenetic-state transitions driven by stochastic GRN dynamics between cell division events. A unified mathematical framework naturally emerges by superposing these two mechanisms. 

To this end, let $f(\vec{x}, t)$ denote the probability density of epigenetic states for the fully coupled system. Extracting the kinematic structure from \eqref{eq:evl_f}, we define the \emph{population renewal operator} $\mathcal{R}[f; Q_{\mathrm{tot}}](\vec{x}, t)$ (written as $\mathcal{R}[f]$ when the dependence is clear) as
\begin{equation}
\begin{aligned}
\mathcal{R}[f; Q_{\mathrm{tot}}](\vec{x}, t) = &\frac{2}{Q_{\mathrm{tot}}(t)} \int_\Omega M(\vec{y}, t)\bigl[p(\vec{x},\vec{y}) - f(\vec{x}, t)\bigr]\,d\vec{y}\\
&{}  - f(\vec{x}, t) \bigl[r(\vec{x}, t) - \bar{r}_{f}(t)\bigr].
\end{aligned}
\end{equation}
This operator governs the exact rate of change in the epigenetic probability density contributed purely by cellular turnover. The first term dictates discrete epigenetic redistribution upon mitosis, while the second term captures a replicator-like continuous selection. Combining \eqref{eq:FP_epi} with this population renewal operator yields the unified integro-PDE system:
\begin{equation}
\begin{aligned}
	\frac{\partial f(\vec{x}, t)}{\partial t} = &\underbrace{\nabla \cdot \bigl(\tilde{\vec{D}}(\vec{x})\,\nabla f\bigr) - \nabla \cdot \bigl(\tilde{\vec{F}}(\vec{x})\,f\bigr)}_{\text{Between-division continuous GRN dynamics}} \\
	&\; + \underbrace{\mathcal{R}[f; Q_{\mathrm{tot}}](\vec{x}, t)}_{\text{Division-driven discrete redistribution}}.
\end{aligned}
	\label{mod1}
\end{equation}
Mathematically, equation \eqref{mod1} replaces the phenomenological scalar birth--death terms used in classical population balance models~\cite{weinreb2018,Briggs:2018hg,shi2022energy} with $\mathcal{R}[f; Q_{\mathrm{tot}}]$ (see \emph{SI Appendix} C.3), formally integrating single-cell gene regulation with population-level macroscopic evolution. The mathematically exact and biologically interpretable operator $\mathcal{R}[f; Q_{\mathrm{tot}}]$ provides a mechanistic representation of both the absolute mitotic influx and imperfect epigenetic inheritance through $p(\vec{x}, \vec{y})$, while maintaining the properly normalized balance between cells entering and leaving the regenerative pool.

Equation \eqref{mod1} is supplemented with the no-flux condition 
\begin{equation}
\label{eq:zero_flux}
\bigl(\tilde{\vec{D}}(\vec{x})\,\nabla f - \tilde{\vec{F}}(\vec{x})\,f\bigr)\cdot \vec{n}(\vec{x}) = 0,\quad \vec{x}\in \partial \Omega,
\end{equation}
where $\vec{n}(\vec{x})$ is the unit normal vector on $\partial \Omega$. 
The delayed population equation \eqref{eq:macro_Q} is supplemented with the nonnegative history
\begin{equation}
\label{eq:Q_initial_history}
Q(\vec{x}, s) = Q_{\mathrm{in}}(\vec{x}, s) \geq 0, \quad (\vec{x}, s) \in \Omega \times [-\tau_{\max}, 0],
\end{equation}
where $\tau_{\max} = \max_{\vec{x}\in\Omega}\tau(\vec{x})$. Accordingly, the initial quiescent population 
\begin{equation}
\label{eq:Qtot_ini}
Q_{\mathrm{tot}}(0) = \int_\Omega Q_{\mathrm{in}}(\vec{x}, 0) d\, \vec{x},
\end{equation}
and $Q_{\mathrm{tot}}(0) > 0$. The initial epigenetic probability density can be given by normalizing the initial cell-number density  
\begin{equation}
\label{eq:f_initial_condition}
f(\vec{x}, 0) = \frac{Q_{\mathrm{in}}(\vec{x}, 0)}{Q_{\mathrm{tot}}(0)}.
\end{equation}
Here, we emphasize that \eqref{mod1} is a modeling \emph{superposition}: the continuous GRN dynamics (the Fokker--Planck term) and the discrete population redistribution (the renewal operator $\mathcal{R}[f; Q_{\mathrm{tot}}]$) act additively on $f$, so that for $t>0$ the normalized density $f(\vec{x},t)$ no longer equals $Q(\vec{x},t)/Q_{\mathrm{tot}}(t)$ despite the initial probability density \eqref{eq:f_initial_condition}. The renewal operator remains well defined because the mitotic completion flux $M(\vec{y},t)$ and the total cell number $Q_{\mathrm{tot}}(t)$ involved in its formulation are explicitly supplied by the absolute cell-number density $Q(\vec{x},t)$, which evolves independently through \eqref{eq:macro_Q}. Thus $\mathcal{R}[f; Q_{\mathrm{tot}}]$ acts on $f$ but is driven by $Q$, and, because it integrates to zero on the probability simplex by construction (see Proposition~\ref{prop:simplex_invariance} below), it preserves the normalization of $f$.

In summary, the complete model consists of the renewal equation for the cell-number density $Q(\vec{x}, t)$ in \eqref{eq:macro_Q}, the definition of the total cell number $Q_{\mathrm{tot}}(t)$ in \eqref{eq:total_population}, and the unified equation for the epigenetic probability density $f(\vec{x}, t)$ in \eqref{mod1}. The initial and boundary conditions are given by \eqref{eq:zero_flux}--\eqref{eq:f_initial_condition}. 

\subsection{The dynamic Waddington landscape and its non-equilibrium decomposition}

Building upon the evolution equation for the epigenetic probability density $f(\vec{x},t)$, we define a dynamic Waddington landscape as
\begin{equation}
\label{eq:landscape_def}
U(\vec{x}, t) = -\log f(\vec{x}, t).
\end{equation}
The mathematical well-definedness of this definition is rigorously established in our subsequent theoretical analysis. In this formulation, valleys of $U$ correspond to highly populated cell states, whereas barrier heights characterize the relative separation between cell states in probability space. Unlike classical steady-state landscapes \cite{Wang:2008aa,wang2011landscape}, the landscape $U$ co-evolves with the underlying population distribution and can therefore be dynamically reshaped by both intracellular fluctuations and population-level regenerative and selective dynamics.

To disentangle intrinsic epigenetic-state dynamics from population-level regulation, we introduce a non-equilibrium decomposition of the drift $\tilde{\vec{F}}(\vec{x})$ using the reference density $f_0(\vec{x}, t)$ generated by the same epigenetic diffusion but with $\mathcal{R}[f; Q_{\mathrm{tot}}]\equiv 0$ (see \emph{SI Appendix} D.1):
\begin{equation}
\label{eq:decomposition}
\tilde{\vec{F}}(\vec{x}) = -\tilde{\vec{D}}(\vec{x})\nabla U(\vec{x},t) -\tilde{\vec{D}}(\vec{x})\nabla V(\vec{x},t) + \tilde{\vec{F}}_{\mathrm{curl}}(\vec{x},t),
\end{equation}
where the population-induced potential $V(\vec{x}, t)$ is defined as
\begin{equation}
\label{eq:main_text_V_def}
V(\vec{x},t)=\log\!\left[\frac{f(\vec{x},t)}{f_0(\vec{x},t)}\right].
\end{equation}
This decomposition assigns distinct biological roles to each component. $U$ represents the occupancy-based dynamic landscape, identifying the instantaneous stability structure of cell states. $V$ acts as a \emph{selective potential} that quantifies how proliferation, survival, and differentiation bias the distribution relative to the no-proliferation baseline. Consequently, high-$V$ regions denote states with a relative net population-level advantage (e.g., enhanced self-renewal or survival), while low-$V$ regions signify states undergoing relative depletion (e.g., terminal differentiation or exhaustion). The residual term $\tilde{\vec{F}}_{\mathrm{curl}}$ represents the non-gradient component of the intrinsic epigenetic dynamics, which satisfies $\nabla\cdot (f_0 \tilde{\vec{F}}_{\mathrm{curl}}) = -\partial_t f_0$.

The proposed decomposition further reveals the dynamic coupling between these potentials (see \emph{SI Appendix} D.2). By introducing the operator $\mathcal{L}_{U}\varphi = \nabla \cdot (\tilde{\vec{D}} \nabla \varphi) -\tilde{\vec{D}}\nabla U \cdot \nabla \varphi$, we obtain
\begin{equation}
\label{eq:main_text_dynamic_V}
\mathcal{L}_{U}V = \frac{\partial_t f-\mathcal{R}[f; Q_{\mathrm{tot}}] + \nabla\cdot (f\tilde{\vec{F}}_{\mathrm{curl}})}{f}.
\end{equation}
Using $U(\vec{x}, t) = -\log f(\vec{x}, t)$, this relation can be rewritten equivalently to highlight the temporal deformation of the landscape as 
\begin{equation}
\label{eq:main_text_dynamic_UV}
\partial_t U + \mathcal{L}_{U}V = - \frac{\mathcal{R}[f; Q_{\mathrm{tot}}]}{f} + \frac{1}{f} \nabla \cdot (f \tilde{\vec{F}}_{\mathrm{curl}}).
\end{equation}
Equation \eqref{eq:main_text_dynamic_UV} demonstrates that the time-evolution of the Waddington topography, together with its selective modulation via $V$, is explicitly driven by population regulation and the residual non-equilibrium flux. Crucially, $\mathcal{R}[f; Q_{\mathrm{tot}}]$ enters as a direct source term, indicating that population regulation does not merely scale the total cell number but actively remodels the landscape topology.

In the steady-state limit, this balance reduces to (see \emph{SI Appendix} D.3)
\begin{equation}
\label{eq:main_text_V_ss}
\mathcal{L}_{U^\ast}V^\ast - \tilde{\vec{F}}_{\mathrm{curl}}^\ast\cdot\nabla V^\ast = -\frac{\mathcal{R}[f^\ast; Q_{\mathrm{tot}}^\ast]}{f^\ast}.
\end{equation}
This steady relation clarifies the origin of the selective potential: $V^\ast$ is determined by the net population regulation represented by the operator $\mathcal{R}[f^\ast; Q_{\mathrm{tot}}^\ast]$, relative to the intrinsic epigenetic dynamics captured by the reference landscape. In this sense, our framework extends the energy-landscape decomposition of Shi et al. \cite{shi2022energy} to a dynamic, mechanistically grounded, and probability-conserving population framework. Detailed derivations and technical conditions are provided in \emph{SI Appendix} D.

\section{Mathematical analysis}
\label{sec:analysis}

We analyze the mathematical properties of the coupled model introduced in Section~\ref{sec:model}. The main results concern global well-posedness, preservation of nonnegativity, invariance of the probability simplex, and the existence and uniqueness of equilibrium population distribution. Because the population-renewal equation contains state-dependent delays, the problem is formulated as a delay evolution system with a nonlocal renewal term.

Throughout this section, we assume that $\Omega\subset\mathbb{R}^m$ is a bounded connected domain with $C^2$ boundary. The epigenetic-state density $Q$ is defined on $\Omega$, whereas the delayed history is prescribed on the interval $[-\tau_{\max},0]$.

\subsection{Functional setting and assumptions}
\label{sec:functional_setting}

Before stating the main results, we specify the epigenetic state domain $\Omega\subset\mathbb{R}^m$ as a bounded set with sufficiently regular boundary, and impose the following assumptions:
\begin{enumerate}
\item[(A1)]
The drift $\tilde{\vec{F}}:\bar{\Omega}\to\mathbb{R}^m$ and the diffusion tensor $\tilde{\vec{D}}:\bar{\Omega}\to\mathbb{R}^{m\times m}$ satisfy
$$
\tilde{\vec{F}}\in C^1(\bar{\Omega};\mathbb{R}^m), \qquad \tilde{\vec{D}} \in C^1(\bar{\Omega};\mathbb{R}^{m\times m}).
$$
\item[(A2)]
The diffusion tensor is uniformly elliptic; that is, there exists a constant $\lambda_D>0$ such that
\begin{equation}
\boldsymbol{\zeta}^{\top} \tilde{\vec{D}}(\vec{x}) \boldsymbol{\zeta} \geq \lambda_D|\boldsymbol{\zeta}|^2, \qquad \vec{x}\in\bar{\Omega}, \quad \boldsymbol{\zeta}\in\mathbb{R}^m.
\label{eq:uniform_ellipticity}
\end{equation}
\item[(A3)] The kinetic rates $\beta$, $\kappa$, $\mu$, and $\xi$ are nonnegative and bounded on their respective domains. Specifically, $\kappa$, $\mu$, and $\xi$ are continuous on $\bar{\Omega}$, and $\xi$ is non-trivial, i.e., $\xi(\vec{x})\not\equiv 0$. Moreover, the re-entry rate
$$
\beta:[0,\infty)\times\bar{\Omega}\to[0,\infty)
$$
is continuous in $(c,\vec{x})$ and locally Lipschitz continuous in $c$ uniformly with respect to $\vec{x}\in\bar{\Omega}$. In particular, there exist finite constants $\beta_{\max}, \kappa_{\max}>0$ such that
\begin{equation}
0\leq\beta(c,\vec{x})\leq\beta_{\max} \qquad \text{for all } c\geq0,\ \text{ and } \vec{x}\in\bar{\Omega},
\label{eq:beta_uniform_bound}
\end{equation}
and
\begin{equation}
0\leq \kappa(\vec{x}) \leq \kappa_{\max}\qquad \text{for all } \vec{x}\in \bar{\Omega}.
\end{equation}
\item[(A4)]
The cell-cycle duration $\tau$ is continuous on $\bar{\Omega}$ and satisfies
\begin{equation}
0<\tau_{\min} \leq\tau(\vec{x}) \leq\tau_{\max}<\infty, \qquad \vec{x}\in\bar{\Omega}.
\label{eq:delay_bounds}
\end{equation}
\item[(A5)]
The epigenetic inheritance kernel $p:\Omega\times\Omega\to[0,\infty)$ is measurable, bounded on $\Omega\times \Omega$, and satisfies the normalization condition
\begin{equation}
\int_\Omega p(\vec{x},\vec{y})\,d\vec{x}=1
\qquad
\text{for all }\vec{y}\in\Omega.
\label{eq:kernel_normalization_analysis}
\end{equation}
\item[(A6)]
The cytokine concentration is determined by the population through the bounded linear functional
\begin{equation}
c[Q](t) = \int_\Omega \xi(\vec{x})Q(\vec{x},t)\,d\vec{x}.
\label{eq:cytokine_population_feedback}
\end{equation}
Consequently, it satisfies the Lipschitz property
\begin{equation}
|c[Q_1](t)-c[Q_2](t)| \leq \|\xi\|_{L^\infty(\Omega)} \|Q_1(\cdot,t)-Q_2(\cdot,t)\|_{L^1(\Omega)}.
\label{eq:cytokine_feedback_lipschitz}
\end{equation}
\item[(A7)] The prescribed initial history $Q_{\mathrm{in}}$ satisfies
\begin{equation}
Q_{\mathrm{in}} \in C\bigl([-\tau_{\max},0];L^1(\Omega)\bigr),\qquad Q_{\mathrm{in}}(\cdot,s)\geq0 \quad\text{almost everywhere in }\Omega,
\label{eq:initial_history_assumption}
\end{equation}
with a strictly positive initial total cell number
\begin{equation}
Q_{\mathrm{tot}}(0) := \int_\Omega Q_{\mathrm{in}}(\vec{x},0)\,d\vec{x} > 0.
\label{eq:positive_initial_population}
\end{equation}
Furthermore, there exists a constant $C_0 \geq 1$ such that 
\begin{equation}
Q_{\mathrm{in}}(\vec{x}, s)\leq C_0Q_{\mathrm{in}}(\vec{x}, 0),\quad \text{for a.e. } \vec{x}\in \Omega\ \text{and\ all }s\in [-\tau_{\max}, 0]. 
\label{eq:history_comparability}
\end{equation}
The initial epigenetic density $f_{\mathrm{in}}$ is a probability density on $\Omega$, satisfying
\begin{equation}
f_{\mathrm{in}}(\vec{x})\geq0, \quad \text{a.e. in } \Omega, \qquad \int_\Omega f_{\mathrm{in}}(\vec{x})\,d\vec{x}=1.
\label{eq:normalized_initial_density_analysis}
\end{equation}
\end{enumerate}

For later use, define the survival factor
\begin{equation}
\sigma(\vec{x}) := \exp\bigl[-\mu(\vec{x})\tau(\vec{x})\bigr].
\label{eq:survival_factor}
\end{equation}
Since $\mu$ and $\tau$ are nonnegative, we have
\begin{equation}
0<\sigma(\vec{x})\leq1.
\label{eq:survival_factor_bound}
\end{equation}

To specify the dependence on the population $Q(\vec{x}, t)$, we define the renewal gain operator as
\begin{equation}
\label{eq:renewal_gain_operator}
\begin{aligned}
\mathcal{K}[Q](\vec{x},t) ={}& 2\int_\Omega M[Q_{\tau(\vec{y})}](\vec{y},t) p(\vec{x},\vec{y})\,d\vec{y},\\
M[Q_{\tau(\vec{y})}](\vec{y},t) ={}& \beta\bigl(c[Q_{\tau(\vec{y})}],\vec{y}\bigr) Q_{\tau(\vec{y})} \sigma(\vec{y}),\qquad Q_{\tau(\vec{y})} = Q(\vec{y}, t - \tau(\vec{y})).
\end{aligned}
\end{equation}
Thus, equation \eqref{eq:macro_Q} can be written as
\begin{equation}
\frac{\partial Q}{\partial t} = -r[Q](\vec{x},t)Q+\mathcal{K}[Q], \qquad r[Q](\vec{x},t) = \beta\bigl(c[Q](t),\vec{x}\bigr)+\kappa(\vec{x}).
\label{eq:abstract_Q_equation}
\end{equation}

\subsection{Well-posedness of the population-renewal equation}
\label{sec:global_wellposedness}

We first establish global existence, uniqueness, and positivity for the population-renewal equation.

\begin{theorem}[Global well-posedness of the renewal equation]
\label{thm:Q_global_wellposedness}
Assume {\rm(A3)--(A7)}. Then, for every $T>0$, equation \eqref{eq:macro_Q} admits a unique mild solution
\begin{equation}
Q\in C\bigl([-\tau_{\max},T];L^1(\Omega)\bigr),
\label{eq:Q_global_solution_space}
\end{equation}
which agrees with the prescribed history on $[-\tau_{\max},0]$. Furthermore, the solution depends continuously on the initial history $Q_{\mathrm{in}}$. If the initial history is nonnegative, then
\begin{equation}
Q(\cdot,t)\geq0
\quad\text{almost everywhere in }\Omega, \qquad t\in[0,T].
\label{eq:Q_global_nonnegative}
\end{equation}
Moreover, the total cell number satisfies
\begin{equation}
Q_{\mathrm{tot}}(t) \leq \left(Q_{\mathrm{tot}}(0) + 2 C_{\mathrm{hist}}\right)e^{2 \beta_{\max} t}, \qquad t\in[0,T],
\label{eq:global_L1_bound}
\end{equation}
where
$$
C_{\mathrm{hist}} = \int_\Omega \int_{-\tau(\vec{y})}^0 \beta(c[Q_{\mathrm{in}}](s), \vec{y}) Q_{\mathrm{in}}(\vec{y}, s)\, ds\,d\vec{y} \geq 0.
$$ 
In particular, the solution exists globally in time in $L^1(\Omega)$.
\end{theorem}

\begin{proof}
Since $r[Q](\vec{x}, t)\geq0$, the mild solution, i.e.\ the continuous function satisfying the variation-of-constants (integral) form of the equation, is, for $t\in[0, \tau_{\min}]$, given by
\begin{equation}
\begin{aligned}
Q(\vec{x},t) ={}& Q_{\mathrm{in}}(\vec{x},0) \exp\left( -\int_0^t r[Q](\vec{x},s)\,ds \right)\\
&+ \int_0^t \exp\left( -\int_s^t r[Q](\vec{x},\sigma)\,d\sigma \right) \mathcal{K}[Q](\vec{x},s)\,ds.
\end{aligned}
\label{eq:Q_variation_constants}
\end{equation}
For $t\in[0, \tau_{\min}]$, all delayed values appearing in $\mathcal{K}[Q](\vec{x}, t)$ are determined by the prescribed initial history. Under assumptions {\rm(A3)--(A7)} the right-hand side of \eqref{eq:Q_variation_constants} defines a locally Lipschitz operator on $C([0, \tau_{\min}]; L^1(\Omega))$. Hence, for $0 < T_1 < \tau_{\min}$ sufficiently small, the standard Picard fixed-point argument yields a unique mild solution on $[0, T_1]$, with continuous dependence on $Q_{\mathrm{in}}$ intrinsically guaranteed by the contraction mapping principle. The nonnegativity of $Q$ follows directly from \eqref{eq:Q_variation_constants}, since the initial history, the renewal term, and the exponential factors are nonnegative. 

The solution can then be continued successively by the method of steps. More precisely, once the solution has been constructed up to some time $T$, the delayed terms on a sufficiently short subsequent interval are determined by the already constructed history. Applying the same local fixed-point argument on each such interval propagates this local well-posedness onto a maximal interval of existence $[0, T_{\max})$. We will later show via a priori estimates that $T_{\max} = \infty$, thereby ensuring the global existence of the solution.

It remains to derive an a priori estimate in $L^1(\Omega)$. Since $Q\geq 0$, 
$$
Q_{\mathrm{tot}}(t) = \int_{\Omega} Q(\vec{x}, t) d\, \vec{x} = \|Q(\cdot,t)\|_{L^1(\Omega)}.
$$
Integrating \eqref{eq:macro_Q} over $\Omega$ gives, using Tonelli's theorem and the normalization of the inheritance kernel \eqref{eq:kernel_normalization_analysis}, for almost every $t \in [0, T_{\max})$, 
\begin{equation}
\frac{d Q_{\mathrm{tot}}}{dt} = -\int_\Omega \bigl[\beta(c[Q](t), \vec{x}) + \kappa(\vec{x})\bigr] Q(\vec{x},t)\,d\vec{x} + 2\int_\Omega M(\vec{y}, t)\, d\vec{y}.
\label{eq:total_mass_balance_equation}
\end{equation}

Integrating \eqref{eq:total_mass_balance_equation} from $0$ to $t$, we obtain
\begin{equation}
Q_{\mathrm{tot}}(t) = Q_{\mathrm{tot}}(0) -\int_0^t\!\!\int_\Omega \bigl[\beta(c[Q](s),\vec{x})+\kappa(\vec{x})\bigr] Q(\vec{x},s)\,d\vec{x}\,ds + 2\int_0^t\!\!\int_\Omega M(\vec{y},s)\,d\vec{y}\,ds.
\label{eq:integrated_total_mass_balance}
\end{equation}

By conservation of the age-structured cycling compartment, the cumulative mitotic output over $[0,t]$ is bounded by the cumulative entry flux, up to a possible contribution from cells already in the cycling compartment at $t=0$. Thus,
\begin{equation}
\int_0^t\!\!\int_\Omega M(\vec{y},s)\,d\vec{y}\,ds \leq C_{\mathrm{hist}} +\int_0^t\!\!\int_\Omega \beta(c[Q](s),\vec{y})Q(\vec{y},s)\,d\vec{y}\,ds,
\label{eq:cumulative_mitotic_flux_bound}
\end{equation}
where 
$$C_{\mathrm{hist}} = \int_\Omega \int_{-\tau(\vec{y})}^0 \beta(c[Q_{\mathrm{in}}](s), \vec{y}) Q_{\mathrm{in}}(\vec{y}, s)\, ds\,d\vec{y} \geq 0$$ 
denotes the finite cumulative contribution from the initial cycling population or, equivalently, from the prescribed prehistory. 

Since $0\leq\beta(c,\vec{x})\leq\beta_{\max}$, it follows that
$$
\int_0^t\!\!\int_\Omega \beta(c[Q](s),\vec{y})Q(\vec{y},s)\,d\vec{y}\,ds \leq \beta_{\max}\int_0^t Q_{\mathrm{tot}}(s)\,ds.
$$
Therefore, after discarding the nonpositive loss term in
\eqref{eq:integrated_total_mass_balance},
\begin{equation}
Q_{\mathrm{tot}}(t) \leq Q_{\mathrm{tot}}(0)+2C_{\mathrm{hist}} +2\beta_{\max}\int_0^t Q_{\mathrm{tot}}(s)\,ds.
\label{eq:total_mass_integral_inequality}
\end{equation}
Gronwall's inequality yields
\begin{equation}
Q_{\mathrm{tot}}(t) \leq \bigl(Q_{\mathrm{tot}}(0)+2C_{\mathrm{hist}}\bigr) e^{2\beta_{\max}t},
\label{eq:delayed_mass_gronwall_bound}
\end{equation}
for every \(t\in[0,T_{\max})\). Consequently,
\begin{equation}
\|Q(\cdot,t)\|_{L^1(\Omega)} = Q_{\mathrm{tot}}(t) \leq \bigl(Q_{\mathrm{tot}}(0)+2C_{\mathrm{hist}}\bigr) e^{2\beta_{\max}t}, \qquad t<T_{\max}.
\label{eq:final_global_L1_bound}
\end{equation}

The local continuation criterion states that if
$T_{\max}<\infty$, then
$$
\limsup_{t\uparrow T_{\max}} \|Q(\cdot,t)\|_{L^1(\Omega)} = \infty.
$$
However, \eqref{eq:final_global_L1_bound} gives the uniform bound
$$
\sup_{0\leq t<T_{\max}} \|Q(\cdot,t)\|_{L^1(\Omega)} \leq \bigl(Q_{\mathrm{tot}}(0)+2C_{\mathrm{hist}}\bigr)e^{2\beta_{\max}T_{\max}} < \infty.
$$
This contradiction proves that $T_{\max}=\infty$. Hence the mild solution is global.
\end{proof}

\begin{remark}[Meaning of global well-posedness]
\label{rem:L1_global_wellposedness}
The global well-posedness established above is understood in the $L^1(\Omega)$ sense. In particular, the estimate \eqref{eq:final_global_L1_bound} excludes finite-time blow-up of the total cell number. It does not, by itself, imply a uniform $L^\infty(\Omega)$ or pointwise bound for $Q$. Thus, possible singular behavior of a representative of $Q(\vec{x},t)$ on sets of measure zero is not detected by the $L^1$ theory and does not constitute finite-time blow-up in the phase space $C([0,\infty); L^1(\Omega))$.
\end{remark}

\subsection{Well-posedness of the epigenetic probability density equation}
\label{sec:local_wellposedness}

We now address the time-dependent epigenetic probability density equation \eqref{mod1}. Throughout this subsection, $Q(\vec{x},t)$ denotes the global mild solution of the renewal equation \eqref{eq:macro_Q} furnished by Theorem~\ref{thm:Q_global_wellposedness}. Consequently, the coefficients entering $\mathcal{R}[f; Q_{\mathrm{tot}}]$ are determined a priori by the known population $Q$, and \eqref{mod1} is treated as an evolution equation for $f$ alone with time-dependent, population-fixed coefficients.

Let $\mathcal{L}^\ast$ denote the no-flux realization on $L^1(\Omega)$ of the Fokker--Planck operator
\begin{equation}
\mathcal{L}^\ast f=\nabla\cdot\bigl(\tilde{\vec{D}}\nabla f\bigr)-\nabla\cdot\bigl(\tilde{\vec{F}}f\bigr),
\label{eq:FP_operator_analysis}
\end{equation}
subject to the boundary condition \eqref{eq:zero_flux}. Under {\rm(A1)--(A2)}, $\mathcal{L}^\ast$ generates a positive $C_0$-contraction semigroup $\{e^{t\mathcal{L}^\ast}\}_{t\geq0}$ on $L^1(\Omega)$, which is a standard consequence of the uniform ellipticity and regularity of the coefficients \cite{Bogachev:2015aa}.

To expose the structure of the renewal operator, we rewrite $\mathcal{R}[f; Q_{\mathrm{tot}}]$ in terms of the population-determined birth rate
\begin{equation}
A(t)=\frac{2}{Q_{\mathrm{tot}}(t)}\int_\Omega M(\vec{y},t)\,d\vec{y},
\label{eq:f_coefficients}
\end{equation}
and the population-determined source
\begin{equation}
S(\vec{x},t)=\frac{2}{Q_{\mathrm{tot}}(t)}\int_\Omega M(\vec{y},t)\,p(\vec{x},\vec{y})\,d\vec{y}.
\label{eq:f_source}
\end{equation}
Here $M(\vec{y},t)$ is the mitotic completion flux determined by the delayed history of $Q$ (see \eqref{eq:renewal_gain_operator}). With these definitions, the renewal operator in \eqref{eq:evl_f} takes the semilinear form
\begin{equation}
\mathcal{R}[f;Q_{\mathrm{tot}}](\vec{x},t)=S(\vec{x},t)-\bigl[A(t)+r(\vec{x},t)\bigr]f(\vec{x},t)+\bar{r}_f(t)\,f(\vec{x},t),
\label{eq:renewal_rewritten}
\end{equation}
so that the epigenetic probability density equation \eqref{mod1} reads as the abstract semilinear evolution equation
\begin{equation}
\frac{\partial f}{\partial t}=\mathcal{L}^\ast f+S(\cdot,t)-\bigl[A(t)+r(\cdot,t)\bigr]f+\bar{r}_f(t)\,f,\qquad f(\cdot,0)=f_{\mathrm{in}}.
\label{eq:abstract_f_equation}
\end{equation}

To formally establish the well-posedness, we need to control the time-dependent coefficients driving \eqref{eq:abstract_f_equation}. First, we establish an upper bound for the delayed population mass.

\begin{lemma}
\label{lem:delay_mass_estimate}
Assume {\rm(A1)--(A7)}, and let $Q$ be the global mild solution of
\eqref{eq:macro_Q}. Set
$$
r_{\max}:=\beta_{\max}+\kappa_{\max}.
$$
Then, for almost every $\vec{x}\in\Omega$ and all $t\geq0$,
\begin{equation}
Q\bigl(\vec{x},t-\tau(\vec{x})\bigr) \leq C_{\tau}Q(\vec{x},t),
\label{eq:pointwise_delay_comparison}
\end{equation}
where
\begin{equation}
C_{\tau} := C_0 e^{r_{\max}\tau_{\max}}.
\label{eq:C_tau}
\end{equation}
Consequently, the delayed mass satisfies
\begin{equation}
\int_{\Omega}
Q\bigl(\vec{x},t-\tau(\vec{x})\bigr)\,d\vec{x} \leq C_{\tau}Q_{\mathrm{tot}}(t),
\qquad t\geq0.
\label{eq:delay_mass_bound}
\end{equation}
\end{lemma}

\begin{proof}
Since the renewal term in \eqref{eq:macro_Q} is nonnegative, and the total loss rate is bounded above by $r_{\max}$, the mild representation of $Q$ yields the comparison estimate
\begin{equation}
Q(\vec{x},t) \geq e^{-r_{\max}(t-s)}Q(\vec{x},s), \qquad 0\leq s\leq t.
\label{eq:Q_lower_comparison}
\end{equation}

First, suppose that $t-\tau(\vec{x})\geq0$. Taking $s=t-\tau(\vec{x})$ in \eqref{eq:Q_lower_comparison} gives
$$
Q\bigl(\vec{x},t-\tau(\vec{x})\bigr) \leq e^{r_{\max}\tau(\vec{x})}Q(\vec{x},t) \leq e^{r_{\max}\tau_{\max}}Q(\vec{x},t).
$$

Next, suppose that $t-\tau(\vec{x})<0$. In this case, \eqref{eq:history_comparability} in {\rm (A7)} yields
$$
Q\bigl(\vec{x},t-\tau(\vec{x})\bigr) = Q_{\mathrm{in}}\bigl(\vec{x},t-\tau(\vec{x})\bigr) \leq C_0 Q_{\mathrm{in}}(\vec{x}, 0).
$$
On the other hand, applying \eqref{eq:Q_lower_comparison} with $s=0$ yields
$$
Q_{\mathrm{in}}(\vec{x}, 0) \leq e^{r_{\max}t} Q(\vec{x}, t) \leq e^{r_{\max}\tau_{\max}} Q(\vec{x}, t).
$$
Combining these two inequalities gives
$$
Q\bigl(\vec{x}, t-\tau(\vec{x})\bigr) \leq C_0 e^{r_{\max}\tau_{\max}}Q(\vec{x},t) = C_{\tau} Q(\vec{x}, t).
$$
Thus \eqref{eq:pointwise_delay_comparison} holds for all $t \geq 0$ and almost every $\vec{x} \in \Omega$. Integrating \eqref{eq:pointwise_delay_comparison} over $\Omega$ gives
\eqref{eq:delay_mass_bound}.
\end{proof}

\begin{theorem}[Local well-posedness of the epigenetic probability density equation]
\label{thm:f_local_wellposedness}
Assume {\rm(A1)--(A7)}, let $\mathcal{L}^\ast$ be as in \eqref{eq:FP_operator_analysis}, and let $Q$ be the global mild solution of \eqref{eq:macro_Q} given by Theorem~\ref{thm:Q_global_wellposedness}. Let $f_{\mathrm{in}}\in L^1(\Omega)$. Then there exists a time $T_\ast>0$ and a unique mild solution
\begin{equation}
f\in C\bigl([0,T_\ast]; L^1(\Omega)\bigr)
\label{eq:f_local_solution}
\end{equation}
of \eqref{eq:abstract_f_equation} with $f(\cdot,0)=f_{\mathrm{in}}$. 
\end{theorem}

\begin{proof}
By Theorem~\ref{thm:Q_global_wellposedness}, $Q\in C([-\tau_{\max},T];L^1(\Omega))$, which implies that the map $t\mapsto Q_{\mathrm{tot}}(t)=\|Q(\cdot,t)\|_{L^1(\Omega)}$ is continuous. Since $Q_{\mathrm{tot}}(0)>0$ by {\rm(A7)}, there exist a local time $T_0\in(0,T]$ and a constant $\delta>0$ such that
\begin{equation}
Q_{\mathrm{tot}}(t)\geq\delta>0,\qquad t\in[0,T_0].
\label{eq:positive_Qtot_interval}
\end{equation}
This strictly positive lower bound guarantees that the division by $Q_{\mathrm{tot}}(t)$ in the definitions of $S(\cdot, t)$ and $A(t)$ is well defined on $[0, T_0]$.

Applying Lemma~\ref{lem:delay_mass_estimate}, we can obtain a uniform bound for these coefficients that is strictly independent of $\delta$. By {\rm(A3)}, \eqref{eq:delay_mass_bound}, and $e^{-\mu(\vec{y})\tau(\vec{y})} \leq 1$, we have
\begin{align}
\int_{\Omega}M(\vec{y},t)\,d\vec{y} &= \int_{\Omega} \beta\bigl(c(t-\tau(\vec{y})),\vec{y}\bigr) Q\bigl(\vec{y},t-\tau(\vec{y})\bigr) e^{-\mu(\vec{y})\tau(\vec{y})}\,d\vec{y}\nonumber\\
&\leq \beta_{\max} \int_{\Omega} Q\bigl(\vec{y},t-\tau(\vec{y})\bigr)\,d\vec{y} \nonumber\\
&\leq \beta_{\max}C_{\tau}Q_{\mathrm{tot}}(t).
\label{eq:M_total_bound}
\end{align}
Consequently, the normalized birth rate $A(t)$ admits a uniform upper bound:
\begin{equation}
\label{eq:A_bound}
A(t) = \frac{2}{Q_{\mathrm{tot}}}\int_\Omega M(\vec{y}, t)\, d\vec{y} \leq 2 \beta_{\max} C_\tau:=\bar{A},\quad t\geq 0.
\end{equation}
Recalling the normalization condition $\int_\Omega p(\vec{x}, \vec{y})\, d\vec{x} = 1$ from {\rm(A5)}, it follows immediately that 
\begin{equation}
\|S(\cdot,t)\|_{L^1(\Omega)} = A(t) \leq \bar{A},\quad \|r(\cdot, t)\|_{L^\infty(\Omega)} \leq r_{\max},\qquad t \in [0, T_0].
\label{eq:S_A_uniform_bound}
\end{equation}
Thus, the functions $A$, $S$, and $r$ belong to $C([0, T_0]; \mathbb{R})$, $C([0, T_0]; L^1(\Omega))$, and $C([0, T_0]; L^\infty(\Omega))$ respectively, and are uniformly bounded on $[0, T_0]$.

Fix $\rho=\|f_{\mathrm{in}}\|_{L^1}+1$. We seek a final time $T_{\ast}\in (0, T_0]$ such that the closed ball
$$
B_\rho=\Bigl\{f\in C\bigl([0,T_\ast];L^1(\Omega)\bigr):\ \sup_{t\in[0,T_\ast]}\|f(t)\|_{L^1(\Omega)}\leq\rho\Bigr\}
$$
is invariant under the mapping
\begin{equation}
\Phi[f](t)=e^{t\mathcal{L}^\ast}f_{\mathrm{in}} + \int_0^t e^{(t-s)\mathcal{L}^\ast}\,\mathcal{G}(f)(\cdot,s)\,ds,
\label{eq:f_fixed_point_map}
\end{equation}
where
\begin{equation}
\mathcal{G}(f)(\cdot,s)=S(\cdot,s) - \bigl[A(s)+r(\cdot,s)\bigr]f(\cdot,s)+\bar{r}_f(s)\,f(\cdot,s).
\label{eq:f_nonlinearity}
\end{equation}

We verify the local Lipschitz property of $\mathcal{G}$. Let $f_1, f_2\in L^1(\Omega)$ with $\|f_i\|_{L^1}\leq\rho$. Because
$$
\bar{r}_{f_1}f_1-\bar{r}_{f_2}f_2 = \bar{r}_{f_1}(f_1-f_2) + \bigl(\bar{r}_{f_1} - \bar{r}_{f_2}\bigr)f_2,
$$
and
$$
\bigl|\bar{r}_{f_1}-\bar{r}_{f_2}\bigr|\leq\|r(\cdot,s)\|_{L^\infty}\|f_1-f_2\|_{L^1},\qquad
\bigl|\bar{r}_{f_1}\bigr|\leq\|r(\cdot,s)\|_{L^\infty}\rho,
$$
we obtain
\begin{equation}
\|\mathcal{G}(f_1)-\mathcal{G}(f_2)\|_{L^1}\leq L_\rho\,\|f_1-f_2\|_{L^1},
\label{eq:f_lipschitz}
\end{equation}
with $L_\rho = \bar{A} + r_{\max}(1+2\rho)$, which is independent of $s$.

Next, we show that $\Phi(B_\rho)\subseteq B_\rho$ for $T_\ast$ sufficiently small. Using $A(s)\leq \bar{A}$ and $\bar{r}_f(s)\leq r_{\max}\rho$ for $f\in B_\rho$, it follows that
$$
\|\Phi[f](t)\|_{L^1}\leq\|f_{\mathrm{in}}\|_{L^1}+\int_0^t\|\mathcal{G}(f)(s)\|_{L^1}\,ds
\leq\|f_{\mathrm{in}}\|_{L^1}+T_\ast\Bigl[\bar{A} + \bigl(\bar{A} + r_{\max}\bigr)\rho + r_{\max}\rho^2\Bigr].
$$
Since $\rho > \|f_{\mathrm{in}}\|_{L^1}$, choosing $T_\ast>0$ sufficiently small fulfills the invariance condition.

Finally, we establish the contraction property. Equip $C([0,T_\ast];L^1(\Omega))$ with the equivalent norm
$$
\|f\|_{\lambda}=\sup_{t\in[0,T_\ast]}e^{-\lambda t}\|f(t)\|_{L^1},\qquad \lambda>0.
$$
The operator $\mathcal{L}^\ast$ acts as a conservative Fokker--Planck operator under no-flux boundary conditions. By standard parabolic theory, it generates a strongly continuous Markovian semigroup, ensuring $e^{t \mathcal{L}^\ast}$ is a contraction on $L^1(\Omega)$, meaning $\|e^{t\mathcal{L}^\ast}f\|_{L^1}\leq\|f\|_{L^1}$. Utilizing this contractivity along with \eqref{eq:f_lipschitz}, we have, for $f, g\in B_\rho$,
$$
\|\Phi[f](t)-\Phi[g](t)\|_{L^1}\leq\int_0^t L_\rho\|f(s)-g(s)\|_{L^1}\,ds
\leq L_\rho\int_0^t e^{\lambda s}\,ds\,\|f-g\|_\lambda
\leq\frac{L_\rho e^{\lambda t}}{\lambda}\|f-g\|_\lambda,
$$
whence
\begin{equation}
\|\Phi[f]-\Phi[g]\|_{\lambda}\leq\frac{L_\rho}{\lambda}\|f-g\|_\lambda.
\label{eq:f_contraction}
\end{equation}
Taking $\lambda>L_\rho$ ensures that $\Phi$ is a strict contraction on $B_\rho$ under the $\lambda$-norm. By Banach's fixed-point theorem, $\Phi$ admits a unique fixed point $f\in B_\rho$, yielding the unique mild solution of \eqref{eq:abstract_f_equation} on $[0, T_\ast]$.
\end{proof}

Define the probability simplex on $\Omega$ as
\begin{equation}
\label{eq:probability_simplex}
\mathcal{P}(\Omega) = \left\{g\in L^1(\Omega): g\geq0\ \text{a.e.},\ \int_\Omega g(\vec{x})\,d\vec{x}=1 \right\}.
\end{equation}

Before establishing global well-posedness, we first demonstrate a critical \emph{a priori} property of the system: as long as a local solution exists, it strictly preserves the probability simplex.

\begin{proposition}[Invariance of the probability simplex]
\label{prop:simplex_invariance}
Assume {\rm(A1)--(A7)}, and let $f$ be the mild solution of \eqref{mod1} on its maximal interval of local existence $[0, T_{\max})$ given by Theorem~\ref{thm:f_local_wellposedness}. If $f_{\mathrm{in}} \in \mathcal{P}(\Omega)$, then
$$
f(\cdot,t)\in\mathcal{P}(\Omega), \qquad t\in [0, T_{\max}).
$$
\end{proposition}

\begin{proof}
We first prove the nonnegativity of $f$. Fix any $T \in (0, T_{\max})$. Since $f\in C([0, T]; L^1(\Omega))$, the map $t\mapsto \bar{r}_f(t)$ is continuous. Using the bounds on $r$ and $A$ from \eqref{eq:A_bound} and \eqref{eq:S_A_uniform_bound}, we have
$$
|\bar{r}_f(t)| \leq \| r(\cdot, t)\|_{L^\infty(\Omega)} \|f(\cdot, t)\|_{L^1(\Omega)} < \infty,\qquad t\in [0, T].
$$
Consequently, $q(\vec{x}, t) := A(t) + r(\vec{x}, t) - \bar{r}_f(t)$ belongs to $L^\infty(\Omega)$ for each $t\in [0, T]$, and
$$
\alpha(t) := \|q(\cdot, t)\|_{L^\infty(\Omega)}
$$
is bounded on $[0, T]$. In particular, $\alpha(t) - q(\vec{x}, t)\geq 0$ for almost every $(\vec{x}, t)\in \Omega\times[0, T]$.

Using the definition of $\mathcal{R}[f; Q_{\mathrm{tot}}]$, equation \eqref{mod1} can be rearranged as
\begin{equation}
\partial_t f = \bigl(\mathcal{L}^\ast-\alpha(t)I\bigr)f + \bigl[\alpha(t)-q(\cdot,t)\bigr]f + S(\cdot,t).
\label{eq:f_positivity_decomposition}
\end{equation}
By the nonnegativity of $M$ and the inheritance kernel $p$, together with $Q_{\mathrm{tot}}(t)>0$, we have
$$
S(\vec{x},t) = \frac{2}{Q_{\mathrm{tot}}(t)} \int_\Omega M(\vec{y},t)p(\vec{x},\vec{y})\,d\vec{y} \geq 0.
$$
Define the evolution operator
$$
U_\alpha(t,s) = \exp\left(-\int_s^t\alpha(\sigma)\,d\sigma\right) e^{(t-s)\mathcal{L}^\ast}, \qquad 0\leq s\leq t\leq T.
$$
Since the Fokker--Planck semigroup $e^{t\mathcal{L}^\ast}$ preserves positivity and the scalar exponential prefactor is strictly positive, $U_\alpha(t,s)$ is also positivity preserving on $L^1(\Omega)$. The mild formulation associated with \eqref{eq:f_positivity_decomposition} reads
\begin{equation}
f(t) = U_\alpha(t,0)f_{\mathrm{in}} + \int_0^t U_\alpha(t,s) \left(\bigl[\alpha(s)-q(\cdot,s)\bigr]f(s) + S(\cdot,s) \right)ds.
\label{eq:f_positive_duhamel}
\end{equation}

The bounded multiplication operator $h\mapsto \bigl[\alpha(t)-q(\cdot,t)\bigr]h$ is positive, and $S(\cdot,t)\geq 0$. Therefore, considering the Picard iterates for \eqref{eq:f_positive_duhamel} initialized at
$$
f^{(0)}(t) = U_\alpha(t,0)f_{\mathrm{in}} + \int_0^t U_\alpha(t,s)S(\cdot,s)\,ds \geq 0,
$$
an induction argument shows that $f^{(k)}(t) \geq 0$ almost everywhere for each $k \geq 0$. Since $f^{(k)}$ converges to the unique mild solution $f$ in $C([0,T]; L^1(\Omega))$, positivity is preserved in the limit:
$$
f(\cdot,t) \geq 0 \quad\text{a.e. in }\Omega,\qquad t\in[0,T].
$$

Next, we establish the conservation of total probability. Set 
$$m(t) := \int_\Omega f(\vec{x},t)\,d\vec{x}.$$ 
Because $\mathcal{L}^\ast$ is conservative under the no-flux boundary conditions, we have $\int_\Omega \mathcal{L}^\ast h\,d\vec{x} = 0$ for any $h$ in the domain of the operator, and thus $\int_\Omega e^{t\mathcal{L}^\ast}h\,d\vec{x} = \int_\Omega h\,d\vec{x}$. Integrating the mild formula \eqref{eq:f_fixed_point_map} over $\Omega$ shows that $m(t)$ is differentiable on $[0, T_{\max})$. Integrating \eqref{mod1} directly over $\Omega$ and using the normalization $\int_\Omega p(\vec{x},\vec{y})\,d\vec{x} = 1$ yields
$$
\int_\Omega S(\vec{x},t)\,d\vec{x} = A(t), \qquad \int_\Omega r(\vec{x},t)f(\vec{x},t)\,d\vec{x} = \bar{r}_f(t).
$$
Consequently,
\begin{align}
\frac{dm}{dt} &= \int_\Omega S(\vec{x},t)\,d\vec{x} - A(t)\int_\Omega f(\vec{x},t)\,d\vec{x} - \int_\Omega r(\vec{x},t)f(\vec{x},t)\,d\vec{x} + \bar{r}_f(t)\int_\Omega f(\vec{x},t)\,d\vec{x} \nonumber\\
&= A(t)\bigl(1 - m(t)\bigr) - \bar{r}_f(t)\bigl(1 - m(t)\bigr) \nonumber\\
&= \bigl(A(t) - \bar{r}_f(t)\bigr)\bigl(1 - m(t)\bigr), \qquad t\in [0, T_{\max}).
\label{eq:mass_ode}
\end{align}
Since $f_{\mathrm{in}} \in \mathcal{P}(\Omega)$, we have $m(0) = 1$. The unique solution to this initial value problem is $m(t) \equiv 1$ for all $t\in [0, T_{\max})$. Combining this with the nonnegativity of $f$ demonstrates that $f(\cdot,t) \in \mathcal{P}(\Omega)$ for all $t\in [0, T_{\max})$.
\end{proof}

\begin{remark}[Autonomous probability conservation versus phenomenological definitions]
A key theoretical advantage of this framework is the unconditional preservation of probability normalization. In classical population balance formulations~\cite{weinreb2018,shi2022energy}, Fokker--Planck dynamics are typically coupled with a phenomenological growth term $R(\vec{x})f(\vec{x}, t)$. However, this structure inherently violates the conservative property of the Fokker--Planck equation (see \emph{SI Appendix} C.2), precluding its fundamental validity for tracking non-equilibrium transient dynamics. To salvage probability normalization, prior models were forced to impose artificial global constraints (e.g., $\int_\Omega R(\vec{x}) f_{\mathrm{ss}}(\vec{x})\,d\vec{x} = 0$) that hold only at steady state. By contrast, because our population renewal operator $\mathcal{R}[f; Q_{\mathrm{tot}}]$ is rigorously derived across scales from the underlying mass-conserving flux \eqref{eq:macro_Q}, its integral over $\Omega$ satisfies
$$
\int_\Omega \mathcal{R}[f; Q_{\mathrm{tot}}](\vec{x},t)\,d\vec{x} = \bigl(A(t) - \bar{r}_f(t)\bigr)\left(1 - \int_\Omega f(\vec{x}, t)\,d\vec{x}\right),
$$
which vanishes whenever $\int_{\Omega} f(\vec{x}, t)\, d\vec{x} = 1$. This invariant property ensures that the coupled formulation autonomously guarantees probability conservation across any dynamic, non-equilibrium trajectory.
\end{remark}

\begin{theorem}[Global well-posedness under bounded state-dependent delay]
\label{thm:f_global_wellposedness}
Assume {\rm(A1)--(A7)}, and let $Q$ be the global mild solution of \eqref{eq:macro_Q} given by Theorem~\ref{thm:Q_global_wellposedness}. If $f_{\mathrm{in}} \in \mathcal{P}(\Omega)$, then equation \eqref{mod1} admits a unique global mild solution
\begin{equation}
f\in C\bigl([0,\infty);L^1(\Omega)\bigr)
\label{eq:f_global_solution}
\end{equation}
with $f(\cdot,0)=f_{\mathrm{in}}$, and $f(\cdot,t)\in\mathcal{P}(\Omega)$ for every $t\geq0$.
\end{theorem}

\begin{proof}
Set $r_{\max}:=\beta_{\max}+\kappa_{\max}$. By the total mass balance \eqref{eq:total_mass_balance_equation}, the nonnegativity of the renewal term, and the bounds $\beta\leq\beta_{\max}$, $\kappa\leq\kappa_{\max}$, we obtain
$$
\frac{dQ_{\mathrm{tot}}}{dt} \geq -r_{\max}Q_{\mathrm{tot}},
$$
which implies $Q_{\mathrm{tot}}(t) \geq Q_{\mathrm{tot}}(0)e^{-r_{\max}t} >0$ for all $t\geq0$. Therefore, all normalized coefficients appearing in \eqref{mod1} are well defined globally.

As established in the proof of Theorem~\ref{thm:f_local_wellposedness}, the functions $A(t)$, $S(\cdot, t)$, and $r(\cdot, t)$ are uniformly bounded on any finite time interval $[0, T]$.

By Proposition~\ref{prop:simplex_invariance}, the $L^1$-norm of the solution is strictly conserved, with $\|f(\cdot,t)\|_{L^1(\Omega)} \equiv 1$. Because the coefficient bounds \eqref{eq:A_bound}--\eqref{eq:S_A_uniform_bound} are uniform and independent of $f$, the maximal length of the local existence interval at any starting time, as provided by Theorem~\ref{thm:f_local_wellposedness}, is bounded uniformly away from zero. A standard continuation argument then extends the solution to the entire interval $[0,\infty)$ without finite-time blow-up. The uniqueness follows directly by successively applying the local uniqueness property over natural consecutive intervals.
\end{proof}

\begin{remark}[Biological relevance of the initial history condition]
The history compatibility condition \eqref{eq:history_comparability} in {\rm (A7)} is naturally satisfied in biologically meaningful contexts. It ensures that the population density does not exhibit singular, unbounded fluctuations within the finite observational memory window $[-\tau_{\max}, 0]$. In experimental setups or generic physiological processes, initial cell distributions typically emerge from continuous environments. For example, if the system is identically stationary in the past (i.e., $Q_{\mathrm{in}}(\vec{x},s) = Q_{\mathrm{in}}(\vec{x},0)$ for all $s \in [-\tau_{\max}, 0]$), the condition holds trivially with $C_0 = 1$. More generally, any continuous and strictly positive initial distribution over a compact epigenetic space automatically guarantees the existence of such a finite constant.
\end{remark}

\begin{remark}[Well-definedness of the epigenetic landscape]
Since the obtained probability density satisfies $f(\vec{x},t) > 0$ almost everywhere, the epigenetic landscape defined by $U(\vec{x},t):= - \log f(\vec{x}, t)$ is well defined on $\Omega$ almost everywhere. For any null set where $f$ vanishes, one may extend the definition by setting $U = \infty$. This convention is biologically intuitive as it represents states with infinite potential energy barriers that are impossible to occupy. Moreover, this extension does not affect any subsequent integral-based analyses or the broader dynamic formulation.
\end{remark}
   
\subsection{Equilibria of the coupled system}
\label{sec:equilibria}

We next analyze the equilibria of the coupled system. In contrast to
the time-dependent normalization used in deriving the equation for
$f$, the stationary densities $Q^\ast$ and $f^\ast$ are treated as
distinct state variables. Accordingly, a nontrivial equilibrium is
defined as a triple $(Q^\ast,f^\ast,c^\ast)$ such that
$$
Q^\ast\geq0,\quad c^\ast = \int_\Omega \xi(\vec{x})Q^\ast(\vec{x})\,d\vec{x},\quad f^\ast\in\mathcal{P}(\Omega),
$$
and 
$$Q_{\mathrm{tot}}^\ast = \int_\Omega Q^\ast(\vec{x})\,d\vec{x}>0.$$

\subsubsection{Stationary population-renewal profile}
\label{sec:stationary_population_profile}

At equilibrium, the population density becomes time-independent, and the delayed state coincides with the stationary state. The mitotic flux therefore reduces to 
\begin{equation}
M^\ast(\vec{y}) = \beta(c^\ast,\vec{y}) Q^\ast(\vec{y}) e^{-\mu(\vec{y})\tau(\vec{y})}.
\label{eq:stationary_mitotic_flux}
\end{equation}
Consequently, the stationary population-renewal equation reads
\begin{equation}
\label{eq:stationary_population_equation}
\begin{aligned}
0={}& -\bigl[\beta(c^\ast,\vec{x})+\kappa(\vec{x})\bigr] Q^\ast(\vec{x})\\
&+ 2\int_\Omega \beta(c^\ast,\vec{y}) e^{-\mu(\vec{y})\tau(\vec{y})} Q^\ast(\vec{y}) p(\vec{x},\vec{y})\,d\vec{y}.
\end{aligned}
\end{equation}

\begin{lemma}[Positivity and compactness of the renewal operator]
\label{lem:compact_positive_Tc}
Fix $c\geq 0$ and define the integral operator
\begin{equation}
\label{eq:stationary_population_operator}
(\mathcal{T}_c\varphi)(\vec{x}) = \frac{2}{\beta(c,\vec{x})+\kappa(\vec{x})} \int_\Omega \beta(c,\vec{y}) e^{-\mu(\vec{y})\tau(\vec{y})} p(\vec{x},\vec{y})\varphi(\vec{y})\,d\vec{y}, \qquad \varphi\in L^1(\Omega).
\end{equation}
Assume that $\Omega$ is bounded, $\beta(c,\vec{x})+\kappa(\vec{x})>0$ for a.e. $\vec{x}\in\Omega$, and the kernel
$$
K_c(\vec{x},\vec{y}) := \frac{2\beta(c,\vec{y}) e^{-\mu(\vec{y})\tau(\vec{y})}p(\vec{x},\vec{y})} {\beta(c,\vec{x})+\kappa(\vec{x})}
$$
is continuous on $\bar{\Omega}\times\bar{\Omega}$ and bounded there. Then $\mathcal{T}_c$ is a positive compact operator on $L^1(\Omega)$.

If, in addition,
\begin{equation}
K_c(\vec{x},\vec{y})>0\qquad \text{for almost every } (\vec{x},\vec{y})\in\Omega\times\Omega,
\label{eq:strict_positivity_Kc}
\end{equation}
then $\mathcal{T}_c$ is irreducible on $L^1(\Omega)$.
\end{lemma}

\begin{proof}
The positivity is immediate. If $\varphi\geq 0$ almost everywhere, all factors in the definition of $\mathcal{T}_c$ are nonnegative, implying
$$
(\mathcal{T}_c\varphi)(\vec{x})\geq 0 \quad\text{for a.e. }\vec{x}\in\Omega.
$$

To prove the compactness, note that the assumptions imply $K_c\in C(\bar{\Omega}\times\bar{\Omega})\cap L^\infty(\Omega\times\Omega)$.
Therefore, $\mathcal{T}_c$ is an integral operator with a continuous kernel on the bounded domain $\Omega$. Standard compactness results then imply that $\mathcal{T}_c: L^1(\Omega)\to L^1(\Omega)$ is compact. More precisely, a bounded set in $L^1(\Omega)$ is mapped by $\mathcal{T}_c$ into an equicontinuous and uniformly bounded family in $C(\bar{\Omega})$, which is relatively compact in $L^1(\Omega)$ by the Arzel\`a--Ascoli theorem and the continuous embedding $C(\bar{\Omega})\hookrightarrow L^1(\Omega)$.

It remains to prove irreducibility under \eqref{eq:strict_positivity_Kc}. Let $\varphi \in L^1(\Omega)$ be a nonnegative function with $\|\varphi\|_{L^1}>0$. The support set
$$
\Omega_\varphi:=\{\vec{y}\in\Omega:\varphi(\vec{y})>0\}
$$
thus has strictly positive Lebesgue measure. By \eqref{eq:strict_positivity_Kc},
$$
K_c(\vec{x},\vec{y})\varphi(\vec{y})>0
$$
for almost every $(\vec{x},\vec{y})\in\Omega\times \Omega_\varphi$. Fubini's theorem implies that, for almost every $\vec{x}\in\Omega$,
$$
(\mathcal{T}_c\varphi)(\vec{x}) = \int_\Omega K_c(\vec{x},\vec{y})\varphi(\vec{y})\,d\vec{y} \geq \int_{\Omega_\varphi} K_c(\vec{x},\vec{y})\varphi(\vec{y})\,d\vec{y} >0.
$$
Thus, $\mathcal{T}_c$ is positivity improving: it maps any non-trivial nonnegative function to a function that is strictly positive almost everywhere on $\Omega$.

To verify irreducibility, suppose that $L^1(A)\subset L^1(\Omega)$ is a nonzero closed ideal invariant under $\mathcal{T}_c$, for some measurable set $A\subset\Omega$. If this ideal is proper (neither $\{0\}$ nor $L^1(\Omega)$), the structural properties of ideals in $L^p$ spaces dictate that both $A$ and its complement $\Omega\setminus A$ must have strictly positive measure. For the indicator function $\boldsymbol{1}_A$, we have
$$
\boldsymbol{1}_A\in L^1(A), \qquad \boldsymbol{1}_A\geq0, \qquad \|\boldsymbol{1}_A\|_{L^1} > 0.
$$
The positivity improving property yields $\mathcal{T}_c\boldsymbol{1}_A>0$ almost everywhere in $\Omega$. In particular, $\mathcal{T}_c\boldsymbol{1}_A>0$ on $\Omega\setminus A$, an event of positive measure. This implies $\mathcal{T}_c\boldsymbol{1}_A \notin L^1(A)$, contradicting the assumed invariance of $L^1(A)$. Hence, $\mathcal{T}_c$ has no nontrivial closed invariant ideals and is therefore irreducible.
\end{proof}

By the definition of $\mathcal{T}_c$ in Lemma~\ref{lem:compact_positive_Tc}, equation \eqref{eq:stationary_population_equation} is equivalent to the eigenvalue problem
\begin{equation}
Q^\ast = \mathcal{T}_{c^\ast}Q^\ast.
\label{eq:stationary_population_eigenproblem}
\end{equation}
Thus, a nonzero stationary cell-number density is a nonnegative eigenfunction of the feedback-dependent renewal operator $\mathcal{T}_{c^\ast}$ associated with the eigenvalue one.

\begin{theorem}[Existence of a self-consistent stationary population]
\label{thm:cytokine_compatible_stationary_population}
Assume that the hypotheses of Lemma~\ref{lem:compact_positive_Tc}, including the strict positivity condition \eqref{eq:strict_positivity_Kc}, hold for every cytokine concentration $c\in[c_0,c_1]$, where $0\leq c_0<c_1$. Assume further that the map $c\longmapsto \rho(\mathcal{T}_c)$ is continuous on $[c_0,c_1]$ and satisfies the crossing condition
\begin{equation}
\rho(\mathcal{T}_{c_0})>1, \qquad \rho(\mathcal{T}_{c_1})<1.
\label{eq:stationary_spectral_crossing}
\end{equation}
Then there exists a critical feedback level $c^\ast\in(c_0,c_1)$ such that
\begin{equation}
\rho(\mathcal{T}_{c^\ast})=1.
\label{eq:stationary_unit_spectral_radius}
\end{equation}
Consequently, there exists a stationary cell-number density $Q^\ast \in L^1(\Omega)$ that is strictly positive almost everywhere, satisfying the stationary population renewal equation
\begin{equation}
\mathcal{T}_{c^\ast}Q^\ast=Q^\ast,
\label{eq:stationary_population_equation_at_cstar}
\end{equation}
as well as the cytokine self-consistency condition
\begin{equation}
c^\ast =  \int_\Omega \xi(\vec{x})Q^\ast(\vec{x})\,d\vec{x}.
\label{eq:equilibrium_cytokine_signal_at_cstar}
\end{equation}
Moreover, the spatial shape of the stationary profile $Q^\ast$ is uniquely determined.
\end{theorem}

\begin{proof}
Define the continuous function
$$
F(c):=\rho(\mathcal{T}_c)-1,\qquad c\in[c_0,c_1].
$$
The crossing condition \eqref{eq:stationary_spectral_crossing} implies $F(c_0)>0$ and $F(c_1)<0$. By the intermediate value theorem, there exists a root $c^\ast\in(c_0,c_1)$ such that $F(c^\ast)=0$, implying $\rho(\mathcal{T}_{c^\ast})=1$.

By Lemma~\ref{lem:compact_positive_Tc}, the operator $\mathcal{T}_{c^\ast}$ is positive, compact, and irreducible on $L^1(\Omega)$. Since its spectral radius $\rho(\mathcal{T}_{c^\ast}) = 1$ is non-zero, the Krein--Rutman theorem implies that $\rho(\mathcal{T}_{c^\ast})$ is a simple principal eigenvalue of $\mathcal{T}_{c^\ast}$. The corresponding eigenfunction $\phi_{c^\ast}\in L^1(\Omega)$ can be chosen to be strictly positive almost everywhere on $\Omega$, satisfying
$$
\mathcal{T}_{c^\ast}\phi_{c^\ast} = \phi_{c^\ast}.
$$

To construct the stationary density $Q^\ast$, we scale this principal eigenfunction to satisfy the cytokine feedback constraint. Since the sensitivity weight function $\xi(\vec{x})$ is non-trivial and nonnegative, and $\phi_{c^\ast}$ is strictly positive, the integral $\int_\Omega \xi(\vec{x})\phi_{c^\ast}(\vec{x})\,d\vec{x}$ is strictly positive. This scalar guarantees that the amplitude term
\begin{equation}
a^\ast = \frac{c^\ast}{\displaystyle \int_\Omega \xi(\vec{x})\phi_{c^\ast}(\vec{x})\,d\vec{x}}
\label{eq:stationary_population_amplitude}
\end{equation}
is well defined and strictly positive (as $c^\ast > c_0 \geq 0$). 

Define the scaled density $Q^\ast := a^\ast\phi_{c^\ast}$, the linearity of $\mathcal{T}_{c^\ast}$ immediately gives
$$
\mathcal{T}_{c^\ast}Q^\ast = a^\ast\mathcal{T}_{c^\ast}\phi_{c^\ast} = a^\ast\phi_{c^\ast} = Q^\ast.
$$
Evaluating the cytokine concentration produced by this density yields
$$
\int_\Omega \xi(\vec{x})Q^\ast(\vec{x})\,d\vec{x} = a^\ast \int_\Omega \xi(\vec{x})\phi_{c^\ast}(\vec{x})\,d\vec{x} = c^\ast.
$$
Thus, $Q^\ast$ self-consistently generates the cytokine concentration $c^\ast$ required for its own stationarity. 

Finally, the uniqueness of the stationary profile (up to a scalar multiple determined by $c^\ast$) follows directly from the simplicity of the principal eigenvalue established by the Krein--Rutman theorem.
\end{proof}

To operationalize Theorem~\ref{thm:cytokine_compatible_stationary_population}, we need to ensure that the spectral radius $\rho(\mathcal{T}_c)$ crosses unity. The following proposition provides explicit, verifiable bounds on $\rho(\mathcal{T}_c)$ to guarantee this crossing without explicitly computing the entire spectrum. 

\begin{proposition}[Explicit sufficient conditions for spectral crossing]
\label{prop:explicit_spectral_crossing}
Assume that the hypotheses of Lemma~\ref{lem:compact_positive_Tc} hold for every $c\geq0$. Suppose that the inheritance kernel is normalized such that
$$
p(\vec{x},\vec{y})\geq0,\qquad \int_\Omega p(\vec{x},\vec{y})\,d\vec{x}=1
$$
for almost every $\vec{y}\in\Omega$. Define the state-dependent reproduction factor
\begin{equation}
G_c(\vec{y}) := 2\beta(c,\vec{y})\sigma(\vec{y})\int_\Omega\frac{p(\vec{x},\vec{y})}{\beta(c, \vec{x}) + \kappa(\vec{x})}\,d\vec{x},\qquad \text{with }\sigma(\vec{x}) = e^{-\mu(\vec{x})\tau(\vec{x})},
\label{eq:state_dependent_reproduction_factor}
\end{equation}
and let
\begin{equation}
L(c) := \operatorname*{ess\,inf}_{\vec{y}\in\Omega}G_c(\vec{y}), \qquad U(c) := \operatorname*{ess\,sup}_{\vec{y}\in\Omega}G_c(\vec{y}).
\label{eq:explicit_spectral_bounds}
\end{equation}
Then we have the spectral bounds
\begin{equation}
L(c)\leq\rho(\mathcal{T}_c)\leq U(c) \qquad\text{for any }c\geq0.
\label{eq:two_sided_spectral_bounds}
\end{equation}
Consequently, if there exist constants $0\leq c_0<c_1$ such that
\begin{equation}
L(c_0)>1,\qquad U(c_1)<1,
\label{eq:explicit_crossing_conditions}
\end{equation}
then the strict endpoint spectral inequalities hold:
\begin{equation}
\rho(\mathcal{T}_{c_0})>1,\qquad \rho(\mathcal{T}_{c_1})<1.
\label{eq:endpoint_spectral_inequalities}
\end{equation}
\end{proposition}

\begin{proof}
Fix $c\geq0$. For any nonnegative function $\varphi\in L^1(\Omega)$, Tonelli's theorem gives
\begin{align*}
\|\mathcal{T}_c\varphi\|_{L^1} & = \int_\Omega\int_\Omega \frac{2\beta(c,\vec{y})\sigma(\vec{y})p(\vec{x},\vec{y})}  {\beta(c, \vec{x}) + \kappa(\vec{x})}\varphi(\vec{y})\,d\vec{y}\,d\vec{x} \\
&= \int_\Omega G_c(\vec{y})\varphi(\vec{y})\,d\vec{y}
\\
&\leq U(c)\|\varphi\|_{L^1}.
\end{align*}
By the positivity of $\mathcal{T}_c$, the inequality $|\mathcal{T}_c\varphi| \leq \mathcal{T}_c|\varphi|$ holds for all $\varphi\in L^1(\Omega)$. It follows that $ \|\mathcal{T}_c\varphi\|_{L^1} \leq U(c)\|\varphi\|_{L^1}$, which implies the operator norm bound $\|\mathcal{T}_c\|_{\mathcal{L}(L^1)}\leq U(c)$. Therefore,
\begin{equation}
\rho(\mathcal{T}_c) \leq \|\mathcal{T}_c\|_{\mathcal{L}(L^1)} \leq U(c).
\label{eq:proof_spectral_upper_bound}
\end{equation}

For the lower bound, consider the adjoint operator $\mathcal{T}_c^\ast: L^\infty(\Omega)\to L^\infty(\Omega)$ defined by
$$
(\mathcal{T}_c^\ast \psi)(\vec{y}) = \int_\Omega K_c(\vec{x},\vec{y})\psi(\vec{x})\,d\vec{x}.
$$
Applying this to the constant function $\boldsymbol{1}$, Fubini's theorem yields
$$
(\mathcal{T}_c^\ast \boldsymbol{1})(\vec{y}) = G_c(\vec{y}).
$$
The definition of $L(c)$ thus guarantees
\begin{equation}
\mathcal{T}_c^\ast \boldsymbol{1} \geq L(c)\boldsymbol{1} \qquad\text{almost everywhere in }\Omega.
\label{eq:adjoint_lower_inequality}
\end{equation}
Since $\mathcal{T}_c^\ast $ is a positive operator, iterating \eqref{eq:adjoint_lower_inequality} produces
$$
(\mathcal{T}_c^\ast )^n\boldsymbol{1} \geq L(c)^n\boldsymbol{1} \qquad (n\geq1).
$$
Noting that $\|\boldsymbol{1}\|_{L^\infty}=1$, we obtain $\|(\mathcal{T}_c^\ast )^n\|_{\mathcal{L}(L^\infty)} \geq L(c)^n$.
The spectral radius formula then implies
$$
\rho(\mathcal{T}_c^\ast ) = \lim_{n\to\infty} \|(\mathcal{T}_c^\ast )^n\|_{\mathcal{L}(L^\infty)}^{1/n} \geq L(c).
$$
Since an operator and its adjoint share the same spectrum, $\rho(\mathcal{T}_c)=\rho(\mathcal{T}_c^\ast )\geq L(c)$. Combining this with \eqref{eq:proof_spectral_upper_bound} verifies \eqref{eq:two_sided_spectral_bounds}. The inequalities in \eqref{eq:endpoint_spectral_inequalities} then follow immediately from condition \eqref{eq:explicit_crossing_conditions}.
\end{proof}

\begin{remark}[Simplified parameter bounds]
\label{rem:simplified_spectral_bounds}
Evaluating $L(c)$ and $U(c)$ directly requires integrating the specific inheritance kernel $p$. We can relax this requirement by uniformly bounding the removal rate. Define
$$
\underline{r}(c) := \operatorname*{ess\,inf}_{\vec{x}\in\Omega} (\beta(c, \vec{x}) + \kappa(\vec{x})), \qquad \bar{r}(c) := \operatorname*{ess\,sup}_{\vec{x}\in\Omega} (\beta(c, \vec{x}) + \kappa(\vec{x})).
$$
If $0<\underline{r}(c)\leq\bar{r}(c)<\infty$, the normalization of $p$ yields
$$
\frac{1}{\bar{r}(c)} \leq \int_\Omega \frac{p(\vec{x},\vec{y})}{\beta(c, \vec{x}) + \kappa(\vec{x})}\,d\vec{x} \leq \frac{1}{\underline{r}(c)}
$$
for almost every $\vec{y}\in\Omega$. Consequently, the spectral bounds can be simplified to:
\begin{eqnarray*}
L(c) \geq &\frac{2}{\bar{r}(c)} \operatorname*{ess\,inf}_{\vec{y}\in\Omega} \{\beta(c,\vec{y})\sigma(\vec{y})\},\\
U(c) \leq &\frac{2}{\underline{r}(c)} \operatorname*{ess\,sup}_{\vec{y}\in\Omega} \{\beta(c,\vec{y})\sigma(\vec{y})\}.
\end{eqnarray*}
Thus, the explicit endpoint conditions in Proposition~\ref{prop:explicit_spectral_crossing} are satisfied if
$$
\frac{2}{\bar{r}(0)} \operatorname*{ess\,inf}_{\vec{y}\in\Omega} \{\beta(0,\vec{y})\sigma(\vec{y})\}>1,
$$
and, for some sufficiently large $c_1$,
$$
\frac{2}{\underline{r}(c_1)} \operatorname*{ess\,sup}_{\vec{y}\in\Omega} \{\beta(c_1,\vec{y})\sigma(\vec{y})\}<1.
$$
In particular, if the base exit rate satisfies $\operatorname*{ess\,inf}_{\vec{x}\in\Omega}\kappa(\vec{x})>0$
while the maximum proliferation rate decays to zero ($\operatorname*{ess\,sup}_{\vec{x}\in\Omega}\beta(c,\vec{x}) \to 0$) as $c\to \infty$, then $U(c)\to 0$ since $0<\sigma(\vec{x})\leq1$. Hence, the upper endpoint $c_1$ can always be chosen sufficiently large.
\end{remark}

\begin{remark}[Finite-step growth criterion]
\label{rem:finite_step_growth_criterion}
The condition $L(c_0)>1$ requires uniform one-step growth across the entire state space. A weaker, albeit less explicit, alternative is the existence of an integer $m\geq1$, a non-trivial function $\varphi_0\in L^1(\Omega)$ with $\varphi_0\geq0$, and a constant $\eta>1$ such that
\begin{equation}
\mathcal{T}_{c_0}^{m}\varphi_0 \geq \eta\varphi_0 \qquad\text{almost everywhere in }\Omega.
\label{eq:finite_step_growth_condition}
\end{equation}
By positivity, iterating this inequality yields $\mathcal{T}_{c_0}^{km}\varphi_0 \geq \eta^k\varphi_0$ for $k\geq1$, and the spectral radius formula gives $\rho(\mathcal{T}_{c_0}) \geq \eta^{1/m}>1$. Therefore, \eqref{eq:finite_step_growth_condition} may serve as a substitute when the single-generation lower bound $L(c_0)>1$ is too stringent. Biologically, this criterion allows population growth to emerge through repeated transitions across phenotypic states, accommodating scenarios where certain individual states do not actively proliferate within a single cell cycle.
\end{remark}

To apply the intermediate value theorem in Theorem~\ref{thm:cytokine_compatible_stationary_population}, we must ensure that the spectral radius varies continuously with the feedback signal. This continuity is established in the following proposition. 

\begin{proposition}[Continuity of the spectral radius]
\label{prop:continuity_spectral_radius}
Let $I\subset[0,\infty)$ be a compact interval. Assume that the inheritance kernel satisfies
$$
p(\vec{x},\vec{y})\geq0, \qquad \int_\Omega p(\vec{x},\vec{y})\,d\vec{x}=1
$$
for almost every $\vec{y}\in\Omega$. Suppose there exist constants $0 < \underline{r} \leq \bar{r}< \infty$ such that the combined transition rate satisfies
\begin{equation}
\label{eq:bound_beta_kappa}
\underline{r} \leq \beta(c,\vec{x})+\kappa(\vec{x}) \leq \bar{r} 
\end{equation}
for almost every $\vec{x}\in\Omega$ and all $c\in I$. Assume further that the map $c\mapsto \beta(c, \cdot)$ is continuous from $I$ into $L^\infty(\Omega)$, meaning
\begin{equation}
\lim_{c'\to c} \operatorname*{ess\,sup}_{\vec{x}\in\Omega} \left|\beta(c',\vec{x})-\beta(c,\vec{x})\right| = 0 \qquad\text{for every }c\in I.
\label{eq:uniform_continuity_beta}
\end{equation}
Then the map $c\longmapsto\mathcal{T}_c$ is continuous from $I$ into
$\mathcal{L}(L^1(\Omega))$ with respect to the operator norm. Consequently, the spectral radius mapping
$c\longmapsto\rho(\mathcal{T}_c)$ is continuous on $I$.
\end{proposition}

\begin{proof}
For any $c,c'\in I$, the integral kernel of the difference operator
$\mathcal{T}_c-\mathcal{T}_{c'}$ is $K_c(\vec{x},\vec{y}) - K_{c'}(\vec{x},\vec{y})$. For any nonnegative $\varphi\in L^1(\Omega)$, Tonelli's theorem gives
\begin{align*}
\|(\mathcal{T}_c-\mathcal{T}_{c'})\varphi\|_{L^1} &\leq \int_\Omega \left[ \int_\Omega |K_c(\vec{x},\vec{y})-K_{c'}(\vec{x},\vec{y})| \,d\vec{x} \right] \varphi(\vec{y})\,d\vec{y}.
\end{align*}
Using \eqref{eq:bound_beta_kappa}, we algebraically bound the difference of the kernels by
\begin{align*}
\left| \frac{\beta(c,\vec{y})} {\beta(c,\vec{x})+\kappa(\vec{x})} - \frac{\beta(c',\vec{y})} {\beta(c',\vec{x})+\kappa(\vec{x})} \right|
\leq \frac{|\beta(c,\vec{y})-\beta(c',\vec{y})|}{\underline{r}} + \frac{\beta(c',\vec{y})}
{\underline{r}^2} |\beta(c,\vec{x})-\beta(c',\vec{x})|.
\end{align*}
Since $\beta$ is a continuous $L^\infty$-valued map on the compact interval $I$, it is uniformly bounded, implying the existence of a constant $\beta_{\max}<\infty$ such that $0\leq\beta(c,\vec{x})\leq \beta_{\max}$. Integrating over $\vec{x}$ and applying the normalization of $p$ alongside with $e^{-\mu(\vec{y})\tau(\vec{y})}\leq 1$, we obtain
\begin{align*}
\int_\Omega |K_c(\vec{x},\vec{y})-K_{c'}(\vec{x},\vec{y})| \,d\vec{x} \leq 2 \left(\frac{1}{\underline{r}} + \frac{\beta_{\max}}{\underline{r}^2} \right) \|\beta(c,\cdot)-\beta(c',\cdot)\|_{L^\infty}.
\end{align*}
Therefore, the operator norm satisfies
\begin{equation}
\|\mathcal{T}_c-\mathcal{T}_{c'}\|_{\mathcal{L}(L^1)} \leq C_I \|\beta(c,\cdot)-\beta(c',\cdot)\|_{L^\infty},
\label{eq:operator_norm_continuity},
\end{equation}
where $C_I = 2 \left(\underline{r}^{-1} + \beta_{\max}\underline{r}^{-2} \right)$. By \eqref{eq:uniform_continuity_beta}, $\|\mathcal{T}_c-\mathcal{T}_{c'}\|_{\mathcal{L}(L^1)} \to 0$ as $c'\to c$,
establishing the norm-continuity of $c\mapsto\mathcal{T}_c$.

Because the spectral radius is continuous with respect to the operator
norm in any Banach algebra, we conclude that $\rho(\mathcal{T}_{c'}) \to \rho(\mathcal{T}_c)$. Continuous dependence of the spectral radius on $c$ follows.
\end{proof}

Assuming the local uniform bounds $0 < \underline{r}_I \leq \bar{r}_I < \infty$ hold on every compact interval $I\subset[0,\infty)$ and $c\mapsto \beta(c, \cdot)$ is continuous in $L^\infty(\Omega)$, Proposition~\ref{prop:continuity_spectral_radius} guarantees that $c\mapsto\rho(\mathcal{T}_c) $ is continuous globally on $[0,\infty)$. Combining this continuity with the bounds from Proposition~\ref{prop:explicit_spectral_crossing} and the principal eigenvalue framework of Theorem~\ref{thm:cytokine_compatible_stationary_population} yields the following comprehensive existence criterion.

\begin{theorem}[Explicit sufficient conditions for a stationary cell-number density]
\label{thm:explicit_stationary_population_existence}
Let $\Omega\subset\mathbb{R}^m$ be a bounded measurable epigenetic state space. Assume (A1)--(A7) are satisfied, alongside the following specific conditions:
\begin{enumerate}
\item There exists $\underline{r}>0$ such that $\beta(c, \vec{x}) + \kappa(\vec{x}) \geq \underline{r} > 0$ for almost every $\vec{x}\in \Omega$ and all $c \geq 0$.
\item The renewal kernel $K_c(\vec{x}, \vec{y})$ is continuous and bounded on $\bar{\Omega}\times \bar{\Omega}$.
\item The state-dependent reproduction bounds $L(c)$ and $U(c)$, defined by 
$$
G_c(\vec{y}) = 2\beta(c, \vec{y}) e^{-\mu(\vec{y})\tau(\vec{y})}\int_{\Omega} \frac{p(\vec{x}, \vec{y})}{\beta(c, \vec{x}) + \kappa(\vec{x})}d \vec{x},
$$
$$
L(c) := \operatorname*{ess\,inf}_{\vec{y}\in\Omega}G_c(\vec{y}), \qquad U(c) := \operatorname*{ess\,sup}_{\vec{y}\in\Omega}G_c(\vec{y}),
$$
satisfy $L(c_0) > 1$ and $U(c_1) < 1$ for some parameters $0 \leq c_0 < c_1$.
\item The map $c\mapsto \beta(c, \cdot)$ is continuous from $[0, \infty)$ into $L^\infty(\Omega)$.
\end{enumerate}
Then there exists a stationary cell-number density $Q^\ast \in L^1(\Omega)$, strictly positive almost everywhere, resolving both the population renewal equation \eqref{eq:stationary_population_equation} and the cytokine self-consistency condition 
$$
c^\ast  = \int_{\Omega} \xi(\vec{x}) Q^\ast (\vec{x}) d\, \vec{x}.
$$
\end{theorem}

In the special case where the cell population parameters are homogeneous across the epigenetic state space, the integral equation structurally collapses into a scalar constraint, guaranteeing both the existence and uniqueness of the stationary density. This result is predicated on a fundamental biological viability condition that the intrinsic mitotic yield $2 e^{-\mu\tau}$ must strictly exceed unity. 

\begin{theorem}[Existence and uniqueness in the epigenetically homogeneous case]
\label{thm:homogeneous_spectral_crossing}
Assume that the cell population parameters are independent of the epigenetic state $\vec{x}$:
$$
\beta(c,\vec{x})\equiv\beta(c), \qquad \kappa(\vec{x})\equiv\kappa, \qquad \mu(\vec{x})\equiv\mu, \qquad \tau(\vec{x})\equiv\tau,
$$
where $\kappa>0$, $\mu\geq0$, and $\tau>0$. Incorporating the biological viability requirement, assume the expected mitotic yield satisfies
\begin{equation}
2 e^{-\mu \tau} > 1. 
\end{equation}
Additionally, let the inheritance kernel be properly normalized such that $p(\vec{x},\vec{y}) \geq 0$ and $\int_\Omega p(\vec{x},\vec{y})\,d\vec{x}=1$ for almost every $\vec{y}\in\Omega$. 

If the proliferation rate $\beta:[0,\infty)\to(0,\infty)$ is continuous, strictly decreasing, and satisfies the boundary limits
\begin{equation}
\beta(0) > \frac{\kappa}{2 e^{-\mu \tau} - 1},\qquad \lim_{c\to\infty}\beta(c)=0,
\label{eq:beta_decay}
\end{equation}
then there exists a unique critical feedback level $c^\ast > 0$ such that
\begin{equation}
R(c^\ast):=\frac{2\beta(c^\ast)e^{-\mu\tau}} {\beta(c^\ast)+\kappa}=1.
\label{eq:critical_feedback_condition}
\end{equation}
The underlying stationary population-renewal equation at $c=c^\ast$ admits a stationary density $Q^\ast$, strictly positive almost everywhere, supporting the cytokine self-consistency condition. 

Furthermore, if $p$ is continuous and $p(\vec{x}, \vec{y})>0$ on $\Omega \times \Omega$, this stationary configuration $Q^\ast $ is unique.
\end{theorem}

\begin{proof}
We first explicitly characterize the spectral radius as $\rho(\mathcal{T}_c) = R(c)$ for every $c \geq 0$. Under the homogeneous assumptions, the renewal operator reduces to  
$$
(\mathcal{T}_c\varphi)(\vec{x}) = R(c)(\mathcal{P}\varphi)(\vec{x}),\qquad\text{where}\quad
(\mathcal{P}\varphi)(\vec{x}) = \int_{\Omega} p(\vec{x}, \vec{y}) \varphi(\vec{y}) d\vec{y}.
$$
For any nonnegative $\varphi\in L^1(\Omega)$, Tonelli's theorem and the normalization of $p$ give
\begin{align*}
\|\mathcal{P}\varphi\|_{L^1} &= \int_\Omega\int_\Omega p(\vec{x},\vec{y})\varphi(\vec{y}) \,d\vec{y}\,d\vec{x}= \int_\Omega\varphi(\vec{y}) \left(\int_\Omega p(\vec{x},\vec{y})\,d\vec{x}\right)d\vec{y}=\|\varphi\|_{L^1}.
\end{align*}
Positivity implies that the operator norms are simply  $\|\mathcal{P}\|_{\mathcal{L}(L^1)}=1$ and $\|\mathcal{T}_c\|_{\mathcal{L}(L^1)}=R(c)$.

By duality, the adjoint operator acting on the uniform state $\boldsymbol{1}$ yields
$$
(\mathcal{P}^\ast \boldsymbol{1})(\vec{y}) = \int_\Omega p(\vec{x},\vec{y})\,d\vec{x} =1, \implies 
\mathcal{T}_c^\ast \boldsymbol{1} = R(c)\boldsymbol{1}.
$$
Thus, $R(c)$ is an eigenvalue of $\mathcal{T}_c^\ast$, rendering $\rho(\mathcal{T}_c) = \rho(\mathcal{T}_c^\ast ) \geq R(c)$. Combined with the upper bound $\rho(\mathcal{T}_c) \leq\|\mathcal{T}_c\|=R(c)$, we exactly evaluate the spectral radius:
\begin{equation}
\rho(\mathcal{T}_c) = R(c) = \frac{2\beta(c)e^{-\mu\tau}}{\beta(c)+\kappa}.
\end{equation}

Since $\beta(c)>0$ and $\kappa>0$, the algebraic map $\beta\mapsto (2\beta e^{-\mu\tau})/(\beta+\kappa)$ is strictly increasing. Because the physiological model dictates that $\beta(c)$ is continuous and strictly decreasing with respect to $c$, the mapping $c\mapsto R(c)$ is likewise continuous and strictly
decreasing on $[0,\infty)$.

The boundary conditions in \eqref{eq:beta_decay} translate to $R(0)>1$ and $\lim_{c\to\infty}R(c)=0$. The intermediate value theorem guarantees exactly one root $c^\ast>0$ where $R(c^\ast)=1$. Thus, $\rho(\mathcal{T}_{c^\ast})=R(c^\ast)=1$. The existence of the self-consistent, strictly positive density $Q^\ast$ supporting $c^\ast$ then follows immediately from Theorem~\ref{thm:cytokine_compatible_stationary_population}. 

Finally, under the strict positivity condition $p(\vec{x}, \vec{y})>0$ and $p\in C(\bar{\Omega}, \bar{\Omega})$, Lemma \ref{lem:compact_positive_Tc} confirms that $\mathcal{P}$ is both compact and strictly irreducible. By the Krein--Rutman theorem, the principal eigenspace is exactly one-dimensional, meaning the spatial profile $\varphi^{\ast}$ is unique up to scalar multiplication. Fixing this underdetermined scalar through the cytokine loop constraint $c^\ast = \int_\Omega \xi(\vec{x}) Q^\ast(\vec{x})\,d\vec{x}$ definitively locks in the amplitude, establishing the uniqueness of the full stationary cell-number density $Q^\ast$.
\end{proof}

\subsubsection{Stationary epigenetic probability density}
\label{sec:stationary_landscape_density}

Let $Q^\ast \in L^1(\Omega)$ be a nontrivial stationary cell-number density established in Theorem~\ref{thm:cytokine_compatible_stationary_population}, with total population $Q_{\mathrm{tot}}^\ast = \int_\Omega Q^\ast(\vec{x})\,d\vec{x} > 0$ and associated equilibrium cytokine concentration $c^\ast$. To formulate the steady-state problem in the probability simplex $\mathcal{P}(\Omega)$ defined in \eqref{eq:probability_simplex}, we introduce the state-dependent mitotic rate $M^\ast$ and cell removal rate $r^\ast$ evaluated at equilibrium:
\begin{equation}
M^\ast(\vec{y}) = \beta(c^\ast,\vec{y})Q^\ast(\vec{y}) e^{-\mu(\vec{y})\tau(\vec{y})}, \qquad r^\ast(\vec{x}) = \beta(c^\ast,\vec{x})+\kappa(\vec{x}).
\label{eq:stationary_renewal_coefficients}
\end{equation}
For any normalized state $f\in\mathcal{P}(\Omega)$, the ensemble mean removal rate is given by
\begin{equation}
\bar{r}_f^\ast = \int_\Omega r^\ast(\vec{y})f(\vec{y})\,d\vec{y}.
\label{eq:stationary_mean_fitness}
\end{equation}
The stationary renewal operator naturally reduces to 
\begin{equation}
\mathcal{R}^\ast[f](\vec{x}) := \frac{2}{Q_{\mathrm{tot}}^\ast} \int_\Omega M^\ast(\vec{y}) \bigl[p(\vec{x},\vec{y})-f(\vec{x})\bigr]\,d\vec{y} -f(\vec{x}) \bigl[r^\ast(\vec{x})-\bar{r}_f^\ast\bigr].
\label{eq:stationary_renewal_operator}
\end{equation}
The stationary epigenetic probability density $f^\ast$ is thus governed by the integro-differential equation
\begin{equation}
\mathcal{L}^\ast f^\ast+\mathcal{R}^\ast[f^\ast]=0
\qquad\text{in }\Omega,
\label{eq:stationary_landscape_equation}
\end{equation}
where the underlying advection-diffusion is captured by the Fokker--Planck operator
\begin{equation}
\mathcal{L}^\ast f = \nabla\cdot\mathcal{J}_f, \qquad \mathcal{J}_f = \tilde{\mathbf{D}}(\vec{x})\nabla f - \tilde{\vec{F}}(\vec{x})f.
\label{eq:stationary_fokker_planck_operator}
\end{equation}
This system is supplemented by the no-flux boundary condition
\begin{equation}
\mathcal{J}_{f^\ast}\cdot\vec{n} = 0 \qquad\text{on }\partial\Omega,
\label{eq:stationary_landscape_zero_flux}
\end{equation}
and the fundamental probability constraints
\begin{equation}
f^\ast\geq0, \qquad \int_\Omega f^\ast(\vec{x})\,d\vec{x}=1.
\label{eq:stationary_landscape_normalization}
\end{equation}

The inherent mass-conservation property of both the Fokker--Planck and renewal terms (Proposition~\ref{prop:simplex_invariance}) guarantees that
$\int_\Omega\left(\mathcal{L}^\ast f+\mathcal{R}^\ast[f]\right)d\vec{x}=0$
for any admissible no-flux state $f$. This structural conservation ensures that the stationary governing equation is inherently compatible with the probability simplex constraint \eqref{eq:stationary_landscape_normalization}.

\begin{theorem}[Existence of a stationary epigenetic probability density]
\label{thm:stationary_landscape_existence}
Assume that the domain $\Omega\subset\mathbb{R}^m$ is bounded with a $C^2$ boundary, the diffusion matrix $\tilde{\vec{D}}\in C^1(\bar{\Omega}; \mathbb{R}^{m\times m})$ is symmetric and uniformly elliptic, and the drift field $\tilde{\vec{F}}\in C^1(\bar{\Omega}; \mathbb{R}^m)$. Let $\mathcal{L}^\ast$ denote the no-flux realization of the Fokker--Planck operator on $L^1(\Omega)$. Suppose this realization generates a positive conservative contraction semigroup on $L^1(\Omega)$, and that its resolvent is compact at some point $\eta_1 > 0$ on the positive real axis; namely, $R_{\eta_1}:=(\eta_1 I - \mathcal{L}^\ast)^{-1}$ is a compact operator on $L^1(\Omega)$. Then the stationary landscape problem \eqref{eq:stationary_landscape_equation}--\eqref{eq:stationary_landscape_normalization} admits at least one $f^\ast\in\mathcal{P}(\Omega)$.
\end{theorem}

\begin{proof}
Let $m^\ast=\int_\Omega M^\ast(\vec{y})\,d\vec{y}$, and $a^\ast=2m^\ast/Q_{\mathrm{tot}}^\ast$. Because the no-flux realization of $\mathcal{L}^\ast$ generates a contraction semigroup on $L^1(\Omega)$, the resolvent $R_\eta=(\eta I-\mathcal{L}^\ast)^{-1}$ is well-defined, positive, and bounded for all $\eta>0$. Through the standard resolvent identity, the assumed compactness of $R_{\eta_1}$ guarantees that $R_{\eta}$ remains compact for every $\eta > 0$.

To formulate a fixed-point mapping, we pick a sufficiently large spectral shift
$$
\eta>a^\ast+\|r^\ast\|_{L^\infty(\Omega)},
$$
and define the nonlinear operator map on $\mathcal{P}(\Omega)$ by
\begin{equation}
\mathscr{S}^\ast(g) = R_\eta\bigl[\eta g + \mathcal{R}^\ast[g]\bigr].
\label{eq:stationary_landscape_resolvent_map}
\end{equation}

We first verify that $\mathscr{S}^\ast$ maps $\mathcal{P}(\Omega)$ into itself. Substituting the definition of $\mathcal{R}^\ast$, the forcing term expands as
\begin{equation}
\label{eq:stationary_resolvent_rhs}
\eta g + \mathcal{R}^\ast[g] = \frac{2}{Q_{\mathrm{tot}}^\ast}\int_\Omega M^\ast(\vec{y})p(\cdot,\vec{y})\,d\vec{y} + \bigl(\eta - a^\ast - r^\ast + \bar{r}_g^\ast\bigr)g,
\end{equation}
where $\bar{r}_g^\ast =\int_\Omega r^\ast(\vec{y})g(\vec{y})\,d\vec{y}$. The inheritance integral is globally nonnegative. Since $r^\ast\geq 0$ and $\bar{r}_g^\ast\geq0$, the parameter constraint on $\eta$ ensures 
$$
\eta - a^\ast - r^\ast(\vec{x}) + \bar{r}_g^\ast \geq \eta - a^\ast - \|r^\ast\|_{L^\infty(\Omega)}>0.
$$
Hence $\eta g + \mathcal{R}^\ast[g]\geq0$, which implies $\mathscr{S}^\ast(g)\geq0$ due to the positivity of the resolvent $R_\eta$. 

Letting $f=\mathscr{S}^\ast(g)$, integrating the identity $(\eta I - \mathcal{L}^\ast)f = (\eta g + \mathcal{R}^\ast[g])$ across the domain $\Omega$, and applying the no-flux boundary condition yields
$$
\eta\int_\Omega f\,d\vec{x} = \eta\int_\Omega g\,d\vec{x} + \int_\Omega \mathcal{R}^\ast[g]\,d\vec{x} + \int_\Omega \mathcal{L}^\ast f\,d\vec{x} = \eta \cdot 1 + 0 + 0 = \eta.
$$
Thus, $\int_\Omega f\,d\vec{x}=1$, confirming that $\mathscr{S}^\ast(\mathcal{P}(\Omega)) \subseteq \mathcal{P}(\Omega)$.

Next, we establish continuity. Consider arbitrary $g_1,g_2\in\mathcal{P}(\Omega)$ and denote $h=g_1-g_2$. The strict inheritance term cancels dynamically, giving
$$
\mathcal{R}^\ast[g_1] - \mathcal{R}^\ast[g_2] = -a^\ast h-r^\ast h + \bar{r}_{g_1}^\ast h + \bigl(\bar{r}_{g_1}^\ast - \bar{r}_{g_2}^\ast\bigr)g_2.
$$
Using the elementary bound $\left|\bar{r}_{g_1}^\ast -  \bar{r}_{g_2}^\ast\right| \leq \|r^\ast\|_{L^\infty(\Omega)} \|g_1-g_2\|_{L^1(\Omega)}$, it acts locally Lipschitz, demonstrating that $\mathcal{R}^\ast$ is continuous on $\mathcal{P}(\Omega)$ under the $L^1$ topology. Because $R_\eta$ is a bounded linear operator, the composite map $\mathscr{S}^\ast$ inherits this $L^1$ continuity.

Furthermore, leveraging the mass conservation $\int_\Omega \mathcal{R}^\ast[g](\vec{x})\,d\vec{x}=0$ and local positivity, the $L^1$-norm evaluates explicitly as
$$
\|\eta g + \mathcal{R}^\ast[g]\|_{L^1(\Omega)} = \int_\Omega \bigl(\eta g+\mathcal{R}^\ast[g]\bigr)(\vec{x})\,d\vec{x} = \eta.
$$
Consequently, the source image $\{\eta g+\mathcal{R}^\ast[g]: g\in\mathcal{P}(\Omega)\}$
is uniformly bounded in $L^1(\Omega)$. The compactness of $R_\eta$ forces the target set $\mathscr{S}^\ast(\mathcal{P}(\Omega))$ to be relatively compact in $L^1(\Omega)$.

Since $\mathscr{S}^\ast$ constitutes a continuous, compact self-map on the closed convex set $\mathcal{P}(\Omega)$, Schauder's fixed-point theorem asserts the existence of at least one target density $f^\ast\in\mathcal{P}(\Omega)$ satisfying $\mathscr{S}^\ast(f^\ast)=f^\ast$. Rearranging \eqref{eq:stationary_landscape_resolvent_map} directly recovers $\mathcal{L}^\ast f^\ast+\mathcal{R}^\ast[f^\ast]=0$, finalizing the existence proof.
\end{proof}

\begin{theorem}[Uniqueness of the stationary epigenetic probability density]
\label{thm:stationary_landscape_uniqueness}
Suppose $Q^\ast$ is fixed, and let $\mathcal{H}\subseteq L^1(\Omega)$ be a Hilbert space admitting at least one stationary solution $f^\ast \in \mathcal{P}(\Omega)\cap \mathcal{H}$. Assume there exists a constant $\lambda>0$ such that, for every pair $f_1, f_2\in\mathcal{P}(\Omega)\cap\mathcal{H}$, the coupled operator system exhibits the strict functional dissipativity
\begin{equation}
\left\langle
\mathcal{L}^\ast(f_1-f_2) + \mathcal{R}^\ast[f_1] - \mathcal{R}^\ast[f_2], f_1-f_2 \right\rangle_{\mathcal{H}} \leq - \lambda\|f_1-f_2\|_{\mathcal{H}}^2.
\label{eq:stationary_landscape_strict_dissipativity}
\end{equation}
Then the stationary landscape equation possesses exactly one solution in the functional branch $\mathcal{P}(\Omega)\cap\mathcal H$.
\end{theorem}

\begin{proof}
Let $f_1^\ast, f_2^\ast\in\mathcal{P}(\Omega)\cap\mathcal{H}$ be two arbitrary stationary solutions. Subtracting the respective stationary equations yields the residual balance
$$
\mathcal{L}^\ast(f_1^\ast - f_2^\ast) + \mathcal{R}^\ast[f_1^\ast] - \mathcal{R}^\ast[f_2^\ast]=0.
$$
Taking the inner products against $f_1^\ast - f_2^\ast$ in $\mathcal{H}$ and invoking the dissipativity condition \eqref{eq:stationary_landscape_strict_dissipativity} directly enforces
$$
0 = \left\langle\mathcal{L}^\ast(f_1^\ast-f_2^\ast) + \mathcal{R}^\ast[f_1^\ast] - \mathcal{R}^\ast[f_2^\ast], f_1^\ast - f_2^\ast \right\rangle_{\mathcal{H}} \leq - \lambda\|f_1^\ast-f_2^\ast\|_{\mathcal{H}}^2.
$$
Because $\lambda$ is strictly positive, the norm identically vanishes, forcing $f_1^\ast=f_2^\ast$.
\end{proof}

\begin{remark}[Spectrum of strict dissipativity]
Condition \eqref{eq:stationary_landscape_strict_dissipativity} represents a strict continuous dissipativity requirement for the integrated operator network $\mathcal{L}^\ast + \mathcal{R}^\ast$ within the designated $\mathcal{H}$-inner product. In practical modeling, $\mathcal{H}$ is normally chosen as $L^2(\Omega)$ or a correctly weighted Sobolev space. Verifying this condition in applications amounts to estimating the diffusion, drift, and nonlocal terms separately to obtain a globally negative-definite quadratic form.
\end{remark}

\subsubsection{Equilibrium of the full coupled model}
\label{sec:full_equilibrium}

Synthesizing the preceding analytical framework, a globally consistent, nontrivial equilibrium of the full coupled cross-scale model is defined by the triple $(Q^\ast, f^\ast, c^\ast)$ satisfying the hierarchical system
\begin{equation}
\begin{cases}
\mathcal{T}_{c^\ast}Q^\ast = Q^\ast,\quad Q^\ast_{\mathrm{tot}} = \displaystyle \int_\Omega Q^\ast(\vec{x}) \, d\vec{x},\\[3mm]
c^\ast=\displaystyle\int_\Omega \xi(\vec{x})Q^\ast(\vec{x})\,d\vec{x},\\[3mm]
\mathcal{L}^{\ast}f^\ast + \mathcal{R}^\ast[f^\ast]=0, \\[2mm]
f^\ast\geq0,\qquad \displaystyle\int_\Omega f^\ast(\vec{x})\,d\vec{x}=1.
\end{cases}
\label{eq:full_equilibrium_system}
\end{equation}
This system is structurally decoupled in a top-down manner. The first two integral constraints uniquely resolve the macroscopic population steady state---locking in the cell number density profile $Q^\ast$, the total cell number $Q^\ast_{\mathrm{tot}}$, and the auto-regulatory cytokine concentration $c^\ast$. The subsequent integro-differential equation then governs the stationary epigenetic probability density $f^\ast$ through the dynamically updated Fokker--Planck operator. 

Importantly, due to the continuous cell population selection (state-dependent proliferation and apoptosis) across the epigenetic state space, the rigorous probability density generally deviates from the simple normalization of the population density functional:
\begin{equation}
f^\ast(\vec{x}) \neq \frac{Q^\ast(\vec{x})}{Q_{\mathrm{tot}}^\ast}.
\label{eq:equilibrium_nonidentification}
\end{equation}
This fundamental non-identification encapsulates the breaking of local equivalence between absolute abundance and relative likelihood in open, proliferating systems.

The existence and uniqueness of this global equilibrium follow from the sequential analyses established in the preceding sections.

\begin{theorem}[Global existence and uniqueness of the full equilibrium triple]
\label{thm:full_equilibrium_existence_uniqueness}
Suppose the hypotheses of Theorem~\ref{thm:cytokine_compatible_stationary_population} hold, and the critical feedback level $c^\ast$ ensuring the spectral crossing $\rho(\mathcal{T}_{c^\ast})=1$ is unique. Further assume that the principal eigenvalue of $\mathcal{T}_{c^\ast}$ is simple, and that the stationary landscape problem \eqref{eq:stationary_landscape_equation} satisfies the strict dissipativity condition for uniqueness (Theorem~\ref{thm:stationary_landscape_uniqueness}). Then, the full coupled system admits exactly one equilibrium triple $(Q^\ast, f^\ast, c^\ast)$, satisfying $c^\ast = \int_\Omega \xi(\vec{x}) Q^\ast(\vec{x})\, d \vec{x}$.
\end{theorem}

\begin{proof}
By Theorem~\ref{thm:cytokine_compatible_stationary_population} and the uniqueness assumption of the spectral crossing, the critical cytokine level $c^\ast$ is uniquely fixed. The simplicity of the principal eigenvalue $\rho(\mathcal{T}_{c^\ast})=1$ ensures that the spatial profile of the eigenfunction is unique up to a scalar multiplier. As established in Theorem~\ref{thm:cytokine_compatible_stationary_population}, the cytokine self-consistency condition $\int_\Omega \xi(\vec{x})Q^\ast(\vec{x})\,d\vec{x}=c^\ast$ unambiguously isolates this scalar amplitude, thereby uniquely determining the macroscopic density profile $Q^\ast$.

With the environmental variables $(Q^\ast, c^\ast)$ completely specified, they rigorously parameterize the stationary renewal operator $\mathcal{R}^\ast$. The existence and uniqueness of the corresponding epigenetic probability density $f^\ast\in\mathcal{P}(\Omega)$ then follow directly from Theorems~\ref{thm:stationary_landscape_existence} and \ref{thm:stationary_landscape_uniqueness}. Thus, the complete equilibrium triple $(Q^\ast, f^\ast, c^\ast)$ is uniquely determined.
\end{proof}

Finally, having rigorously established the micro-scale invariant probability density $f^\ast(\vec{x})\in \mathcal{P}(\Omega)$, we note that $f^\ast > 0$ almost everywhere on $\Omega$ since it satisfies a uniformly elliptic stationary problem coupled with a positivity-improving nonlocal term. Following our established convention to set $U^\ast = \infty$ on any null set where $f^\ast$ vanishes, the stationary, non-equilibrium Waddington landscape of the cellular population is precisely quantified via the generalized Boltzmann--Gibbs projection:
\begin{equation}
U^\ast(\vec{x}) = -\log f^\ast(\vec{x}).
\label{eq:stationary_waddington_landscape}
\end{equation}
This mathematically closes the theoretical formulation, mapping the complex, trait-driven stochastic dynamics of single cells directly to a unified energetic topology governing tissue-level homeostasis.

\section{Applications: Demonstrations in canonical gene circuits}

Having established the theoretical framework, we next investigate how cell proliferation reshapes the topology of the dynamic Waddington landscape in representative gene-circuit models. For notational simplicity, we write $\mathcal{R}[f]$ for the population renewal operator $\mathcal{R}[f; Q_{\mathrm{tot}}]$, with its dependence on $Q_{\mathrm{tot}}$ understood. We consider canonical one- and two-gene circuits as minimal models to examine how population renewal modifies attractor stability, transition-barrier heights, and the directionality bias of cell-fate transitions. As shown below, proliferation is not merely a mechanism for regulating population size; it also acts as a phenotype-selective force that reconfigures the dynamic Waddington landscape in a circuit-dependent manner.

\subsection{Single-gene circuit in fate decision}
\begin{figure}[htbp]
\centering
	\includegraphics[width=0.9\linewidth, clip, bb={110  285 1100 1000}]{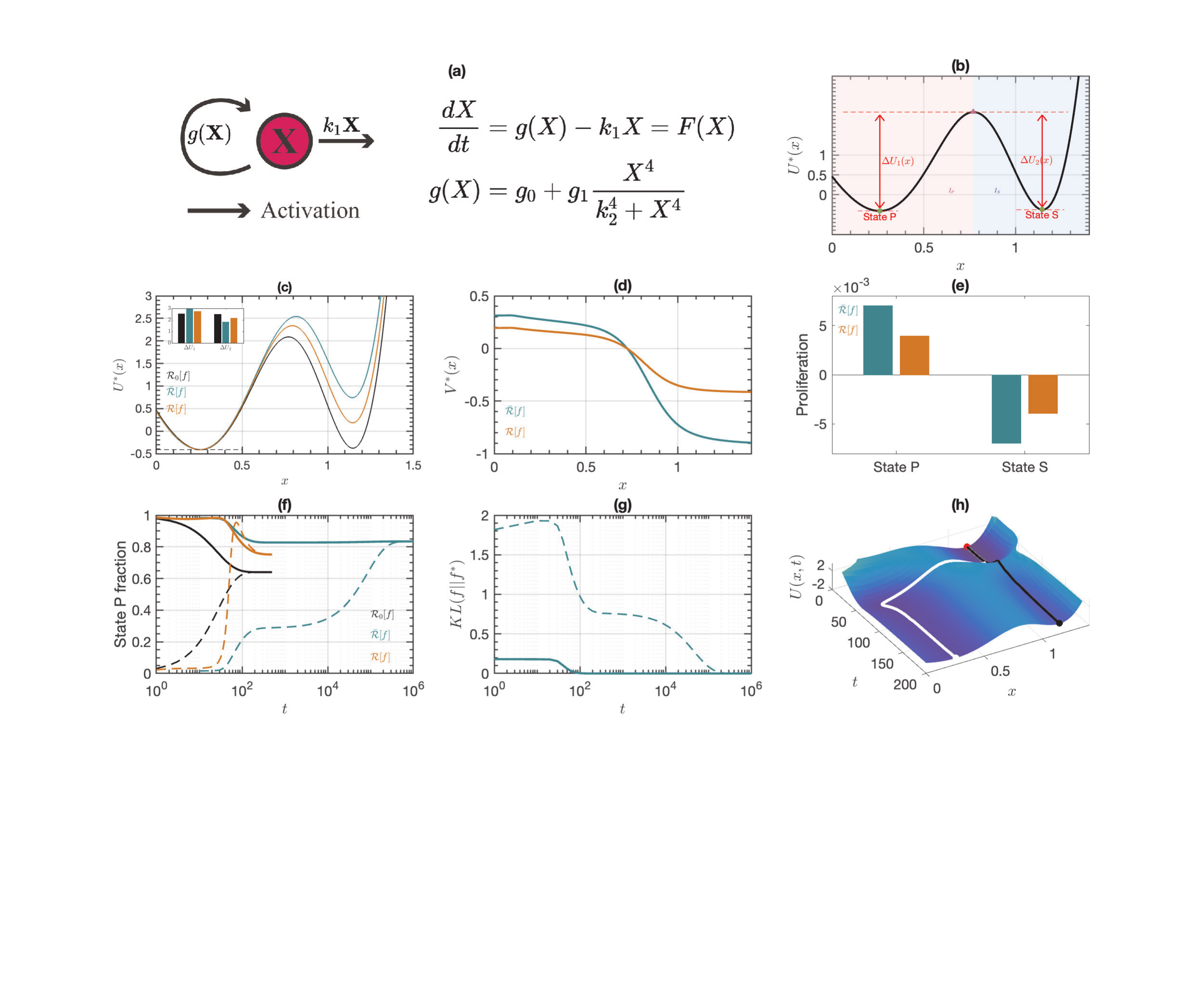}	
	\caption{\textbf{Cell-fate decisions and the dynamic Waddington landscape in a single-gene circuit.} (a) Schematic of the single-gene positive-feedback circuit. (b) Reference landscape $U^\ast(x)$ generated by the intrinsic GRN dynamics in the absence of population regulation ($\mathcal{R}_0[f] = 0$). The two local minima correspond to the inherent epigenetic attractors. (c) Steady-state Waddington landscapes $U^\ast(x)$ under three distinct population renewal operators. The left minima are vertically aligned to highlight topological deviations; the inset compares the reshaped transition barrier heights ($\Delta U$). (d) Profiles of the equilibrium selective potential $V^\ast(x)$. (e) Net population regulation within each attractor basin, computed as the integral of $\mathcal{R}[f]$ over the respective domains. The contrasting source-sink dynamics functionally identify State S as stem-like and State P as progenitor-like. (f) Temporal evolution of the State P fraction starting from two homogeneous initial conditions: 100\% State S (dashed lines) and 100\% State P (solid lines). Both trajectories converge to a stable homeostatic state, demonstrating the capacity for spontaneous regeneration of the stem-like state from an initially progenitor-only population. (g) Temporal evolution of the Kullback--Leibler divergence $\mathrm{KL}(f(x, t) \| f^*(x))$ under the population renewal operator $\bar{\mathcal{R}}[f]$, initialized from the same homogeneous states as in (f). (h) Spatiotemporal evolution of the dynamic landscape $U(x, t)$ illustrating the spontaneous differentiation process. The system, initialized as a pure stem-cell pool (State S), autonomously relaxes into a steady-state distribution spanning both attractors without modifying any external parameters. Two characteristic trajectories $x(t)$, obtained by minimizing the time-integrated landscape functional $\int U(x(t), t)\, dt$, depict continuous maintenance within State S (black) and the transition path toward State P (white). This illustrates the autonomous remodeling of the landscape driven by state-dependent cell proliferation.}
	\label{fig2}
\end{figure}

We first consider a single-gene autoregulatory circuit with positive feedback \cite{ferrell2012} (Fig.~\ref{fig2}a), yielding the integro-PDE:
\begin{equation}
\frac{\partial f(x, t)}{\partial t} = \frac{\partial}{\partial x}\!\left(\tilde{D}(x)\frac{\partial f}{\partial x}\right) - \frac{\partial}{\partial x}\!\left(\tilde{F}(x)f\right) + \mathcal{R}[f],
\label{mod3}
\end{equation}
where the transformed drift and diffusion are
$$
\tilde{F}(x)  =\frac{F(e^{x}-1)}{e^{x}} + \frac{D_0}{e^{2x}}, \qquad
\tilde{D}(x) =\frac{D_0}{e^{2x}},
$$
with $F(X) = g_0 + g_1 X^4/(k_2^4 + X^4) - k_1 X$ representing the Hill-type synthesis-degradation balance. To isolate the role of macroscopic proliferation, we compare the landscape topology under three operator definitions:
\begin{enumerate}
\item[(i)] $\mathcal{R}_0[f] = 0$: pure autonomous GRN dynamics, serving as the single-cell baseline;
\item[(ii)] $\bar{\mathcal{R}}[f]$: a state-independent renewal operator capturing uniform proliferation and removal;
\item[(iii)] $\mathcal{R}[f]$: the full mechanistic operator, incorporating the state-dependent re-entry rate $\beta(c, x)$ and removal rate $\kappa(x)$. Here, $x$ characterizes epigenetic stemness: larger $x$ denotes slowly cycling stem-like cells with low rates of both proliferation and terminal differentiation, whereas smaller $x$ represents rapidly proliferating progenitor-like cells subject to strong removal fluxes. 
\end{enumerate}
Details of the model derivation, parameterization, boundary treatment, finite-difference discretization, and numerical validation are provided in \emph{SI Appendix} E.1.

Evaluating the intrinsic GRN dynamics in the absence of population regulation ($\mathcal{R}_0[f] = 0$) under the chosen parameter regime yields a classical bistable landscape $U^\ast(x)$ (Fig.~\ref{fig2}b). The two local minima correspond to intrinsic epigenetic attractors, whose distinct $x$-coordinates define two phenotypic states, State P and State S. This bistable landscape provides the epigenetic scaffold upon which population regulation subsequently acts.

To examine how population renewal reshapes the intrinsic GRN landscape, Fig.~\ref{fig2}c compares the steady-state Waddington landscapes $U^\ast(x)$ generated in the absence of population renewal and under the two renewal operators. For a direct comparison of the transition barriers, we vertically align the landscapes at the minimum associated with State P and evaluate the corresponding barrier heights $\Delta U$. These comparisons show how the coupling between cellular kinetotypic dynamics and the epigenetic state space modifies the stability structure of the attractors.

Both $\bar{\mathcal{R}}[f]$ and $\mathcal{R}[f]$ induce the same qualitative remodeling relative to the non-renewing baseline. In both cases, the barrier associated with State P increases, whereas the barrier associated with State S decreases. Population renewal therefore enhances the relative stability of the State P attractor while making transitions away from the State S attractor more accessible. The two operators differ primarily in the magnitude of this effect: the changes in both barrier heights are more pronounced under $\bar{\mathcal{R}}[f]$ than under the fully state-dependent operator $\mathcal{R}[f]$. Thus, although the two renewal mechanisms generate different quantitative landscape profiles, they produce the same qualitative reorganization of attractor stability. Further changes in the kinetic and inheritance parameters lead to additional quantitative modifications of the equilibrium landscape (see \emph{SI Appendix} E.2). 

The corresponding equilibrium selective potentials $V^\ast(x)$ are shown in Fig.~\ref{fig2}d. For both renewal operators, $V^\ast(x)$ decreases monotonically across the epigenetic state space, indicating a systematic population-level bias rather than a localized effect confined to one of the attractor basins. This decrease is more pronounced under $\bar{\mathcal{R}}[f]$, consistent with the stronger remodeling of the steady-state landscape observed in Fig.~\ref{fig2}c. 

The basin-integrated population regulation in Figure~\ref{fig2}e further clarifies how this bias is generated. The highly proliferative State P exhibits a net positive population renewal ($\mathcal{R}[f] > 0$), mathematically acting as a \emph{kinetic source} for the probability density. In contrast, the State S basin exhibits a net negative renewal ($\mathcal{R}[f] < 0$), acting as a \emph{kinetic sink}. Although this terminology stems strictly from the continuous density flux of the system, it naturally captures the biological reality of tissue homeostasis. The kinetic ``sink'' (State S) corresponds to a slowly cycling stem-like pool that serves as the actual lineage origin, whereas the kinetic ``source'' (State P) represents a rapidly expanding transit-amplifying progenitor compartment. The magnitudes of both effects are larger under $\bar{\mathcal{R}}[f]$, consistent with its stronger selective potential and more pronounced barrier remodeling. These results indicate that despite both attractors being stable states of the intrinsic GRN landscape, they assume asymmetric and cooperative roles in maintaining the non-equilibrium population dynamics.

The high-$x$ attractor is therefore identified with stem-like cells (\emph{State S}). Its negative basin-integrated population regulation indicates that it is not the primary site of net population expansion. Instead, its biological role is associated with maintaining a relatively plastic stem-like compartment that can give rise to downstream cell states. The reduced transition barrier on the State S side of the landscape is consistent with this role, because it facilitates state transitions out of the stem-like basin. By contrast, the low-$x$ attractor corresponds to progenitor-like cells (\emph{State P}). Analogous to a transit-amplifying compartment, this state functions as a net population source and supports local population expansion. Its increased transition barrier indicates that the progenitor-like state is more strongly retained in the steady-state landscape, while the population-level flux from State S provides a mechanism for its continual replenishment.

Taken together, Figs.~\ref{fig2}c--e show that population renewal influences the dynamic Waddington landscape through two coupled effects: it changes the barrier structure that governs phenotypic transitions and imposes a state-dependent bias in local population regulation. At steady state, the total population distribution remains unchanged, but this macroscopic stationarity is maintained by continuous redistribution of cells between phenotypic states. The resulting balance between state transitions, net population sources, and net population sinks provides a non-equilibrium mechanism for tissue homeostasis. This mechanism, and its dependence on the cellular renewal operator, cannot be inferred from the intrinsic single-cell GRN dynamics alone.

We next examine the macroscopic population dynamics generated by the three renewal operators considered above. Figure~\ref{fig2}f shows the temporal evolution of the State P fraction from two extreme initial conditions, corresponding to populations composed entirely of State S and State P cells. For all three operators, the State P fraction eventually approaches the same long-time value for the two initial conditions considered. The renewal operators nevertheless produce distinct transient dynamics. Under $\mathcal{R}_0[f]$, the State P fraction relaxes relatively rapidly toward its asymptotic value. The fully coupled operator $\mathcal{R}[f]$ produces a similar qualitative relaxation, although the transient trajectory differs quantitatively. Nevertheless, a much slower relaxation is observed under the state-independent operator $\bar{\mathcal{R}}[f]$ when the initial population consists entirely of State S cells. In this case, the State P fraction first increases to an intermediate value of approximately $0.3$ and remains near this level for an extended period before slowly increasing to its final value of approximately $0.8$. This long-lived intermediate regime is indicated by the green dashed curve in Fig.~\ref{fig2}f. In contrast, the green solid curve, corresponding to an initial population composed entirely of State P cells, approaches the same asymptotic value without exhibiting a comparably pronounced plateau.

Figure~\ref{fig2}g further characterizes the slow transient under $\bar{\mathcal{R}}[f]$ through the Kullback--Leibler divergence
\begin{equation}
D_{\mathrm{KL}}\!\left(f(\cdot,t)\,\middle\|\,f^*\right) =\int_{\Omega} f(x,t)\log\frac{f(x,t)}{f^*(x)}\,dx,
\end{equation}
where $f^\ast$ denotes the common steady-state distribution approached from both initial conditions under $\bar{\mathcal{R}}[f]$. For the all-State-P initial condition, shown by the solid curve, the divergence decreases rapidly toward zero, indicating direct relaxation toward $f^\ast$. In contrast, for the all-State-S initial condition, shown by the dashed curve, the divergence remains at a nonzero intermediate plateau for an extended period before eventually decaying to zero. This plateau coincides with the slow intermediate regime observed in the State P fraction and provides a distribution-level signature of a metastable regime rather than a distinct steady state.

Together, Figs.~\ref{fig2}f and~\ref{fig2}g distinguish the uniqueness of the asymptotic distribution from the timescale required to reach it. For all three renewal operators, the State P fraction converges to the same operator-specific long-time value from the two extreme initial conditions. For $\bar{\mathcal{R}}[f]$, the decay of the Kullback--Leibler divergence further confirms that the complete population distributions converge to the same steady state $f^\ast$. Population renewal can nevertheless strongly modify the transient pathway and generate long-lived intermediate population configurations. In particular, the metastable regime under $\bar{\mathcal{R}}[f]$ reflects a delayed redistribution of cells from the initially State-S-dominated population toward the homeostatic mixture. The rapid relaxation from the all-State-P initial condition also demonstrates population-level plasticity: a progenitor-dominated population can regenerate a finite State S compartment before reaching the same homeostatic state. This collective regeneration is consistent with the capacity of progenitor-like populations to restore stem-like populations after perturbation, as reported in mammalian tissues including the intestinal crypt and airway epithelium \cite{Es:2012aa, Tata:2013aa}.

Synthesizing these findings, Fig.~\ref{fig2}h visualizes the spatiotemporal evolution of the dynamic landscape $U(x, t)$ from an initially homogeneous population composed entirely of stem-like cells (State S). Initially, the landscape contains a single valley corresponding to the State S attractor. As the population evolves, a second valley gradually emerges at lower $x$, reflecting the spontaneous accumulation of progenitor-like cells (State P) and the establishment of a second stable population state. Two representative trajectories, obtained by minimizing the path integral $\int U(x, t) d t$, illustrate persistence within the stem-like state and transition toward the progenitor-like state. This process provides a population-level description of spontaneous cell differentiation and direct visualization of the dynamic Waddington landscape. Importantly, the landscape evolves with the actual time $t$ as the population distribution changes, rather than with an externally varied regulatory parameter. Thus, under constant environmental and model parameters, the second valley emerges autonomously from the interplay between intrinsic epigenetic dynamics and population renewal, demonstrating that tissue-level organization and lineage diversification can arise endogenously without externally imposed time-dependent parameter changes.

\subsection{Binary gene circuit in fate decision}

We next apply the framework to a two-gene mutual-inhibition circuit, a canonical model of binary fate decisions (see \emph{SI Appendix} F). Two network topologies, with and without self-activation, are considered to examine how the effects of population renewal depend on the underlying GRN architecture.

Without self-activation, the circuit is bistable, producing two mutually exclusive committed states separated by an unstable intermediate state, as in GATA1/PU.1 and Oct4/Cdx2 regulatory switches \cite{huang2007bifurcation, Masui:2007aa}. With self-activation, the circuit becomes tristable, allowing a central progenitor-like state to coexist with two committed states, consistent with the Nanog/Gata6 and Sox2/Brachyury circuits \cite{Bessonnard:2014aa, Schroter:2015aa, Turner:2014aa, Gouti:2014aa}. This comparison reveals how population renewal reshapes landscape topology, switching plasticity, and developmental directionality across distinct GRN architectures.

\subsubsection{Bistable circuit: Proliferation-induced plasticity and the emergent breathing landscape} 

\begin{figure}[htbp]
\centering
\includegraphics[width=0.9\linewidth, clip, bb={100  150 1100 1000}]{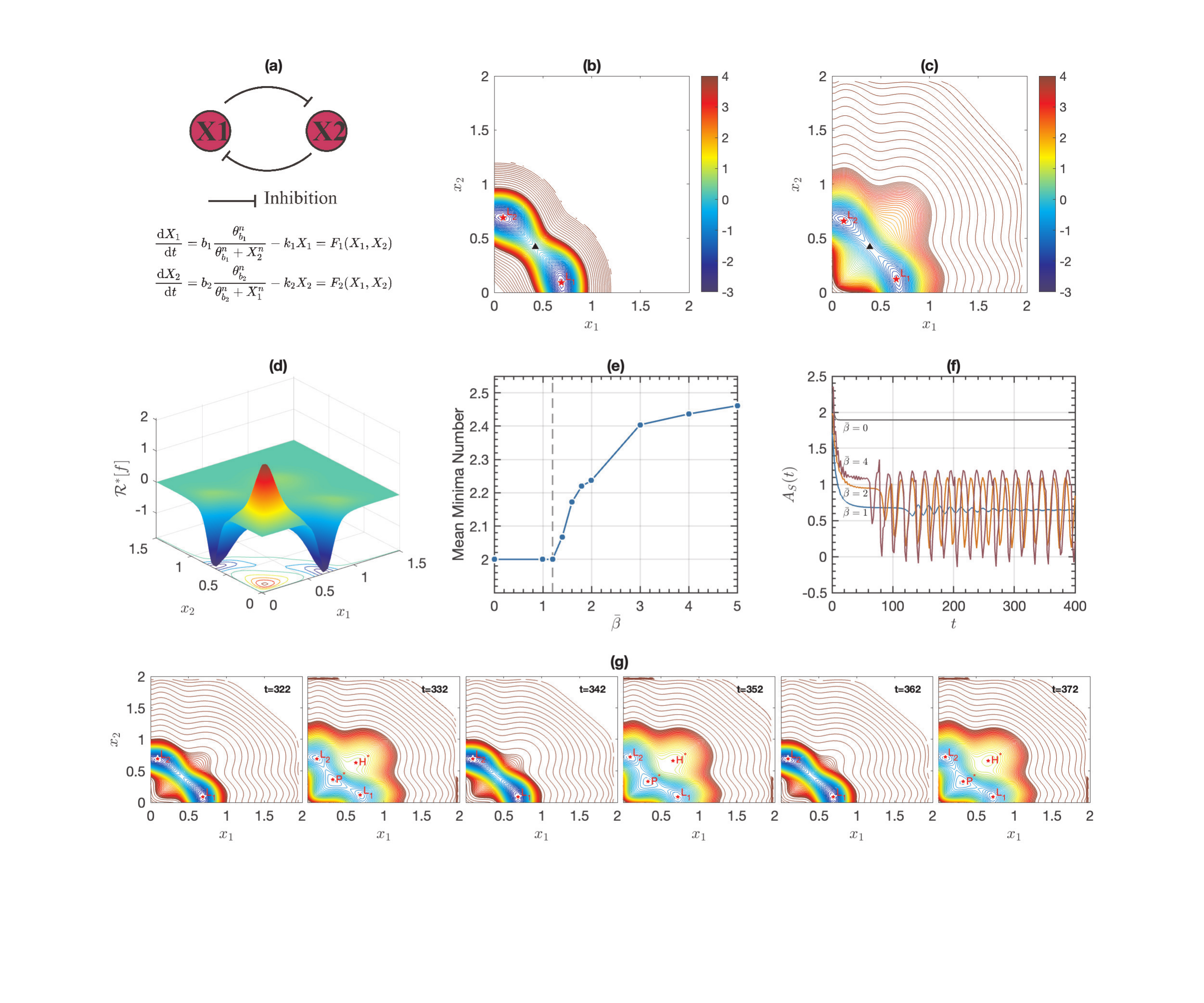} 
\caption{\textbf{Proliferation-induced changes in landscape topology through the emergence of additional attractors in a binary cell-fate circuit.} 
(a) Schematic of the two-gene mutual-inhibition circuit without self-activation. 
(b) Steady-state landscape $U^\ast(\vec{x})$ in the absence of population renewal ($\bar{\beta} = 0$). Two terminal attractors ($L_1$ and $L_2$, red stars) are separated by a central saddle  (black triangle). 
(c) The stationary landscape at a moderately low proliferation rate ($\bar{\beta}=1$). 
(d) Steady-state spatial distribution of the non-equilibrium population renewal operator $\mathcal{R}^\ast[f]$ for $\bar{\beta} = 1$. Negative valleys at the terminal attractors indicate a continuous probability efflux (population depletion), while the positive region near the origin (associated with low gene expression levels) corresponds to a substantial probability influx. This localized accumulation of probability mass provides the kinetic basis for the emergence of proliferation-induced states within the low-expression manifold.
(e) Increasing proliferation induces additional landscape minima. The plot shows the time-averaged number of local minima as a function of $\bar{\beta}$, averaged over $100 \leq t \leq 500$ (\emph{SI Appendix} Fig.~S16).  
(f) Relative intermediate-state depth $A_S(t)$ exhibits a transition from damped relaxation ($\bar{\beta} \leq 1$) to sustained oscillations ($\bar{\beta} \geq 2$), indicating the onset of dynamic attractor remodeling. 
(g) Time-resolved landscape snapshots at $\bar{\beta}=4$. Red stars denote instantaneous local minima. Periodic emergence and disappearance of transient proliferation-induced attractors (tPIAs), labeled $P^\ast $ and $H^\ast $, illustrate continuous landscape reconstruction driven by population renewal.}
\label{fig3}
\end{figure}

We first consider the mutual-inhibition circuit without self-activation (Fig.~\ref{fig3}a). In the absence of population regulation ($\mathcal{R}_0[f]=0$; hereafter denoted by $\bar{\beta} = 0$), the intrinsic GRN generates a bistable landscape with two terminal attractors separated by a saddle point (Fig.~\ref{fig3}b). Direct transitions between the two committed states therefore require crossing the intervening landscape barrier. 

Introducing weak population renewal ($\bar{\beta} = 1$) preserves bistability while substantially reshaping the landscape (Fig.~\ref{fig3}c). At steady state, $\mathcal{R}^\ast[f]$ exports probability from both committed attractors and redistributes it towards the intermediate region, generating a localized source near the saddle point (Fig.~\ref{fig3}d). This redistribution flattens the terminal basins and reduces the transition barrier from $2.31$ to $0.70$ (\emph{SI Appendix}, Fig.~S4), indicating enhanced transdifferentiation plasticity. More importantly, the emergence of this intermediate source suggests that sufficiently strong population renewal could stabilize a previously unstable transient state, a prediction examined below.

Increasing the proliferation rate indeed drives a qualitative transition in the population dynamics. Above a critical threshold ($\bar{\beta}_c \approx 1.2$), the homeostatic steady state is replaced by sustained oscillations in the total cell number $Q_{\mathrm{tot}}(t)$ (\emph{SI Appendix} Fig.~S5), indicating that population renewal overcomes the intrinsic regulatory repulsion and fundamentally alters the landscape dynamics.

Above the proliferation threshold, macroscopic population oscillations periodically reconstruct the Waddington landscape, producing a \emph{breathing phenotypic landscape} in which transient intermediate attractors repeatedly emerge and disappear (\emph{SI Appendix}, Figs.~S6--S15). Temporal analysis reveals that the oscillatory population dynamics alternately erode and restore the intrinsic landscape barriers through a pulsatile non-equilibrium probability flux, thereby driving the cyclic creation and annihilation of these transient attractors. To quantify this evolving topology, we track the number of local landscape minima over time (\emph{SI Appendix} Fig.~S16). The instantaneous landscape synchronizes with the macroscopic population cycle, alternating among bistable, tristable, and tetrastable configurations. Averaging this topological complexity over time further reveals a continuous proliferation-driven phase transition (Fig.~\ref{fig3}e):  the mean number of landscape minima increases progressively with the renewal rate $\bar{\beta}$, demonstrating that population renewal can generate and sustain landscape topologies absent from the intrinsic GRN alone.

The breathing landscape is further quantified by a continuous structural metric, $A_S(t)$, which measures the relative depth of the uncommitted intermediate state with respect to the two committed attractors (Fig.~\ref{fig3}f):
$$A_S(t) = U(S, t) - \frac{U(L_1, t) + U(L_2, t)}{2},$$
where $S$ denotes the minimum of $U$ constrained along the symmetry axis $x_1 = x_2$. In the subcritical regime ($\bar{\beta} = 0, 1$), $A_S(t)$ rapidly converges to a constant, indicating a stationary landscape topology. Above the critical proliferation threshold, however, $A_S(t)$ undergoes a sharp transition to sustained oscillations with a characteristic period of $T\approx 20$, quantitatively confirming the cyclic reconstruction of the intermediate landscape.

Representative snapshots at high proliferation ($\bar{\beta} = 4$) directly visualize the breathing landscape over successive phases of the oscillation cycles (Fig.~\ref{fig3}g and \emph{SI Appendix} Fig.~S17). Throughout the cycle, the two committed attractors ($L_1$ and $L_2$) remain as persistent landscape minima, whereas the intermediate region periodically develops two transient proliferation-induced attractors (tPIAs) during specific phases of the oscillation ($t = 332$ and $352$). These transient basins correspond to the lower-left region of the $(A_S(t), Q_{\mathrm{tot}}(t))$ phase plane, where both the intermediate landscape depth and the total cell number are minimal, indicating a transient depletion state. The two tPIAs, separated by a pronounced landscape barrier, transiently trap cells in distinct progenitor-like configurations. As the oscillation progresses ($t = 322$ and $342$), the localized probability influx dissipates, the transient basins collapse, and the accumulated progenitor population is released toward the terminal attractors, completing one breathing cycle.

The breathing landscape provides a mechanistic framework for understanding proliferation-induced plasticity \cite{Egli:2008aa, Guo:2014aa}. Unlike classical steady-state landscape descriptions, population renewal acts as an intrinsic, phase-dependent driver that continuously remodels the phenotypic landscape.  The periodic emergence and disappearance of tPIAs is reminiscent of the transient windows of plasticity observed during tissue regeneration \cite{Yui:2018aa,Burclaff:2018aa}. Within our framework, these windows arise from oscillatory population-level probability fluxes that transiently lower landscape barriers, enabling reversible expansion of intermediate progenitor-like states without creating permanent additional lineage architecture. This proliferation-induced landscape remodeling is robust, persisting under substantial variations in the proliferation parameter $\bar{\beta}$ ($\pm 20\%$) and the diffusion coefficient $D_0$ ($\pm 50\%$) (\emph{SI Appendix} Fig.~S18). Together, these results demonstrate that proliferation is not merely a source of population expansion but a dynamical mechanism that actively generates transient attractor topology beyond that encoded by the underlying GRN.

\subsubsection{Tristable circuit with intrinsic memory: Tetra-stable coexistence and topological breathing}

\begin{figure}[htbp]
\centering
\includegraphics[width=0.9\linewidth, clip, bb={100, 350, 1100, 1050}]{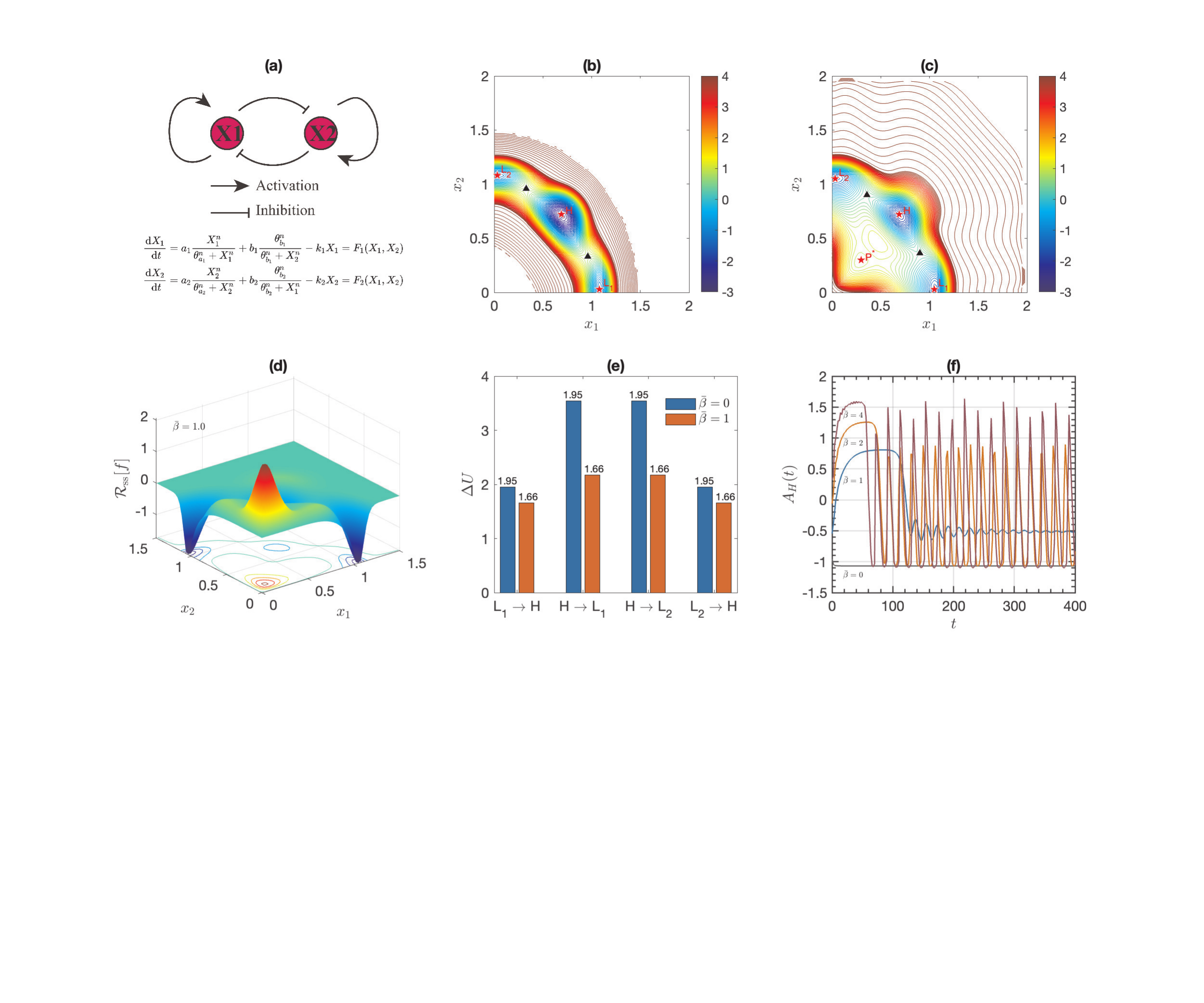}
\caption{\textbf{Proliferation-driven emergence of a tetrastable landscape and rhythmic topological breathing in a gene regulatory network with intrinsic memory.} 
(a) Schematic of the mutual-inhibition circuit with self-activation. 
(b) Steady-state landscape $U^\ast(\vec{x})$ in the absence of population renewal ($\bar{\beta} = 0$). The intrinsic dynamics exhibit a \emph{tristable} topology consisting of two terminal lineage attractors ($L_1$ and $L_2$) and one intrinsic progenitor attractor ($H$). 
(c) Steady-state landscape under moderate population renewal ($\bar{\beta}=1$). A proliferation-induced attractor ($P^\ast $) emerges in the low-expression region while the intrinsic progenitor attractor $H$ is preserved, producing a \emph{tetrastable} landscape.
(d) Spatial distribution of the steady-state population renewal operator $\mathcal{R}^\ast[f]$ at $\bar{\beta} = 1$, showing localized probability influx that stabilizes the proliferation-induced attractor.
(e) Comparison of landscape transition barriers ($\Delta U$) at baseline ($\bar{\beta} = 0$) and under moderate population renewal ($\bar{\beta} = 1$), demonstrating global barrier reduction while preserving the intrinsic progenitor basin. 
(f) Temporal evolution of the intrinsic progenitor landscape depth $A_H(t)$ under increasing population renewal. Above the critical proliferation threshold ($\bar{\beta} = 2, 4$), the tetrastable landscape undergoes sustained limit-cycle oscillations, corresponding to rhythmic topological breathing.}
\label{fig4}
\end{figure}

We next consider the mutual-inhibition circuit with self-activation (Fig.~\ref{fig4}a), a canonical GRN architecture capable of sustaining intrinsic progenitor memory. In the absence of population renewal ($\bar{\beta} = 0$), the steady-state landscape exhibits three stable attractors: two terminal lineage states ($L_1$ and $L_2$) together with an intrinsic progenitor state ($H$) (Fig.~\ref{fig4}b and \emph{SI Appendix} Fig.~S19). Unlike the bistable circuit, where the intermediate state is an unstable saddle, self-activation stabilizes the intermediate progenitor state as a distinct attractor, generating a genuinely tristable landscape that provides an intrinsic reservoir for subsequent fate transitions.

Moderate population renewal ($\bar{\beta} = 1$) transforms the tristable landscape into a tetrastable one by generating an additional proliferation-induced attractor ($P^\ast $) while preserving the intrinsic progenitor state $H$ (Fig.~\ref{fig4}c). This behavior contrasts sharply with the bistable circuit (Fig.~\ref{fig3}g), where population renewal creates only transient intermediate attractors. Here, the intrinsic memory provided by self-activation prevents topological annihilation of the progenitor basin, allowing the proliferation-induced attractor to coexist with the original developmental landscape (\emph{SI Appendix} Fig.~S20).

The tetrastable landscape is maintained by a characteristic source--sink organization of the steady-state population renewal operator $\mathcal{R}^\ast[f]$ (Fig.~\ref{fig4}d), which continuously redistributes probability while globally reshaping the landscape. Specifically, $\mathcal{R}^\ast[f]$ exhibits continuous probability efflux from the three intrinsic attractors ($L_1$, $L_2$, and $H$), balanced by localized influx near the proliferation-induced attractor $P^\ast $. This redistribution stabilizes the additional attractor while simultaneously lowering the global landscape transition barriers (Fig.~\ref{fig4}e). In particular, the reduced differentiation barrier from the intrinsic progenitor state facilitates fate transitions without compromising the progenitor reservoir. Consistent with experimental observations that elevated proliferation promotes developmental plasticity \cite{Ruiz:2011aa}, these results indicate that population renewal is not merely a driver of tissue expansion but an active regulator of landscape topology and cell-fate transitions.

Above the critical proliferation threshold ($\bar{\beta} \geq 2$), the tetrastable landscape undergoes sustained topological breathing rather than converging to a stationary configuration (Fig.~\ref{fig4}f and \emph{SI Appendix} Figs.~S21--S22). To quantify the oscillatory remodeling of the intrinsic progenitor basin, we define the landscape-depth metric $A_H(t)$, analogous to $A_S(t)$ but referenced to the intrinsic progenitor state $H$. As the proliferation rate increases, $A_H(t)$ transitions from damped relaxation to sustained periodic oscillations (Fig.~\ref{fig4}f), demonstrating rhythmic modulation of the intrinsic progenitor landscape.

Projecting the dynamics onto the $(A_H(t), Q_{\mathrm{tot}}(t))$ phase plane reveals a stable limit cycle (\emph{SI Appendix} Fig.~S23), confirming that oscillatory landscape remodeling remains tightly coupled to population renewal throughout the breathing cycle. Unlike the bistable circuit, where proliferation transiently creates and removes the intermediate attractors, self-activation preserves the intrinsic progenitor state while permitting its periodic remodeling. Together, these results show that intrinsic epigenetic memory and population renewal play complementary roles in tissue homeostasis: the former stabilizes the developmental architecture, whereas the latter dynamically reshapes it.

Together, these results establish a dynamical framework in which tissue homeostasis emerges from the interplay between population renewal and intrinsic epigenetic memory. Unlike classical static landscape descriptions, our framework shows that autoregulatory feedback preserves the intrinsic progenitor state while continuous population renewal dynamically remodels the surrounding landscape. This mechanism provides a natural explanation for the sustained progenitor maintenance observed in cyclic hematopoiesis \cite{Morrison:2014aa} and periodic hair follicle regeneration \cite{Gonzales:2017aa}. More broadly, it suggests that resilient stem-cell niches are not simply maintained by static genetic programs, but emerge from the continuous coordination between intracellular regulatory memory and population-level renewal dynamics  \cite{Simons:2011aa, Chen:2014aa}.

\section{Discussion}

The Waddington epigenetic landscape has long served as a central conceptual framework for developmental biology \cite{Waddington:1940, Waddington1957}, yet most quantitative formulations remain rooted in single-cell dynamics and largely detached from the population-level processes that motivated Waddington's original metaphor \cite{Shakiba2022Waddington, Vittadello:2025aa}. Here, we address this gap by defining the landscape $U(\vec{x}, t) = -\log f(\vec{x}, t)$ directly from the evolving epigenetic probability density $f(\vec{x}, t)$, thereby recasting it as a dynamic population-level object. Coupled with the probability-conserving renewed operator $\mathcal{R}[f; Q_{\mathrm{tot}}]$ (aka $\mathcal{R}[f]$ for simplicity), the landscape evolves continuously with proliferation, differentiation, and homeostasis while remaining mechanistically linked to intracellular GRN dynamics.

Our framework complements rather than replaces existing non-equilibrium landscape theories. Wang and colleagues \cite{Wang:2008aa, wang2011landscape, Wang02012015}, together with related studies \cite{AO200963, Zhou:2016gd}, established a thermodynamic decomposition of intracellular dynamics into gradient and curl components. By construction, our formulation recovers this decomposition when population renewal vanishes and the system reaches a steady state. The essential extension is that the landscape is no longer restricted to a stationary property of an isolated cell, but remains well defined throughout transient developmental and regenerative processes driven by continuous population renewal.

Recent population-level formulations have begun incorporating cell birth and death into continuum descriptions of cell-state dynamics. Weinreb et al. \cite{weinreb2018} introduced a transport equation with source and sink terms, and Shi et al. \cite{shi2022energy} further proposed an energy-based differentiation potential that accounts for proliferation. Our framework extends these approaches by replacing scalar birth--death terms with a probability-conserving non-local renewal operator that explicitly represents cell-cycle kinetics and mitotic epigenetic inheritance. The operator preserves global probability while providing a mechanistic description of how proliferation redistributes cells across the epigenetic landscape \cite{Probst:2009aa, Anreiter:2017aa}. This structure also permits analytical questions about stationary distributions and their dependence on the renewal mechanism to be formulated within the same PDE-integral framework.

Emerging from this cross-scale framework, a central finding is that population renewal acts not merely as a quantitative correction to cell abundance, but as a key determinant of phenotypic landscape topology. Rather than introducing a static selective bias \cite{Lander:2009aa}, proliferation functions as a non-equilibrium driving force that remodels transition barriers, reorganizes attractor stability, and, under sufficiently rapid renewal, generates qualitatively new landscape dynamics in a circuit-dependent manner.

The single-gene circuit illustrates this principle by demonstrating that both uniform and state-dependent renewal operators induce the same qualitative landscape remodeling, robustly stabilizing the progenitor state while rendering the stem-like attractor more permissive to outward transitions. This dynamic structural reorganization reveals that cell identity in a renewing tissue is defined not only by its position on the conceptual landscape but also by its functional contribution to population-level probability transport. Within this framework, the progenitor-like state functions as a macroscopic kinetic source that sustains continuous cellular output, whereas the slowly cycling stem-like state acts as a kinetic sink that preserves the lineage origin. Homeostasis therefore emerges from continuous probability circulation across the landscape rather than from static local attractor stability alone, consistent with rapidly renewing tissues such as the hematopoietic system and epithelial compartments \cite{Klein:2011aa,Simons:2011aa}.

The single-gene analysis further illustrates the distinction between the asymptotic steady state and the transient timescale. For each renewal operator, trajectories initiated from entirely State S or entirely State P populations eventually approach the same operator-specific long-time state. Under $\bar{\mathcal{R}}[f]$, however, the all-State-S initial condition produces a long-lived intermediate regime before reaching the asymptotic state. Thus, the renewal operator can strongly modify the transient pathway and the convergence timescale while preserving the long-time population organization. This result highlights how population-level dynamics can exhibit pronounced history dependence over finite observation times even when the asymptotic steady-state distribution is unique. 

Binary gene circuits extend this principle from lineage maintenance to lineage bifurcations \cite{ferrell2012}. In the absence of intrinsic progenitor memory, population renewal does not simply accelerate transitions between committed fates, but transiently creates intermediate phenotypic opportunities by dynamically reshaping the landscape. This proliferation-induced plasticity provides a dynamical explanation for how rapidly expanding tissues transiently explore intermediate cell states without permanently altering lineage architecture. Such behavior is consistent with the transient functional plasticity observed during early embryonic lineage segregation and rapid endothelial expansion \cite{Briggs:2018hg, Schroter:2015aa}.

Above the critical renewal rate, transient plasticity becomes intrinsically rhythmic. Mathematically, oscillations in the macroscopic cell population continuously modulate the probability density, driving the periodic generation and removal of intermediate attractors. This macroscopic driving force produces a ``breathing'' phenotypic landscape, a dynamic behavior fundamentally invisible to conventional models restricted to isolated gene regulatory networks. The breathing phenomenon emerges exclusively from the cross-scale coupling between population-level renewal and intrinsic network dynamics. Such coupling provides a natural explanation for the episodic windows of plasticity observed during regeneration \cite{Burclaff:2018aa}, demonstrating that hyper-proliferation alone can transiently deform phenotypic boundaries without requiring permanent changes to the underlying regulatory circuitry.

Introducing self-activation reveals a general principle: intrinsic GRN memory does not oppose proliferation-induced remodeling, but constrains its long-term outcome. Autoregulatory feedback preserves the core developmental architecture while allowing population renewal to transiently reshape the surrounding landscape, thereby maintaining lineage identity without suppressing phenotypic plasticity. More broadly, these findings suggest that cell-fate regulation in renewing tissues is governed by two complementary processes: intracellular regulatory memory defines the stable architecture of the landscape, whereas population renewal continuously remodels its topology and accessibility. Their interplay provides a dynamical basis for sustained progenitor maintenance in highly regenerative tissues, including cyclic hematopoiesis and hair follicle regeneration \cite{Morrison:2014aa, Gonzales:2017aa}.

Beyond these biological implications, the mathematical analysis establishes the well-posedness of the coupled PDE-integral system and the existence of stationary solutions under proper assumptions. The probability-conserving structure of the renewal operator ensures that the population density remains normalized throughout its evolution, while its non-local coupling to the intracellular dynamics determines how cell-state probability is redistributed across the landscape. These analytical results provide the mathematical foundation for interpreting the dynamic landscapes and their long-time behavior. Notably, for the epigenetically homogeneous case---where the kinetotype is independent of the epigenetic state---the uniqueness of the stationary distribution can be rigorously established. While proving uniqueness for the general state-dependent case remains an open mathematical problem, our numerical simulations of the single-gene models indicate that this uniqueness may hold more broadly. Furthermore, these simulations reveal that even when a unique asymptotic state is reached, different initial conditions and renewal mechanisms can produce substantially different relaxation timescales and transient pathways. Thus, probability conservation and the existence of stationary solutions do not by themselves determine the convergence rate or transient dynamics. Together, the analysis and simulations demonstrate how the renewal mechanism controls both the stationary population organization and the approach to it, providing a basis for analyzing the relationship between long-lived transient regimes and asymptotic steady-state behavior.    

Several fundamental mathematical questions remain open. First, extending the stationary-state analysis to establish rigorous criteria for the uniqueness and global attractivity of the stationary distribution in the general state-dependent case remains a major theoretical challenge. Addressing this issue will likely require developing new entropy-like Lyapunov functionals for non-local renewal operators and advancing the principal spectral theory of coupled integro-differential systems \cite{Su:2023cs}. Second, the recurrent closed orbits observed in the $(A_{\mathrm{top}}(t), Q_{\mathrm{tot}}(t))$ phase plane, where $A_{\mathrm{top}}(t)$ denotes either $A_S(t)$ or $A_H(t)$, suggest that the high-dimensional population dynamics may be organized through a low-dimensional attracting manifold. Deriving a reduced macroscopic system and determining whether topological breathing arises through a Hopf bifurcation would place these numerical observations on a rigorous mathematical foundation. Such a reduction would provide a principled coarse-graining framework for connecting continuum population models with predictive laws of tissue-scale dynamics (see \emph{SI Appendix} G).

Beyond its theoretical implications, our framework provides a natural bridge between mechanistic population dynamics and emerging single-cell data analysis. By defining the epigenetic coordinate $\vec{x}$ as a latent molecular state, the dynamic Waddington landscape becomes compatible with the latent representations increasingly used in modern single-cell machine learning. Within this framework, high-dimensional molecular measurements, such as scRNA-seq profiles, are observable manifestations of an underlying heritable state, whereas the population renewal operator links that latent state to measurable cellular behaviors, including proliferation, differentiation, and cell loss. This connection formalizes the concept of the \emph{kinetotype} \cite{lei2020general} and suggests that cell-state-resolved kinetic measurements are integral to reconstructing developmental landscapes rather than merely complementary annotations.

Realizing this potential will require closer integration between mathematical modeling and quantitative experiments. Because the dynamic landscape $U(\vec{x}, t)$ is defined directly from evolving cell-state distributions, the framework is naturally suited for integration with time-resolved single-cell measurements, including RNA velocity \cite{La-Manno:2018aa} and optimal transport approaches \cite{Sha:2024aa, Peng:2026aa}. Future extensions that incorporate epigenetic memory, spatial organization, and additional layers of cell-state regulation will further broaden its applicability. More generally, our framework illustrates how probabilistic descriptions of cell populations can be integrated with mechanistic gene-regulatory dynamics to establish a predictive, multiscale theory of tissue development and regeneration.

\section{Conclusion}

We have developed a unified PDE-integral framework that establishes the  Waddington epigenetic landscape as a dynamic, probability-conserving description of cell-fate regulation in proliferating populations. By coupling intracellular gene-regulatory dynamics with population renewal, the framework shows that proliferation actively shapes landscape topology rather than merely scaling cell numbers. Population-level kinetics thereby emerge as an integral component of cell-fate regulation, governing attractor stability, transition barriers, and dynamic landscape remodeling. The mathematical analysis further characterizes the stationary population structure and clarifies the distinction between a unique asymptotic steady state and long-lived transient dynamics. More broadly, this work transforms the Waddington landscape from a static conceptual metaphor into a predictive multiscale framework that quantitatively links gene regulation, cell-state dynamics, and tissue homeostasis.

\backmatter

\bmhead{Acknowledgements}

This work was funded by the National Natural Science Foundation of China (NSFC 12331018).

\bmhead{Declarations}
\begin{itemize}
\item Conflict of interest: The authors declare no competing interests.
\item Data and code availability: Simulation code and parameter sets used to generate all numerical results in this manuscript are available at \url{https://github.com/jinzhilei/Dynamic-Waddington-Landscape}. Key parameter values are provided in the SI Appendix. 
\item Use of AI tools: During the preparation of this work, the authors used large language model-based assistants (GPT-5.6 Sol and Gemini 3.1 Pro) for three purposes: (i) discussing the scientific content and the biological interpretation of the results, which sharpened the authors' own understanding of the mathematical results and of the numerical simulations; (ii) auxiliary checking and verification of the mathematical analysis, used to test the authors' own derivations; and (iii) language polishing, expression, and editorial checking of author-written text. The first draft of the manuscript, all mathematical derivations, all numerical computations, and all figures were produced by the authors themselves. All outputs of these tools were reviewed and verified by the authors, were incorporated only where they conformed to the authors' own reasoning, and the authors take full responsibility for the content of the publication. No AI tool is listed as an author, and no text was included without author review.
\end{itemize}

\bibliography{landscape_references}

@preamble{"\providecommand{\noopsort}[1]{}" #
   "\providecommand{\singleletter}[1]{#1}%"}

@article{Lei:2020gd,
	author = {Jinzhi Lei},
	journal = {Sci China Math},
	number = {3},
	pages = {411--424},
	title = {{Evolutionary dynamics of cancer: From epigenetic regulation to cell population dynamics---mathematical model framework, applications, and open problems}},
	volume = {63},
	year = {2020}}

@book{Bogachev:2015aa,
	address = {Providence, RI},
	author = {Vladimir I. Bogachev and Nicolai V. Krylov and Michael R\"{o}ckner and Stanislav V. Shaposhnikov},
	publisher = {American Mathematical Society},
	series = {Mathematical Surveys and Monographs},
	title = {{Fokker--Planck--Kolmogorov Equations}},
	volume = {207},
	year = {2015}}

@article{Hafemeister:2019aa,
	address = {New York Genome Center, 101 6th Ave, New York, 10013, NY, USA. christoph.hafemeister@nyu.edu.; New York Genome Center, 101 6th Ave, New York, 10013, NY, USA. rsatija@nygenome.org.; Center for Genomics and Systems Biology, New York University, 12 Waverly Pl, New York, 10003, NY, USA. rsatija@nygenome.org.},
	auid = {ORCID: 0000-0001-9448-8833},
	author = {Hafemeister, Christoph and Satija, Rahul},
	cois = {The authors declare that they have no competing interests.},
	crdt = {2019/12/25 06:00},
	date = {2019 Dec 23},
	dcom = {20200320},
	dep = {20191223},
	doi = {10.1186/s13059-019-1874-1},
	edat = {2019/12/25 06:00},
	gr = {R35 NS097404/NS/NINDS NIH HHS/United States; R01 MH071679/MH/NIMH NIH HHS/United States; R01 HD096770/HD/NICHD NIH HHS/United States; DP2 HG009623/HG/NHGRI NIH HHS/United States; OT2 OD026673/OD/NIH HHS/United States},
	issn = {1474-760X (Electronic); 1474-7596 (Print); 1474-7596 (Linking)},
	jid = {100960660},
	journal = {Genome Biol},
	jt = {Genome biology},
	language = {eng},
	lid = {10.1186/s13059-019-1874-1 {$[$}doi{$]$}; 296},
	lr = {20231027},
	mh = {Humans; Regression Analysis; *Sequence Analysis, RNA; *Single-Cell Analysis; *Software},
	mhda = {2020/03/21 06:00},
	month = {Dec},
	number = {1},
	oto = {NOTNLM},
	own = {NLM},
	pages = {296},
	phst = {2019/03/17 00:00 {$[$}received{$]$}; 2019/10/30 00:00 {$[$}accepted{$]$}; 2019/12/25 06:00 {$[$}entrez{$]$}; 2019/12/25 06:00 {$[$}pubmed{$]$}; 2020/03/21 06:00 {$[$}medline{$]$}; 2019/12/23 00:00 {$[$}pmc-release{$]$}},
	pii = {10.1186/s13059-019-1874-1; 1874},
	pl = {England},
	pmc = {PMC6927181},
	pmcr = {2019/12/23},
	pmid = {31870423},
	pst = {epublish},
	pt = {Evaluation Study; Journal Article; Research Support, N.I.H., Extramural; Research Support, Non-U.S. Gov't},
	sb = {IM},
	status = {MEDLINE},
	title = {{Normalization and variance stabilization of single-cell RNA-seq data using regularized negative binomial regression}},
	volume = {20},
	year = {2019}}

@article{Stuart:2019aa,
	address = {New York Genome Center, New York, NY, USA.; New York Genome Center, New York, NY, USA. rsatija{\char64}nygenome.org.; Center for Genomics and Systems Biology, New York University, New York, NY, USA. rsatija@nygenome.org.},
	auid = {ORCID: 0000-0002-3044-0897; ORCID: 0000-0001-9448-8833},
	author = {Stuart, Tim and Satija, Rahul},
	cin = {Nat Rev Genet. 2019 May;20(5):249. doi: 10.1038/s41576-019-0118-2. PMID: 30911142},
	crdt = {2019/01/31 06:00},
	date = {2019 May},
	dcom = {20190724},
	doi = {10.1038/s41576-019-0093-7},
	edat = {2019/01/31 06:00},
	issn = {1471-0064 (Electronic); 1471-0056 (Linking)},
	jid = {100962779},
	journal = {Nat Rev Genet},
	jt = {Nature reviews. Genetics},
	language = {eng},
	lid = {10.1038/s41576-019-0093-7 {$[$}doi{$]$}},
	lr = {20220416},
	mh = {Computational Biology/*methods; Data Mining/*statistics \& numerical data; Datasets as Topic; Epigenesis, Genetic; High-Throughput Nucleotide Sequencing; Humans; Proteins/genetics/metabolism; RNA/chemistry/*genetics/metabolism; Single-Cell Analysis/methods/*statistics \& numerical data},
	mhda = {2019/07/25 06:00},
	month = {May},
	number = {5},
	own = {NLM},
	pages = {257--272},
	phst = {2019/01/31 06:00 {$[$}pubmed{$]$}; 2019/07/25 06:00 {$[$}medline{$]$}; 2019/01/31 06:00 {$[$}entrez{$]$}},
	pii = {10.1038/s41576-019-0093-7},
	pl = {England},
	pmid = {30696980},
	pst = {ppublish},
	pt = {Journal Article; Research Support, N.I.H., Extramural; Research Support, Non-U.S. Gov't; Review},
	rn = {0 (Proteins); 63231-63-0 (RNA)},
	status = {MEDLINE},
	title = {{Integrative single-cell analysis}},
	volume = {20},
	year = {2019}}

@article{Wang:2008aa,
	address = {State Key Laboratory of Electroanalytical Chemistry, Changchun Institute of Applied Chemistry, Chinese Academy of Sciences, Changchun, Jilin 130022, People's Republic of China. jin.wang.1@stonybrook.edu},
	author = {Wang, Jin and Xu, Li and Wang, Erkang},
	cois = {The authors declare no conflict of interest.},
	crdt = {2008/08/23 09:00},
	date = {2008 Aug 26},
	dcom = {20081024},
	dep = {20080821},
	doi = {10.1073/pnas.0800579105},
	edat = {2008/08/23 09:00},
	issn = {1091-6490 (Electronic); 0027-8424 (Print); 0027-8424 (Linking)},
	jid = {7505876},
	journal = {Proc Natl Acad Sci USA},
	jt = {Proceedings of the National Academy of Sciences of the United States of America},
	language = {eng},
	lid = {10.1073/pnas.0800579105 {$[$}doi{$]$}},
	lr = {20211020},
	mh = {*Biological Clocks; Cell Cycle; Diffusion; Entropy; Models, Biological; Models, Chemical; *Models, Theoretical},
	mhda = {2008/10/25 09:00},
	month = {Aug},
	number = {34},
	own = {NLM},
	pages = {12271--12276},
	phst = {2008/08/23 09:00 {$[$}pubmed{$]$}; 2008/10/25 09:00 {$[$}medline{$]$}; 2008/08/23 09:00 {$[$}entrez{$]$}; 2009/02/26 00:00 {$[$}pmc-release{$]$}},
	pii = {0800579105; 4179},
	pl = {United States},
	pmc = {PMC2527901},
	pmcr = {2009/02/26},
	pmid = {18719111},
	pst = {ppublish},
	pt = {Journal Article; Research Support, Non-U.S. Gov't},
	sb = {IM},
	status = {MEDLINE},
	title = {{Potential landscape and flux framework of nonequilibrium networks: robustness, dissipation, and coherence of biochemical oscillations}},
	volume = {105},
	year = {2008}}

@article{Chen:2014aa,
	address = {1{$]$} Department of Molecular Carcinogenesis, The University of Texas MD Anderson Cancer Center. [2] Center for Cancer Epigenetics, The University of Texas MD Anderson Cancer Center, Science Park, 1808 Park Road 1C, Smithville, Texas 78957, USA. [3] The University of Texas Graduate School of Biomedical Sciences at Houston, Houston, Texas 77030, USA.; 1{$]$} Department of Molecular Carcinogenesis, The University of Texas MD Anderson Cancer Center. [2] Center for Cancer Epigenetics, The University of Texas MD Anderson Cancer Center, Science Park, 1808 Park Road 1C, Smithville, Texas 78957, USA. [3] The University of Texas Graduate School of Biomedical Sciences at Houston, Houston, Texas 77030, USA.},
	author = {Chen, Taiping and Dent, Sharon Y R},
	crdt = {2013/12/25 06:00},
	date = {2014 Feb},
	dcom = {20140313},
	dep = {20131224},
	doi = {10.1038/nrg3607},
	edat = {2013/12/25 06:00},
	gr = {P30 CA016672/CA/NCI NIH HHS/United States; GM096472/GM/NIGMS NIH HHS/United States; R01 GM067718/GM/NIGMS NIH HHS/United States; R01 GM096472/GM/NIGMS NIH HHS/United States; GM067718/GM/NIGMS NIH HHS/United States},
	issn = {1471-0064 (Electronic); 1471-0056 (Print); 1471-0056 (Linking)},
	jid = {100962779},
	journal = {Nat Rev Genet},
	jt = {Nature reviews. Genetics},
	language = {eng},
	lid = {10.1038/nrg3607 {$[$}doi{$]$}},
	lr = {20220409},
	mh = {Animals; Cell Differentiation/genetics; Cell Lineage/genetics; Chromatin/*genetics/metabolism; Chromatin Assembly and Disassembly/*genetics; *Gene Expression Regulation; Histones/*metabolism; Humans; Models, Genetic; Pluripotent Stem Cells/cytology/metabolism},
	mhda = {2014/03/14 06:00},
	mid = {NIHMS565372},
	month = {Feb},
	number = {2},
	own = {NLM},
	pages = {93--106},
	phst = {2013/12/25 06:00 {$[$}entrez{$]$}; 2013/12/25 06:00 {$[$}pubmed{$]$}; 2014/03/14 06:00 {$[$}medline{$]$}; 2014/08/01 00:00 {$[$}pmc-release{$]$}},
	pii = {nrg3607},
	pl = {England},
	pmc = {PMC3999985},
	pmcr = {2014/08/01},
	pmid = {24366184},
	pst = {ppublish},
	pt = {Journal Article; Research Support, N.I.H., Extramural; Research Support, Non-U.S. Gov't; Review},
	rn = {0 (Chromatin); 0 (Histones)},
	sb = {IM},
	status = {MEDLINE},
	title = {{Chromatin modifiers and remodellers: regulators of cellular differentiation}},
	volume = {15},
	year = {2014}}

@article{Gonzales:2017aa,
	address = {Howard Hughes Medical Institute, Robin Chemers Neustein Laboratory of Mammalian Cell Biology and Development, The Rockefeller University, New York, NY 10065, USA.; Howard Hughes Medical Institute, Robin Chemers Neustein Laboratory of Mammalian Cell Biology and Development, The Rockefeller University, New York, NY 10065, USA. Electronic address: fuchslb@rockefeller.edu.},
	author = {Gonzales, Kevin Andrew Uy and Fuchs, Elaine},
	copyright = {Copyright {\copyright}2017 Elsevier Inc. All rights reserved.},
	crdt = {2017/11/22 06:00},
	date = {2017 Nov 20},
	dcom = {20171219},
	doi = {10.1016/j.devcel.2017.10.001},
	edat = {2017/11/22 06:00},
	gr = {R01 AR027883/AR/NIAMS NIH HHS/United States; R01 AR031737/AR/NIAMS NIH HHS/United States; R01 AR050452/AR/NIAMS NIH HHS/United States; R37 AR027883/AR/NIAMS NIH HHS/United States; HHMI/Howard Hughes Medical Institute/United States},
	issn = {1878-1551 (Electronic); 1534-5807 (Print); 1534-5807 (Linking)},
	jid = {101120028},
	journal = {Dev Cell},
	jt = {Developmental cell},
	language = {eng},
	lid = {S1534-5807(17)30814-6 {$[$}pii{$]$}; 10.1016/j.devcel.2017.10.001 {$[$}doi{$]$}},
	lr = {20220409},
	mh = {Animals; Cell Differentiation/*physiology; Epidermis/pathology; Epithelial Cells/*cytology; Humans; Regeneration/*physiology; Skin/*pathology; Stem Cells/*pathology},
	mhda = {2017/12/20 06:00},
	mid = {NIHMS937808},
	month = {Nov},
	number = {4},
	oto = {NOTNLM},
	own = {NLM},
	pages = {387--401},
	phst = {2017/07/13 00:00 {$[$}received{$]$}; 2017/09/03 00:00 {$[$}revised{$]$}; 2017/10/02 00:00 {$[$}accepted{$]$}; 2017/11/22 06:00 {$[$}entrez{$]$}; 2017/11/22 06:00 {$[$}pubmed{$]$}; 2017/12/20 06:00 {$[$}medline{$]$}; 2018/02/04 00:00 {$[$}pmc-release{$]$}},
	pii = {S1534-5807(17)30814-6},
	pl = {United States},
	pmc = {PMC5797699},
	pmcr = {2018/02/04},
	pmid = {29161590},
	pst = {ppublish},
	pt = {Journal Article; Research Support, N.I.H., Extramural; Research Support, Non-U.S. Gov't; Review},
	sb = {IM},
	status = {MEDLINE},
	title = {{Skin and its regenerative powers: an alliance between stem cells and their niche}},
	volume = {43},
	year = {2017}}

@article{Morrison:2014aa,
	address = {Howard Hughes Medical Institute, Children's Research Institute, Department of Pediatrics, University of Texas Southwestern Medical Center, Dallas, Texas 75390, USA.; Center for Regenerative Medicine, Massachusetts General Hospital, Harvard Stem Cell Institute and the Department of Stem Cell and Regenerative Biology, Harvard University, Cambridge, Massachusetts 02138, USA.},
	author = {Morrison, Sean J and Scadden, David T},
	crdt = {2014/01/17 06:00},
	date = {2014 Jan 16},
	dcom = {20140211},
	doi = {10.1038/nature12984},
	edat = {2014/01/17 06:00},
	gr = {HL096372/HL/NHLBI NIH HHS/United States; R01 HL044851/HL/NHLBI NIH HHS/United States; HL097760/HL/NHLBI NIH HHS/United States; R03 HL096372/HL/NHLBI NIH HHS/United States; HL044851/HL/NHLBI NIH HHS/United States; EB014703/EB/NIBIB NIH HHS/United States; HHMI/Howard Hughes Medical Institute/United States; R01 EB014703/EB/NIBIB NIH HHS/United States; R01 HL097760/HL/NHLBI NIH HHS/United States},
	issn = {1476-4687 (Electronic); 0028-0836 (Print); 0028-0836 (Linking)},
	jid = {0410462},
	journal = {Nature},
	jt = {Nature},
	language = {eng},
	lid = {10.1038/nature12984 {$[$}doi{$]$}},
	lr = {20250529},
	mh = {Animals; Bone Marrow/anatomy \& histology/*physiology; Bone and Bones/cytology; Endothelial Cells/cytology; Hematopoietic Stem Cells/*cytology; Humans; Mesenchymal Stem Cells/cytology; Osteoblasts/cytology; *Stem Cell Niche},
	mhda = {2014/02/12 06:00},
	mid = {NIHMS708815},
	month = {Jan},
	number = {7483},
	own = {NLM},
	pages = {327--334},
	phst = {2013/08/16 00:00 {$[$}received{$]$}; 2013/11/05 00:00 {$[$}accepted{$]$}; 2014/01/17 06:00 {$[$}entrez{$]$}; 2014/01/17 06:00 {$[$}pubmed{$]$}; 2014/02/12 06:00 {$[$}medline{$]$}; 2015/07/24 00:00 {$[$}pmc-release{$]$}},
	pii = {nature12984},
	pl = {England},
	pmc = {PMC4514480},
	pmcr = {2015/07/24},
	pmid = {24429631},
	pst = {ppublish},
	pt = {Journal Article; Research Support, N.I.H., Extramural; Research Support, Non-U.S. Gov't; Review},
	sb = {IM},
	status = {MEDLINE},
	title = {{The bone marrow niche for haematopoietic stem cells}},
	volume = {505},
	year = {2014}}

@article{Burclaff:2018aa,
	address = {Division of Gastroenterology, Departments of Medicine, Pathology \& Immunology, and Developmental Biology, Washington University, St Louis, MO 63110, USA.; Division of Gastroenterology, Departments of Medicine, Pathology \& Immunology, and Developmental Biology, Washington University, St Louis, MO 63110, USA jmills@wustl.edu.},
	auid = {ORCID: 0000-0002-0325-5039; ORCID: 0000-0002-0402-4662},
	author = {Burclaff, Joseph and Mills, Jason C},
	cois = {Competing interestsThe authors declare no competing or financial interests.},
	copyright = {{\copyright}2018. Published by The Company of Biologists Ltd.},
	crdt = {2018/09/02 06:00},
	date = {2018 Aug 31},
	dcom = {20190318},
	dep = {20180831},
	doi = {10.1242/dmm.035071},
	edat = {2018/09/02 06:00},
	gr = {R01 DK094989/DK/NIDDK NIH HHS/United States; P30 CA091842/CA/NCI NIH HHS/United States; R01 DK110406/DK/NIDDK NIH HHS/United States; T32 GM007067/GM/NIGMS NIH HHS/United States; R56 DK110406/DK/NIDDK NIH HHS/United States; R01 DK105129/DK/NIDDK NIH HHS/United States},
	issn = {1754-8411 (Electronic); 1754-8403 (Print); 1754-8403 (Linking)},
	jid = {101483332},
	journal = {Dis Model Mech},
	jt = {Disease models \& mechanisms},
	language = {eng},
	lid = {10.1242/dmm.035071 {$[$}doi{$]$}; dmm035071},
	lr = {20240314},
	mh = {Animals; Carcinogenesis/genetics/*pathology; *Cell Differentiation/genetics; *Cell Plasticity/genetics; Humans; Intestines/*pathology; Skin/*pathology; *Wound Healing/genetics},
	mhda = {2019/03/19 06:00},
	month = {Aug},
	number = {9},
	oto = {NOTNLM},
	own = {NLM},
	phst = {2018/09/02 06:00 {$[$}entrez{$]$}; 2018/09/02 06:00 {$[$}pubmed{$]$}; 2019/03/19 06:00 {$[$}medline{$]$}; 2018/08/31 00:00 {$[$}pmc-release{$]$}},
	pii = {11/9/dmm035071; DMM035071},
	pl = {England},
	pmc = {PMC6177008},
	pmcr = {2018/08/31},
	pmid = {30171151},
	pst = {epublish},
	pt = {Journal Article; Research Support, N.I.H., Extramural; Research Support, Non-U.S. Gov't; Review},
	sb = {IM},
	status = {MEDLINE},
	title = {{Plasticity of differentiated cells in wound repair and tumorigenesis, part II: skin and intestine}},
	volume = {11},
	year = {2018}}

@article{Yui:2018aa,
	address = {BRIC - Biotech Research and Innovation Centre, University of Copenhagen, Ole Maaloes Vej 5, 2200 Copenhagen N, Denmark.; Department of Molecular Medicine, University of Padua School of Medicine, viale Colombo 3, 35126 Padua, Italy.; BRIC - Biotech Research and Innovation Centre, University of Copenhagen, Ole Maaloes Vej 5, 2200 Copenhagen N, Denmark.; BRIC - Biotech Research and Innovation Centre, University of Copenhagen, Ole Maaloes Vej 5, 2200 Copenhagen N, Denmark.; Wellcome - MRC Cambridge Stem Cell Institute, University of Cambridge, Tennis Court Road, Cambridge CB2 1QR, UK.; BRIC - Biotech Research and Innovation Centre, University of Copenhagen, Ole Maaloes Vej 5, 2200 Copenhagen N, Denmark.; BRIC - Biotech Research and Innovation Centre, University of Copenhagen, Ole Maaloes Vej 5, 2200 Copenhagen N, Denmark.; BRIC - Biotech Research and Innovation Centre, University of Copenhagen, Ole Maaloes Vej 5, 2200 Copenhagen N, Denmark.; BRIC - Biotech Research and Innovation Centre, University of Copenhagen, Ole Maaloes Vej 5, 2200 Copenhagen N, Denmark.; BRIC - Biotech Research and Innovation Centre, University of Copenhagen, Ole Maaloes Vej 5, 2200 Copenhagen N, Denmark.; BRIC - Biotech Research and Innovation Centre, University of Copenhagen, Ole Maaloes Vej 5, 2200 Copenhagen N, Denmark.; Department of Gastroenterology, Medical Section, Herlev Hospital, University of Copenhagen, 2730 Herlev, Denmark.; Department of Laboratory Medicine, Division of Translational Cancer Research, Lund University, 223 81 Lund, Sweden.; Department of Gastroenterology and Hepatology, Tokyo Medical and Dental University (TMDU), Bunkyo-ku, Tokyo 113-8519, Japan.; Department of Gastroenterology and Hepatology, Tokyo Medical and Dental University (TMDU), Bunkyo-ku, Tokyo 113-8519, Japan.; Department of Gastroenterology, Medical Section, Herlev Hospital, University of Copenhagen, 2730 Herlev, Denmark.; BRIC - Biotech Research and Innovation Centre, University of Copenhagen, Ole Maaloes Vej 5, 2200 Copenhagen N, Denmark.; Department of Molecular Medicine, University of Padua School of Medicine, viale Colombo 3, 35126 Padua, Italy. Electronic address: piccolo@biouni.pd.it.; BRIC - Biotech Research and Innovation Centre, University of Copenhagen, Ole Maaloes Vej 5, 2200 Copenhagen N, Denmark; Novo Nordisk Foundation Center for Stem Cell Research, Faculty of Health Sciences, University of Copenhagen, 2200 Copenhagen N, Denmark. Electronic address: kim.jensen@bric.ku.dk.},
	author = {Yui, Shiro and Azzolin, Luca and Maimets, Martti and Pedersen, Marianne Terndrup and Fordham, Robert P and Hansen, Stine L and Larsen, Hjalte L and Guiu, Jordi and Alves, Mariana R P and Rundsten, Carsten F and Johansen, Jens V and Li, Yuan and Madsen, Chris D and Nakamura, Tetsuya and Watanabe, Mamoru and Nielsen, Ole H and Schweiger, Pawel J and Piccolo, Stefano and Jensen, Kim B},
	cin = {EMBO J. 2018 Jan 17;37(2):164-166. doi: 10.15252/embj.201798618. PMID: 29282206; Cell Stem Cell. 2018 Jan 4;22(1):7-9. doi: 10.1016/j.stem.2017.12.006. PMID: 29304343},
	copyright = {Copyright {\copyright}2017 The Authors. Published by Elsevier Inc. All rights reserved.},
	crdt = {2017/12/19 06:00},
	date = {2018 Jan 4},
	dcom = {20190114},
	dep = {20171214},
	doi = {10.1016/j.stem.2017.11.001},
	edat = {2017/12/19 06:00},
	gr = {13-1216/AICR{\_}/Worldwide Cancer Research/United Kingdom; 670126/ERC{\_}/European Research Council/International; MC{\_}PC{\_}12009/MRC{\_}/Medical Research Council/United Kingdom},
	issn = {1875-9777 (Electronic); 1934-5909 (Print); 1875-9777 (Linking)},
	jid = {101311472},
	journal = {Cell Stem Cell},
	jt = {Cell stem cell},
	language = {eng},
	lid = {S1934-5909(17)30453-8 {$[$}pii{$]$}; 10.1016/j.stem.2017.11.001 {$[$}doi{$]$}},
	lr = {20250530},
	mh = {Adaptor Proteins, Signal Transducing/*metabolism; Animals; Biomarkers/metabolism; Cell Cycle Proteins; *Cellular Reprogramming; Extracellular Matrix/*metabolism; Fetus/metabolism; Humans; Intestinal Mucosa/*metabolism/*pathology; Mechanotransduction, Cellular; Mice, Inbred C57BL; Phosphoproteins/*metabolism; *Regeneration; Signal Transduction; Transcription, Genetic; Transcriptional Activation/genetics; YAP-Signaling Proteins},
	mhda = {2019/01/15 06:00},
	month = {Jan},
	number = {1},
	oto = {NOTNLM},
	own = {NLM},
	pages = {35--49},
	phst = {2017/03/10 00:00 {$[$}received{$]$}; 2017/09/25 00:00 {$[$}revised{$]$}; 2017/10/31 00:00 {$[$}accepted{$]$}; 2017/12/19 06:00 {$[$}pubmed{$]$}; 2019/01/15 06:00 {$[$}medline{$]$}; 2017/12/19 06:00 {$[$}entrez{$]$}; 2018/01/04 00:00 {$[$}pmc-release{$]$}},
	pii = {S1934-5909(17)30453-8},
	pl = {United States},
	pmc = {PMC5766831},
	pmcr = {2018/01/04},
	pmid = {29249464},
	pst = {ppublish},
	pt = {Journal Article; Research Support, Non-U.S. Gov't},
	rn = {0 (Adaptor Proteins, Signal Transducing); 0 (Biomarkers); 0 (Cell Cycle Proteins); 0 (Phosphoproteins); 0 (YAP-Signaling Proteins); 0 (Yap1 protein, mouse)},
	sb = {IM},
	status = {MEDLINE},
	title = {{YAP/TAZ-dependent reprogramming of colonic epithelium links ECM remodeling to tissue regeneration}},
	volume = {22},
	year = {2018}}

@article{lei2020general,
	address = {Zhou Pei-Yuan Center for Applied Mathematics, MOE Key Laboratory of Bioinformatics, Tsinghua University, Beijing 100084, China. Electronic address: jzlei@tsinghua.edu.cn.},
	author = {Lei, Jinzhi},
	copyright = {Copyright {\copyright}2020 Elsevier Ltd. All rights reserved.},
	crdt = {2020/02/19 06:00},
	date = {2020 May 7},
	dcom = {20210621},
	dep = {20200214},
	doi = {10.1016/j.jtbi.2020.110196},
	edat = {2020/02/19 06:00},
	ein = {J Theor Biol. 2024 Oct 7;593:111896. doi: 10.1016/j.jtbi.2024.111896. PMID: 38992462},
	issn = {1095-8541 (Electronic); 0022-5193 (Linking)},
	jid = {0376342},
	journal = {J Theor Biol},
	jt = {Journal of theoretical biology},
	language = {eng},
	lid = {S0022-5193(20)30051-5 {$[$}pii{$]$}; 10.1016/j.jtbi.2020.110196 {$[$}doi{$]$}},
	lr = {20240711},
	mh = {*Apoptosis; Cell Cycle; Cell Differentiation; Cell Division; Models, Biological; *Stem Cells},
	mhda = {2021/06/22 06:00},
	month = {May},
	oto = {NOTNLM},
	own = {NLM},
	pages = {110196},
	phst = {2019/03/23 00:00 {$[$}received{$]$}; 2019/12/28 00:00 {$[$}revised{$]$}; 2020/02/11 00:00 {$[$}accepted{$]$}; 2020/02/19 06:00 {$[$}pubmed{$]$}; 2021/06/22 06:00 {$[$}medline{$]$}; 2020/02/19 06:00 {$[$}entrez{$]$}},
	pii = {S0022-5193(20)30051-5},
	pl = {England},
	pmid = {32067937},
	pst = {ppublish},
	pt = {Journal Article; Research Support, Non-U.S. Gov't},
	sb = {IM},
	status = {MEDLINE},
	title = {{A general mathematical framework for understanding the behavior of heterogeneous stem cell regeneration}},
	volume = {492},
	year = {2020}}

@article{Su:2023cs,
	author = {Su, Yuan-Hang and Li, Wan-Tong and Lou, Yuan and Wang, Xuefeng},
	doi = {10.1016/j.matpur.2023.06.006},
	issn = {0021-7824},
	journal = {J Math Pures Appl},
	pages = {226--281},
	title = {{Principal spectral theory for nonlocal systems and applications to stem cell regeneration models}},
	volume = {176},
	year = {2023}}

@article{Sha:2024aa,
	address = {Department of Mathematics, University of California, Irvine, Irvine, CA USA. GRID: grid.266093.8. ISNI: 0000 0001 0668 7243; Department of Mathematics, Michigan State University, East Lansing, MI USA. ROR: https://ror.org/05hs6h993. GRID: grid.17088.36. ISNI: 0000 0001 2195 6501; Department of Mathematics, University of California, Irvine, Irvine, CA USA. GRID: grid.266093.8. ISNI: 0000 0001 0668 7243; Department of Mathematics, University of California, Irvine, Irvine, CA USA. GRID: grid.266093.8. ISNI: 0000 0001 0668 7243; Department of Developmental and Cell Biology, University of California, Irvine, Irvine, CA USA. GRID: grid.266093.8. ISNI: 0000 0001 0668 7243; The NSF-Simons Center for Multiscale Cell Fate Research, University of California, Irvine, Irvine, CA USA. GRID: grid.266093.8. ISNI: 0000 0001 0668 7243},
	auid = {ORCID: 0000-0001-5530-734X; ORCID: 0000-0002-7058-6428; ORCID: 0000-0002-8804-3368},
	author = {Sha, Yutong and Qiu, Yuchi and Zhou, Peijie and Nie, Qing},
	cois = {Competing interestsThe authors declare no competing interests.},
	copyright = {{\copyright}The Author(s) 2023, corrected publication 2023.},
	crdt = {2024/01/26 03:53},
	date = {2024},
	dep = {20231130},
	doi = {10.1038/s42256-023-00763-w},
	edat = {2024/01/26 06:43},
	gr = {R01 AR079150/AR/NIAMS NIH HHS/United States; U01 AR073159/AR/NIAMS NIH HHS/United States},
	issn = {2522-5839 (Electronic); 2522-5839 (Print); 2522-5839 (Linking)},
	jid = {101740243},
	journal = {Nat Mach Intell},
	jt = {Nature machine intelligence},
	language = {eng},
	lid = {10.1038/s42256-023-00763-w {$[$}doi{$]$}},
	lr = {20241023},
	mhda = {2024/01/26 06:44},
	number = {1},
	oto = {NOTNLM},
	own = {NLM},
	pages = {25--39},
	phst = {2023/02/08 00:00 {$[$}received{$]$}; 2023/10/25 00:00 {$[$}accepted{$]$}; 2024/01/26 06:44 {$[$}medline{$]$}; 2024/01/26 06:43 {$[$}pubmed{$]$}; 2024/01/26 03:53 {$[$}entrez{$]$}; 2023/11/30 00:00 {$[$}pmc-release{$]$}},
	pii = {763},
	pl = {England},
	pmc = {PMC10805654},
	pmcr = {2023/11/30},
	pmid = {38274364},
	pst = {ppublish},
	pt = {Journal Article},
	status = {PubMed-not-MEDLINE},
	title = {{Reconstructing growth and dynamic trajectories from single-cell transcriptomics data}},
	volume = {6},
	year = {2024}}

@article{Peng:2026aa,
	address = {LMAM and School of Mathematical Sciences, Peking University, Beijing, China.; Center for Machine Learning Research, Peking University, Beijing, China. pjzhou@pku.edu.cn.; Center for Quantitative Biology, Peking University, Beijing, China. pjzhou@pku.edu.cn.; National Engineering Laboratory for Big Data Analysis and Applications, Beijing, China. pjzhou@pku.edu.cn.; AI for Science Institute, Beijing, China. pjzhou{\char64}pku.edu.cn.; LMAM and School of Mathematical Sciences, Peking University, Beijing, China. tieli@pku.edu.cn.; Center for Machine Learning Research, Peking University, Beijing, China. tieli@pku.edu.cn.; National Engineering Laboratory for Big Data Analysis and Applications, Beijing, China. tieli@pku.edu.cn.},
	auid = {ORCID: 0009-0003-4251-0666; ORCID: 0000-0002-4585-2923; ORCID: 0000-0002-2086-1620},
	author = {Peng, Qiangwei and Zhou, Peijie and Li, Tiejun},
	cois = {Competing interests: The authors declare no competing interests.},
	copyright = {{\copyright}2026. The Author(s), under exclusive licence to Springer Nature America, Inc.},
	crdt = {2026/03/13 00:34},
	date = {2026 Mar},
	dcom = {20260313},
	dep = {20260312},
	doi = {10.1038/s41592-026-03010-3},
	edat = {2026/03/13 06:52},
	issn = {1548-7105 (Electronic); 1548-7091 (Linking)},
	jid = {101215604},
	journal = {Nat Methods},
	jt = {Nature methods},
	language = {eng},
	lid = {10.1038/s41592-026-03010-3 {$[$}doi{$]$}},
	lr = {20260313},
	mh = {Animals; *Single-Cell Analysis/methods; Cell Differentiation; Cell Proliferation; Spatio-Temporal Analysis; Cell Movement; *Deep Learning; Drosophila/embryology; Brain/physiology/cytology; *Video Recording/methods; Embryo, Nonmammalian},
	mhda = {2026/03/13 06:53},
	month = {Mar},
	number = {3},
	own = {NLM},
	pages = {542--553},
	phst = {2024/11/01 00:00 {$[$}received{$]$}; 2026/01/26 00:00 {$[$}accepted{$]$}; 2026/03/13 06:53 {$[$}medline{$]$}; 2026/03/13 06:52 {$[$}pubmed{$]$}; 2026/03/13 00:34 {$[$}entrez{$]$}},
	pii = {10.1038/s41592-026-03010-3},
	pl = {United States},
	pmid = {41820580},
	pst = {ppublish},
	pt = {Journal Article},
	sb = {IM},
	status = {MEDLINE},
	title = {{stVCR: spatiotemporal dynamics of single cells}},
	volume = {23},
	year = {2026}}

@article{La-Manno:2018aa,
	address = {Laboratory of Molecular Neurobiology, Department of Medical Biochemistry and Biophysics, Karolinska Institutet, Stockholm, Sweden.; Science for Life Laboratory, Solna, Sweden.; Department of Biomedical Informatics, Harvard Medical School, Boston, MA, USA.; Laboratory of Molecular Neurobiology, Department of Medical Biochemistry and Biophysics, Karolinska Institutet, Stockholm, Sweden.; Science for Life Laboratory, Solna, Sweden.; Laboratory of Molecular Neurobiology, Department of Medical Biochemistry and Biophysics, Karolinska Institutet, Stockholm, Sweden.; Science for Life Laboratory, Solna, Sweden.; Laboratory of Molecular Neurobiology, Department of Medical Biochemistry and Biophysics, Karolinska Institutet, Stockholm, Sweden.; Science for Life Laboratory, Solna, Sweden.; Department of Biomedical Informatics, Harvard Medical School, Boston, MA, USA.; Department of Applied Mathematics, Peter The Great St. Petersburg Polytechnic University, St, Petersburg, Russia.; Department of Biosciences and Nutrition, Karolinska Institutet, Stockholm, Sweden.; Department of Physiology and Pharmacology, Karolinska Institutet, Stockholm, Sweden.; Laboratory of Molecular Neurobiology, Department of Medical Biochemistry and Biophysics, Karolinska Institutet, Stockholm, Sweden.; Science for Life Laboratory, Solna, Sweden.; Laboratory of Molecular Neurobiology, Department of Medical Biochemistry and Biophysics, Karolinska Institutet, Stockholm, Sweden.; Department of Biomedical Informatics, Harvard Medical School, Boston, MA, USA.; Laboratory of Molecular Neurobiology, Department of Medical Biochemistry and Biophysics, Karolinska Institutet, Stockholm, Sweden.; Science for Life Laboratory, Solna, Sweden.; Department of Biomedical Informatics, Harvard Medical School, Boston, MA, USA.; Laboratory of Molecular Neurobiology, Department of Medical Biochemistry and Biophysics, Karolinska Institutet, Stockholm, Sweden.; Department of Biomedical Informatics, Harvard Medical School, Boston, MA, USA.; John van Geest Centre for Brain Repair, Department of Clinical Neurosciences, University of Cambridge, Cambridge, UK.; John van Geest Centre for Brain Repair, Department of Clinical Neurosciences, University of Cambridge, Cambridge, UK.; Division of Neurodegeneration, Department of Neurobiology, Care Sciences and Society, Karolinska Institutet, Stockholm, Sweden.; Laboratory of Molecular Neurobiology, Department of Medical Biochemistry and Biophysics, Karolinska Institutet, Stockholm, Sweden.; Department of Biosciences and Nutrition, Karolinska Institutet, Stockholm, Sweden.; Max Planck Institute for Biophysical Chemistry, Department of Molecular Biology, G{\"o}ttingen, Germany.; Department of Physiology and Pharmacology, Karolinska Institutet, Stockholm, Sweden.; Laboratory of Molecular Neurobiology, Department of Medical Biochemistry and Biophysics, Karolinska Institutet, Stockholm, Sweden. sten.linnarsson@ki.se.; Science for Life Laboratory, Solna, Sweden. sten.linnarsson{\char64}ki.se.; Department of Biomedical Informatics, Harvard Medical School, Boston, MA, USA. peter_kharchenko@hms.harvard.edu.; Harvard Stem Cell Institute, Cambridge, MA, USA. peter_kharchenko@hms.harvard.edu.},
	author = {La Manno, Gioele and Soldatov, Ruslan and Zeisel, Amit and Braun, Emelie and Hochgerner, Hannah and Petukhov, Viktor and Lidschreiber, Katja and Kastriti, Maria E and L{\"o}nnerberg, Peter and Furlan, Alessandro and Fan, Jean and Borm, Lars E and Liu, Zehua and van Bruggen, David and Guo, Jimin and He, Xiaoling and Barker, Roger and Sundstr{\"o}m, Erik and Castelo-Branco, Gon{\c c}alo and Cramer, Patrick and Adameyko, Igor and Linnarsson, Sten and Kharchenko, Peter V},
	cin = {Nature. 2018 Aug;560(7719):434-435. doi: 10.1038/d41586-018-05882-8. PMID: 30127477; Nat Rev Genet. 2018 Nov;19(11):668-669. doi: 10.1038/s41576-018-0049-3. PMID: 30154482; Mol Cell. 2018 Oct 4;72(1):7-9. doi: 10.1016/j.molcel.2018.09.026. PMID: 30290149},
	cois = {The authors declare no competing interests.},
	crdt = {2018/08/10 06:00},
	date = {2018 Aug},
	dcom = {20190220},
	dep = {20180808},
	doi = {10.1038/s41586-018-0414-6},
	edat = {2018/08/10 06:00},
	gr = {108726/WT{\_}/Wellcome Trust/United Kingdom; 261063/ERC{\_}/European Research Council/International; MC{\_}PC{\_}12009/MRC{\_}/Medical Research Council/United Kingdom; R01 HL131768/HL/NHLBI NIH HHS/United States},
	issn = {1476-4687 (Electronic); 0028-0836 (Print); 0028-0836 (Linking)},
	jid = {0410462},
	journal = {Nature},
	jt = {Nature},
	language = {eng},
	lid = {10.1038/s41586-018-0414-6 {$[$}doi{$]$}},
	lr = {20250530},
	mh = {Animals; Brain/*cytology/embryology/metabolism; Cell Lineage/genetics; Chromaffin Cells/cytology/metabolism; Datasets as Topic; Female; Glutamic Acid/metabolism; Hippocampus/cytology/embryology/metabolism; Kinetics; Male; Mice; Neural Crest/cytology/*metabolism; Neurons/*cytology/metabolism; RNA/*analysis/*genetics; RNA Splicing/*genetics; Reproducibility of Results; *Sequence Analysis, RNA; *Single-Cell Analysis; Time Factors; Transcription, Genetic/genetics},
	mhda = {2019/03/21 06:00},
	mid = {EMS78511},
	month = {Aug},
	number = {7719},
	own = {NLM},
	pages = {494--498},
	phst = {2017/12/13 00:00 {$[$}received{$]$}; 2018/07/03 00:00 {$[$}accepted{$]$}; 2018/08/10 06:00 {$[$}pubmed{$]$}; 2019/03/21 06:00 {$[$}medline{$]$}; 2018/08/10 06:00 {$[$}entrez{$]$}; 2019/02/08 00:00 {$[$}pmc-release{$]$}},
	pii = {10.1038/s41586-018-0414-6},
	pl = {England},
	pmc = {PMC6130801},
	pmcr = {2019/02/08},
	pmid = {30089906},
	pst = {ppublish},
	pt = {Journal Article; Research Support, N.I.H., Extramural; Research Support, Non-U.S. Gov't; Research Support, U.S. Gov't, Non-P.H.S.; Validation Study},
	rn = {3KX376GY7L (Glutamic Acid); 63231-63-0 (RNA)},
	sb = {IM},
	status = {MEDLINE},
	title = {{RNA velocity of single cells}},
	volume = {560},
	year = {2018}}

@article{Simons:2011aa,
	address = {Cavendish Laboratory, Department of Physics, J.J. Thomson Avenue, University of Cambridge, Cambridge CB3 0HE, United Kingdom.},
	author = {Simons, Benjamin D and Clevers, Hans},
	copyright = {Copyright {\copyright}2011 Elsevier Inc. All rights reserved.},
	crdt = {2011/06/14 06:00},
	date = {2011 Jun 10},
	dcom = {20110817},
	doi = {10.1016/j.cell.2011.05.033},
	edat = {2011/06/15 06:00},
	issn = {1097-4172 (Electronic); 0092-8674 (Linking)},
	jid = {0413066},
	journal = {Cell},
	jt = {Cell},
	language = {eng},
	lid = {10.1016/j.cell.2011.05.033 {$[$}doi{$]$}},
	lr = {20220408},
	mh = {Adult Stem Cells/*cytology/metabolism; Animals; Cell Division; Cell Lineage; Drosophila/cytology; Humans},
	mhda = {2011/08/19 06:00},
	month = {Jun},
	number = {6},
	own = {NLM},
	pages = {851--862},
	phst = {2011/04/15 00:00 {$[$}received{$]$}; 2011/05/16 00:00 {$[$}revised{$]$}; 2011/05/23 00:00 {$[$}accepted{$]$}; 2011/06/14 06:00 {$[$}entrez{$]$}; 2011/06/15 06:00 {$[$}pubmed{$]$}; 2011/08/19 06:00 {$[$}medline{$]$}},
	pii = {S0092-8674(11)00594-0},
	pl = {United States},
	pmid = {21663791},
	pst = {ppublish},
	pt = {Journal Article; Review},
	sb = {IM},
	status = {MEDLINE},
	title = {{Strategies for homeostatic stem cell self-renewal in adult tissues}},
	volume = {145},
	year = {2011}}

@article{Guo:2014aa,
	address = {Department of Cell Biology, Yale University, New Haven, CT 06520, USA; Yale Stem Cell Center, Yale University, New Haven, CT 06520, USA. Electronic address: shangqin.guo@yale.edu.; Department of Biomedical Engineering, Yale University, New Haven, CT 06520, USA; Department of Cell Biology, Second Military Medical University, Shanghai 200433, China.; Department of Pediatrics, Yale University, New Haven, CT 06520, USA.; Yale Stem Cell Center, Yale University, New Haven, CT 06520, USA; Department of Genetics, Yale University, New Haven, CT 06520, USA.; Department of Cell Biology, Yale University, New Haven, CT 06520, USA; Yale Stem Cell Center, Yale University, New Haven, CT 06520, USA.; Department of Cell Biology, Yale University, New Haven, CT 06520, USA; Yale Stem Cell Center, Yale University, New Haven, CT 06520, USA.; Department of Cell Biology, Yale University, New Haven, CT 06520, USA; Yale Stem Cell Center, Yale University, New Haven, CT 06520, USA.; Yale Stem Cell Center, Yale University, New Haven, CT 06520, USA; Department of Genetics, Yale University, New Haven, CT 06520, USA.; Department of Immunobiology, Yale Flow Cytometry Core Facility, Yale University, New Haven, CT 06520, USA.; Yale Stem Cell Center, Yale University, New Haven, CT 06520, USA; Department of Genetics, Yale University, New Haven, CT 06520, USA; Yale Comprehensive Cancer Center, Yale University, New Haven, CT 06520, USA.; Department of Pediatrics, Yale University, New Haven, CT 06520, USA; Department of Genetics, Yale University, New Haven, CT 06520, USA; Department of Pathology, Yale University, New Haven, CT 06520, USA.; Department of Cell Biology, Yale University, New Haven, CT 06520, USA; Yale Stem Cell Center, Yale University, New Haven, CT 06520, USA; Department of Laboratory Medicine, Yale University, New Haven, CT 06520, USA.; Yale Stem Cell Center, Yale University, New Haven, CT 06520, USA; Department of Biomedical Engineering, Yale University, New Haven, CT 06520, USA; Yale Comprehensive Cancer Center, Yale University, New Haven, CT 06520, USA.; Yale Stem Cell Center, Yale University, New Haven, CT 06520, USA; Department of Genetics, Yale University, New Haven, CT 06520, USA; Yale Comprehensive Cancer Center, Yale University, New Haven, CT 06520, USA.},
	author = {Guo, Shangqin and Zi, Xiaoyuan and Schulz, Vincent P and Cheng, Jijun and Zhong, Mei and Koochaki, Sebastian H J and Megyola, Cynthia M and Pan, Xinghua and Heydari, Kartoosh and Weissman, Sherman M and Gallagher, Patrick G and Krause, Diane S and Fan, Rong and Lu, Jun},
	cin = {Cell. 2014 Feb 13;156(4):631-2. doi: 10.1016/j.cell.2014.01.046. PMID: 24529370},
	copyright = {Copyright {\copyright}2014 Elsevier Inc. All rights reserved.},
	crdt = {2014/02/04 06:00},
	date = {2014 Feb 13},
	dcom = {20140423},
	dep = {20140130},
	doi = {10.1016/j.cell.2014.01.020},
	edat = {2014/02/04 06:00},
	gr = {R01 HL106184/HL/NHLBI NIH HHS/United States; P01GM099130/GM/NIGMS NIH HHS/United States; R01 CA149109/CA/NCI NIH HHS/United States; R01HL106184/HL/NHLBI NIH HHS/United States; R01 DK086267/DK/NIDDK NIH HHS/United States; R01CA149109/CA/NCI NIH HHS/United States; U54 CA143798/CA/NCI NIH HHS/United States; P30 DK0724429/DK/NIDDK NIH HHS/United States; P01 GM099130/GM/NIGMS NIH HHS/United States; K01 DK082982/DK/NIDDK NIH HHS/United States},
	issn = {1097-4172 (Electronic); 0092-8674 (Print); 0092-8674 (Linking)},
	jid = {0413066},
	journal = {Cell},
	jt = {Cell},
	language = {eng},
	lid = {S0092-8674(14)00072-5 {$[$}pii{$]$}; 10.1016/j.cell.2014.01.020 {$[$}doi{$]$}},
	lr = {20211021},
	mh = {Animals; Bone Marrow Cells; Cell Differentiation; *Cellular Reprogramming; Fibroblasts/cytology/metabolism; Gene Knockdown Techniques; Genes, p53; Granulocyte-Macrophage Progenitor Cells/*cytology/metabolism; *Induced Pluripotent Stem Cells; Mice},
	mhda = {2014/04/24 06:00},
	mid = {NIHMS558861},
	month = {Feb},
	number = {4},
	own = {NLM},
	pages = {649--662},
	phst = {2013/06/07 00:00 {$[$}received{$]$}; 2013/11/04 00:00 {$[$}revised{$]$}; 2014/01/10 00:00 {$[$}accepted{$]$}; 2014/02/04 06:00 {$[$}entrez{$]$}; 2014/02/04 06:00 {$[$}pubmed{$]$}; 2014/04/24 06:00 {$[$}medline{$]$}; 2015/02/13 00:00 {$[$}pmc-release{$]$}},
	pii = {S0092-8674(14)00072-5},
	pl = {United States},
	pmc = {PMC4318260},
	pmcr = {2015/02/13},
	pmid = {24486105},
	pst = {ppublish},
	pt = {Journal Article; Research Support, N.I.H., Extramural; Research Support, Non-U.S. Gov't},
	sb = {IM},
	si = {GEO/GSE53074},
	status = {MEDLINE},
	title = {{Nonstochastic reprogramming from a privileged somatic cell state}},
	volume = {156},
	year = {2014}}

@article{Egli:2008aa,
	address = {The Stowers Medical Institute, Harvard Stem Cell Institute and the Department of Stem Cell and Regenerative Biology, Harvard University, Cambridge, Massachusetts, USA.},
	author = {Egli, Dieter and Birkhoff, Garrett and Eggan, Kevin},
	crdt = {2008/06/24 09:00},
	date = {2008 Jul},
	dcom = {20080725},
	doi = {10.1038/nrm2439},
	edat = {2008/06/24 09:00},
	gr = {R01 HD046732/HD/NICHD NIH HHS/United States},
	issn = {1471-0080 (Electronic); 1471-0072 (Print); 1471-0072 (Linking)},
	jid = {100962782},
	journal = {Nat Rev Mol Cell Biol},
	jt = {Nature reviews. Molecular cell biology},
	language = {eng},
	lid = {10.1038/nrm2439 {$[$}doi{$]$}},
	lr = {20211020},
	mh = {Animals; Cell Nucleus/metabolism; *Cloning, Organism/methods; Epigenesis, Genetic; *Gene Expression Regulation; Humans; Mitosis/*physiology; Nuclear Transfer Techniques; Pluripotent Stem Cells/cytology/*physiology; Transcription Factors/*metabolism; Transcription, Genetic},
	mhda = {2008/07/26 09:00},
	mid = {NIHMS1583934},
	month = {Jul},
	number = {7},
	own = {NLM},
	pages = {505--516},
	phst = {2008/06/24 09:00 {$[$}pubmed{$]$}; 2008/07/26 09:00 {$[$}medline{$]$}; 2008/06/24 09:00 {$[$}entrez{$]$}; 2020/05/26 00:00 {$[$}pmc-release{$]$}},
	pii = {nrm2439},
	pl = {England},
	pmc = {PMC7250051},
	pmcr = {2020/05/26},
	pmid = {18568039},
	pst = {ppublish},
	pt = {Journal Article; Research Support, Non-U.S. Gov't; Review},
	rn = {0 (Transcription Factors)},
	sb = {IM},
	status = {MEDLINE},
	title = {{Mediators of reprogramming: transcription factors and transitions through mitosis}},
	volume = {9},
	year = {2008}}

@article{Ruiz:2011aa,
	address = {Gene Expression Laboratory, The Salk Institute for Biological Studies, 10010 North Torrey Pines Road, La Jolla, CA 92037, USA.},
	author = {Ruiz, Sergio and Panopoulos, Athanasia D and Herrer{\'\i}as, A{\'\i}da and Bissig, Karl-Dimiter and Lutz, Margaret and Berggren, W Travis and Verma, Inder M and Izpisua Belmonte, Juan Carlos},
	copyright = {Copyright {\copyright}2011 Elsevier Ltd. All rights reserved.},
	crdt = {2010/12/21 06:00},
	date = {2011 Jan 11},
	dcom = {20110516},
	dep = {20101217},
	doi = {10.1016/j.cub.2010.11.049},
	edat = {2010/12/21 06:00},
	gr = {T32 CA009370/CA/NCI NIH HHS/United States},
	issn = {1879-0445 (Electronic); 0960-9822 (Print); 0960-9822 (Linking)},
	jid = {9107782},
	journal = {Curr Biol},
	jt = {Current biology : CB},
	language = {eng},
	lid = {10.1016/j.cub.2010.11.049 {$[$}doi{$]$}},
	lr = {20250529},
	mh = {Cell Cycle; Cell Differentiation/*physiology; Cell Line; *Cell Proliferation; Embryonic Stem Cells/*cytology/*physiology; Humans},
	mhda = {2011/05/17 06:00},
	mid = {NIHMS254643},
	month = {Jan},
	number = {1},
	own = {NLM},
	pages = {45--52},
	phst = {2010/08/12 00:00 {$[$}received{$]$}; 2010/10/21 00:00 {$[$}revised{$]$}; 2010/11/17 00:00 {$[$}accepted{$]$}; 2010/12/21 06:00 {$[$}entrez{$]$}; 2010/12/21 06:00 {$[$}pubmed{$]$}; 2011/05/17 06:00 {$[$}medline{$]$}; 2012/01/11 00:00 {$[$}pmc-release{$]$}},
	pii = {S0960-9822(10)01515-0},
	pl = {England},
	pmc = {PMC3034649},
	pmcr = {2012/01/11},
	pmid = {21167714},
	pst = {ppublish},
	pt = {Journal Article; Research Support, N.I.H., Extramural; Research Support, Non-U.S. Gov't},
	sb = {IM},
	status = {MEDLINE},
	title = {{A high proliferation rate is required for cell reprogramming and maintenance of human embryonic stem cell identity}},
	volume = {21},
	year = {2011}}

@article{Gouti:2014aa,
	address = {MRC-National Institute for Medical Research, London, United Kingdom.; MRC Centre for Regenerative Medicine, Institute for Stem Cell Research, School of Biological Sciences, University of Edinburgh, Edinburgh, United Kingdom.; MRC Centre for Regenerative Medicine, Institute for Stem Cell Research, School of Biological Sciences, University of Edinburgh, Edinburgh, United Kingdom.; MRC Centre for Regenerative Medicine, Institute for Stem Cell Research, School of Biological Sciences, University of Edinburgh, Edinburgh, United Kingdom.; MRC-National Institute for Medical Research, London, United Kingdom.; MRC Centre for Regenerative Medicine, Institute for Stem Cell Research, School of Biological Sciences, University of Edinburgh, Edinburgh, United Kingdom.; MRC-National Institute for Medical Research, London, United Kingdom.},
	author = {Gouti, Mina and Tsakiridis, Anestis and Wymeersch, Filip J and Huang, Yali and Kleinjung, Jens and Wilson, Valerie and Briscoe, James},
	cin = {Development of Spinal Cord Neurons in Delicate Balance.},
	cois = {The authors have declared that no competing interests exist.},
	crdt = {2014/08/27 06:00},
	date = {2014 Aug},
	dcom = {20150409},
	dep = {20140826},
	doi = {10.1371/journal.pbio.1001937},
	edat = {2014/08/27 06:00},
	gr = {MR/K011200/1/MRC{\_}/Medical Research Council/United Kingdom; U117560541/MRC{\_}/Medical Research Council/United Kingdom; MR/K011200/MRC{\_}/Medical Research Council/United Kingdom; BB/J015539/1/BB{\_}/Biotechnology and Biological Sciences Research Council/United Kingdom},
	issn = {1545-7885 (Electronic); 1544-9173 (Print); 1544-9173 (Linking)},
	jid = {101183755},
	journal = {PLoS Biol},
	jt = {PLoS biology},
	language = {eng},
	lid = {10.1371/journal.pbio.1001937 {$[$}doi{$]$}; e1001937},
	lr = {20260128},
	mh = {Animals; *Body Patterning; Cell Differentiation; Cell Line; Chick Embryo; Embryonic Stem Cells/cytology/metabolism; Fetal Proteins/metabolism; Fibroblast Growth Factors/metabolism; Germ Layers/cytology; Humans; Mesoderm/*cytology; Mice; Neural Stem Cells/*cytology/*metabolism; Spinal Cord/*cytology/*embryology; T-Box Domain Proteins/metabolism; Transcription, Genetic; *Wnt Signaling Pathway; Brachyury Protein},
	mhda = {2015/04/10 06:00},
	month = {Aug},
	number = {8},
	own = {NLM},
	pages = {e1001937},
	phst = {2014/05/28 00:00 {$[$}received{$]$}; 2014/07/21 00:00 {$[$}accepted{$]$}; 2014/08/27 06:00 {$[$}entrez{$]$}; 2014/08/27 06:00 {$[$}pubmed{$]$}; 2015/04/10 06:00 {$[$}medline{$]$}; 2014/08/26 00:00 {$[$}pmc-release{$]$}},
	pii = {PBIOLOGY-D-14-01861},
	pl = {United States},
	pmc = {PMC4144800},
	pmcr = {2014/08/26},
	pmid = {25157815},
	pst = {epublish},
	pt = {Journal Article; Research Support, Non-U.S. Gov't},
	rn = {0 (Fetal Proteins); 62031-54-3 (Fibroblast Growth Factors); 0 (T-Box Domain Proteins); 0 (Brachyury Protein)},
	sb = {IM},
	status = {MEDLINE},
	title = {{In vitro generation of neuromesodermal progenitors reveals distinct roles for wnt signalling in the specification of spinal cord and paraxial mesoderm identity}},
	volume = {12},
	year = {2014}}

@article{Turner:2014aa,
	address = {Department of Genetics, University of Cambridge, Cambridge CB2 3EH, UK.; Department of Genetics, University of Cambridge, Cambridge CB2 3EH, UK Wellcome Trust Centre for Stem Cell Research, University of Cambridge, Cambridge CB2 1QR, UK.; Department of Genetics, University of Cambridge, Cambridge CB2 3EH, UK.; Department of Genetics, University of Cambridge, Cambridge CB2 3EH, UK.; Department of Genetics, University of Cambridge, Cambridge CB2 3EH, UK ama11@hermes.cam.ac.uk.},
	author = {Turner, David A and Trott, Jamie and Hayward, Penelope and Ru{\'e}, Pau and Martinez Arias, Alfonso},
	cois = {Competing interests: The authors have no competing interests to declare.},
	copyright = {{\copyright}2014. Published by The Company of Biologists Ltd.},
	crdt = {2014/06/22 06:00},
	date = {2014 Jun 20},
	dcom = {20140716},
	dep = {20140620},
	doi = {10.1242/bio.20148409},
	edat = {2014/06/22 06:00},
	issn = {2046-6390 (Print); 2046-6390 (Electronic); 2046-6390 (Linking)},
	jid = {101578018},
	journal = {Biol Open},
	jt = {Biology open},
	language = {eng},
	lid = {10.1242/bio.20148409 {$[$}doi{$]$}},
	lr = {20211021},
	mhda = {2014/06/22 06:01},
	month = {Jun},
	number = {7},
	oto = {NOTNLM},
	own = {NLM},
	pages = {614--626},
	phst = {2014/06/22 06:00 {$[$}entrez{$]$}; 2014/06/22 06:00 {$[$}pubmed{$]$}; 2014/06/22 06:01 {$[$}medline{$]$}; 2014/06/20 00:00 {$[$}pmc-release{$]$}},
	pii = {bio.20148409; BIO20148409},
	pl = {England},
	pmc = {PMC4154298},
	pmcr = {2014/06/20},
	pmid = {24950969},
	pst = {epublish},
	pt = {Journal Article},
	status = {PubMed-not-MEDLINE},
	title = {{An interplay between extracellular signalling and the dynamics of the exit from pluripotency drives cell fate decisions in mouse ES cells}},
	volume = {3},
	year = {2014}}

@article{Schroter:2015aa,
	address = {Department of Genetics, University of Cambridge, Downing Street, Cambridge CB2 3EH, UK cs684@cam.ac.uk ama11@cam.ac.uk.; Department of Genetics, University of Cambridge, Downing Street, Cambridge CB2 3EH, UK.; Department of Genetics, University of Cambridge, Downing Street, Cambridge CB2 3EH, UK.; Department of Genetics, University of Cambridge, Downing Street, Cambridge CB2 3EH, UK cs684@cam.ac.uk ama11@cam.ac.uk.},
	author = {Sch\"{o}ter, Christian and Ru\'{e}, Pau and Mackenzie, Jonathan Peter and Martinez Arias, Alfonso},
	cois = {Competing interests The authors declare no competing or financial interests.},
	copyright = {{\copyright}2015. Published by The Company of Biologists Ltd.},
	crdt = {2015/10/30 06:00},
	date = {2015 Dec 15},
	dcom = {20160524},
	dep = {20151028},
	doi = {10.1242/dev.127530},
	edat = {2015/10/30 06:00},
	gr = {WT{\_}/Wellcome Trust/United Kingdom; 250316/ERC{\_}/European Research Council/International},
	issn = {1477-9129 (Electronic); 0950-1991 (Print); 0950-1991 (Linking)},
	jid = {8701744},
	journal = {Development},
	jt = {Development (Cambridge, England)},
	language = {eng},
	lid = {10.1242/dev.127530 {$[$}doi{$]$}},
	lr = {20260127},
	mh = {Animals; Biomarkers/metabolism; *Cell Differentiation; *Cell Lineage; Cells, Cultured; Endoderm/cytology/metabolism; Fibroblast Growth Factors/*metabolism; GATA4 Transcription Factor/metabolism; Gene Expression Regulation; Germ Layers/cytology/metabolism; Luminescent Proteins/metabolism; *MAP Kinase Signaling System; Mice; Models, Biological; Mouse Embryonic Stem Cells/*cytology/*metabolism; Red Fluorescent Protein},
	mhda = {2016/05/25 06:00},
	month = {Dec},
	number = {24},
	oto = {NOTNLM},
	own = {NLM},
	pages = {4205--4216},
	phst = {2015/06/18 00:00 {$[$}received{$]$}; 2015/10/20 00:00 {$[$}accepted{$]$}; 2015/10/30 06:00 {$[$}entrez{$]$}; 2015/10/30 06:00 {$[$}pubmed{$]$}; 2016/05/25 06:00 {$[$}medline{$]$}; 2015/12/15 00:00 {$[$}pmc-release{$]$}},
	pii = {dev.127530; DEV127530},
	pl = {England},
	pmc = {PMC4689219},
	pmcr = {2015/12/15},
	pmid = {26511924},
	pst = {ppublish},
	pt = {Journal Article; Research Support, Non-U.S. Gov't},
	rn = {0 (Biomarkers); 62031-54-3 (Fibroblast Growth Factors); 0 (GATA4 Transcription Factor); 0 (Luminescent Proteins); 0 (Red Fluorescent Protein)},
	sb = {IM},
	status = {MEDLINE},
	title = {{FGF/MAPK signaling sets the switching threshold of a bistable circuit controlling cell fate decisions in embryonic stem cells}},
	volume = {142},
	year = {2015}}

@article{Bessonnard:2014aa,
	address = {Clermont Universit{\'e}, Universit{\'e}d'Auvergne, Laboratoire GReD, Clermont-Ferrand F-63000, France Inserm, UMR1103, Clermont-Ferrand F-63001, France CNRS, UMR6293, Clermont-Ferrand F-63001, France.; Unit{\'e}de Chronobiologie th{\'e}orique, Facult{\'e}des Sciences, Universit{\'e}Libre de Bruxelles (ULB), Campus Plaine, CP 231, Brussels B-1050, Belgium.; Unit{\'e}de Chronobiologie th{\'e}orique, Facult{\'e}des Sciences, Universit{\'e}Libre de Bruxelles (ULB), Campus Plaine, CP 231, Brussels B-1050, Belgium.; Clermont Universit{\'e}, Universit{\'e}d'Auvergne, Laboratoire GReD, Clermont-Ferrand F-63000, France Inserm, UMR1103, Clermont-Ferrand F-63001, France CNRS, UMR6293, Clermont-Ferrand F-63001, France.; Clermont Universit{\'e}, Universit{\'e}d'Auvergne, Laboratoire GReD, Clermont-Ferrand F-63000, France Inserm, UMR1103, Clermont-Ferrand F-63001, France CNRS, UMR6293, Clermont-Ferrand F-63001, France.; Unit{\'e}de Chronobiologie th{\'e}orique, Facult{\'e}des Sciences, Universit{\'e}Libre de Bruxelles (ULB), Campus Plaine, CP 231, Brussels B-1050, Belgium Stellenbosch Institute for Advanced Study (STIAS), Wallenberg Research Center at Stellenbosch University, Stellenbosch 7600, South Africa.; Unit{\'e}de Chronobiologie th{\'e}orique, Facult{\'e}des Sciences, Universit{\'e}Libre de Bruxelles (ULB), Campus Plaine, CP 231, Brussels B-1050, Belgium genevieve.dupont@ulb.ac.be claire.chazaud@udamail.fr.; Clermont Universit{\'e}, Universit{\'e}d'Auvergne, Laboratoire GReD, Clermont-Ferrand F-63000, France Inserm, UMR1103, Clermont-Ferrand F-63001, France CNRS, UMR6293, Clermont-Ferrand F-63001, France genevieve.dupont@ulb.ac.be claire.chazaud@udamail.fr.},
	author = {Bessonnard, Sylvain and De Mot, Laurane and Gonze, Didier and Barriol, Manon and Dennis, Cynthia and Goldbeter, Albert and Dupont, Genevi{\`e}ve and Chazaud, Claire},
	copyright = {{\copyright}2014. Published by The Company of Biologists Ltd.},
	crdt = {2014/09/12 06:00},
	date = {2014 Oct},
	dcom = {20141124},
	dep = {20140910},
	doi = {10.1242/dev.109678},
	edat = {2014/09/12 06:00},
	issn = {1477-9129 (Electronic); 0950-1991 (Linking)},
	jid = {8701744},
	journal = {Development},
	jt = {Development (Cambridge, England)},
	language = {eng},
	lid = {10.1242/dev.109678 {$[$}doi{$]$}},
	lr = {20161125},
	mh = {Animals; Blastocyst Inner Cell Mass/*cytology; Cell Differentiation/*physiology; Cell Lineage/*physiology; Endoderm/cytology; Extracellular Signal-Regulated MAP Kinases/metabolism; GATA6 Transcription Factor/*metabolism; Gene Regulatory Networks/*physiology; Germ Layers/cytology; Homeodomain Proteins/metabolism; In Situ Hybridization, Fluorescence; Indoles; Mice; Microscopy, Confocal; *Models, Biological; Nanog Homeobox Protein; Signal Transduction/genetics/*physiology; Statistics, Nonparametric},
	mhda = {2014/12/15 06:00},
	month = {Oct},
	number = {19},
	oto = {NOTNLM},
	own = {NLM},
	pages = {3637--3648},
	phst = {2014/09/12 06:00 {$[$}entrez{$]$}; 2014/09/12 06:00 {$[$}pubmed{$]$}; 2014/12/15 06:00 {$[$}medline{$]$}},
	pii = {dev.109678},
	pl = {England},
	pmid = {25209243},
	pst = {ppublish},
	pt = {Journal Article; Research Support, Non-U.S. Gov't},
	rn = {0 (GATA6 Transcription Factor); 0 (Gata6 protein, mouse); 0 (Homeodomain Proteins); 0 (Indoles); 0 (Nanog Homeobox Protein); 0 (Nanog protein, mouse); 47165-04-8 (DAPI); EC 2.7.11.24 (Extracellular Signal-Regulated MAP Kinases)},
	sb = {IM},
	status = {MEDLINE},
	title = {{Gata6, Nanog and Erk signaling control cell fate in the inner cell mass through a tristable regulatory network}},
	volume = {141},
	year = {2014}}

@article{Masui:2007aa,
	address = {Laboratory for Pluripotent Cell Studies, RIKEN Center for Developmental Biology, Minatojima-minamimachi 2-2-3, Chu-o-ku, Kobe, Hyogo 650-0047, Japan. masui@ri.imcj.go.jp},
	author = {Masui, Shinji and Nakatake, Yuhki and Toyooka, Yayoi and Shimosato, Daisuke and Yagi, Rika and Takahashi, Kazue and Okochi, Hitoshi and Okuda, Akihiko and Matoba, Ryo and Sharov, Alexei A and Ko, Minoru S H and Niwa, Hitoshi},
	crdt = {2007/05/23 09:00},
	date = {2007 Jun},
	dcom = {20070719},
	dep = {20070521},
	doi = {10.1038/ncb1589},
	edat = {2007/05/23 09:00},
	gr = {Intramural NIH HHS/United States},
	issn = {1465-7392 (Print); 1465-7392 (Linking)},
	jid = {100890575},
	journal = {Nat Cell Biol},
	jt = {Nature cell biology},
	language = {eng},
	lr = {20260128},
	mh = {Animals; Cell Differentiation/*physiology; Cell Line; Cells, Cultured; DNA-Binding Proteins/genetics/*metabolism; Embryonic Development/physiology; Embryonic Stem Cells/*metabolism; Enhancer Elements, Genetic/genetics; Gene Expression Regulation, Developmental/genetics; Homeodomain Proteins/genetics/metabolism; Mice; Mice, Inbred C57BL; Mice, Knockout; Nanog Homeobox Protein; Octamer Transcription Factor-3/genetics/*metabolism; Organic Cation Transport Proteins/genetics/*metabolism; Pluripotent Stem Cells/*metabolism; SOXB1 Transcription Factors; Trans-Activators/genetics/*metabolism; Transcription Factors/genetics; Transcriptional Activation/genetics; Up-Regulation/genetics; Organic Cation Transporter 3},
	mhda = {2007/07/20 09:00},
	month = {Jun},
	number = {6},
	own = {NLM},
	pages = {625--635},
	phst = {2007/02/02 00:00 {$[$}received{$]$}; 2007/04/27 00:00 {$[$}accepted{$]$}; 2007/05/23 09:00 {$[$}pubmed{$]$}; 2007/07/20 09:00 {$[$}medline{$]$}; 2007/05/23 09:00 {$[$}entrez{$]$}},
	pii = {ncb1589},
	pl = {England},
	pmid = {17515932},
	pst = {ppublish},
	pt = {Journal Article; Research Support, N.I.H., Intramural; Research Support, Non-U.S. Gov't},
	rn = {0 (DNA-Binding Proteins); 0 (Homeodomain Proteins); 0 (Nanog Homeobox Protein); 0 (Octamer Transcription Factor-3); 0 (Organic Cation Transport Proteins); 0 (SOXB1 Transcription Factors); 0 (Trans-Activators); 0 (Transcription Factors); 0 (Organic Cation Transporter 3); 0 (Nanog protein, mouse); 0 (Pou5f1 protein, mouse); 0 (Sox2 protein, mouse)},
	sb = {IM},
	status = {MEDLINE},
	title = {Pluripotency governed by Sox2 via regulation of Oct3/4 expression in mouse embryonic stem cells.},
	volume = {9},
	year = {2007}}

@article{Tata:2013aa,
	address = {1{$]$} Center for Regenerative Medicine, Massachusetts General Hospital, 185 Cambridge Street, Boston, Massachusetts 02114, USA [2] Departments of Pediatrics, Massachusetts General Hospital, Boston, Massachusetts 02114, USA [3] Department of Internal Medicine, Pulmonary and Critical Care Unit, Massachusetts General Hospital, Boston, Massachusetts 02114, USA [4] Harvard Stem Cell Institute, Cambridge, Massachusetts 02138, USA.},
	author = {Tata, Purushothama Rao and Mou, Hongmei and Pardo-Saganta, Ana and Zhao, Rui and Prabhu, Mythili and Law, Brandon M and Vinarsky, Vladimir and Cho, Josalyn L and Breton, Sylvie and Sahay, Amar and Medoff, Benjamin D and Rajagopal, Jayaraj},
	cin = {Nature. 2013 Nov 14;503(7475):204-5. doi: 10.1038/nature12706. PMID: 24196710},
	crdt = {2013/11/08 06:00},
	date = {2013 Nov 14},
	dcom = {20131127},
	dep = {20131106},
	doi = {10.1038/nature12777},
	edat = {2013/11/08 06:00},
	gr = {R01 HL118185/HL/NHLBI NIH HHS/United States; P30 DK043351/DK/NIDDK NIH HHS/United States; P30 HL101287/HL/NHLBI NIH HHS/United States; R00 MH086615/MH/NIMH NIH HHS/United States; 5P30HL101287-02/HL/NHLBI NIH HHS/United States},
	issn = {1476-4687 (Electronic); 0028-0836 (Print); 0028-0836 (Linking)},
	jid = {0410462},
	journal = {Nature},
	jt = {Nature},
	language = {eng},
	lid = {10.1038/nature12777 {$[$}doi{$]$}},
	lr = {20220409},
	mh = {Animals; Antineoplastic Agents, Hormonal/pharmacology; *Cell Dedifferentiation; Cell Proliferation/drug effects; Cell Survival; Cells, Cultured; Doxycycline/pharmacology; Epithelial Cells/*cytology/drug effects; Female; Male; Mice, Transgenic; Stem Cells/*cytology/drug effects; Tamoxifen/pharmacology},
	mhda = {2013/12/16 06:00},
	mid = {NIHMS532788},
	month = {Nov},
	number = {7475},
	own = {NLM},
	pages = {218--223},
	phst = {2013/04/04 00:00 {$[$}received{$]$}; 2013/10/17 00:00 {$[$}accepted{$]$}; 2013/11/08 06:00 {$[$}entrez{$]$}; 2013/11/08 06:00 {$[$}pubmed{$]$}; 2013/12/16 06:00 {$[$}medline{$]$}; 2014/05/27 00:00 {$[$}pmc-release{$]$}},
	pii = {nature12777},
	pl = {England},
	pmc = {PMC4035230},
	pmcr = {2014/05/27},
	pmid = {24196716},
	pst = {ppublish},
	pt = {Journal Article; Research Support, N.I.H., Extramural; Research Support, Non-U.S. Gov't},
	rn = {0 (Antineoplastic Agents, Hormonal); 094ZI81Y45 (Tamoxifen); N12000U13O (Doxycycline)},
	sb = {IM},
	status = {MEDLINE},
	title = {{Dedifferentiation of committed epithelial cells into stem cells in vivo}},
	volume = {503},
	year = {2013}}

@article{Es:2012aa,
	address = {Hubrecht Institute for Developmental Biology and Stem Cell Research \& University Medical Centre Utrecht, Uppsalalaan 8, 3584CT Utrecht, Netherlands.; Hubrecht Institute for Developmental Biology and Stem Cell Research \& University Medical Centre Utrecht, Uppsalalaan 8, 3584CT Utrecht, Netherlands.; Hubrecht Institute for Developmental Biology and Stem Cell Research \& University Medical Centre Utrecht, Uppsalalaan 8, 3584CT Utrecht, Netherlands.; Dept. of Physics \& Biology, Massachusetts Institute of Technology, 77 Massachusetts Avenue, Cambridge, MA 02139, USA.; Institute of Medical Biology, 06-06 Immunos, Singapore.; Samuel Lunenfeld Research Institute, Mount Sinai Hospital, 600 University Avenue, Toronto, Ontario M5G 1X5, Canada.; Hubrecht Institute for Developmental Biology and Stem Cell Research \& University Medical Centre Utrecht, Uppsalalaan 8, 3584CT Utrecht, Netherlands.; Hubrecht Institute for Developmental Biology and Stem Cell Research \& University Medical Centre Utrecht, Uppsalalaan 8, 3584CT Utrecht, Netherlands.; Hubrecht Institute for Developmental Biology and Stem Cell Research \& University Medical Centre Utrecht, Uppsalalaan 8, 3584CT Utrecht, Netherlands.; Hubrecht Institute for Developmental Biology and Stem Cell Research \& University Medical Centre Utrecht, Uppsalalaan 8, 3584CT Utrecht, Netherlands.; UMC Utrecht, Dept. of Immunology and Cell Biology, PO BOX 85090, 3508AB Utrecht, Netherlands.; Institute of Medical Biology, 06-06 Immunos, Singapore.; Dept. of Physics \& Biology, Massachusetts Institute of Technology, 77 Massachusetts Avenue, Cambridge, MA 02139, USA.; Hubrecht Institute for Developmental Biology and Stem Cell Research \& University Medical Centre Utrecht, Uppsalalaan 8, 3584CT Utrecht, Netherlands.},
	author = {van Es, Johan H and Sato, Toshiro and van de Wetering, Marc and Lyubimova, Anna and Yee Nee, Annie Ng and Gregorieff, Alex and Sasaki, Nobuo and Zeinstra, Laura and van den Born, Maaike and Korving, Jeroen and Martens, Anton C M and Barker, Nick and van Oudenaarden, Alexander and Clevers, Hans},
	crdt = {2012/09/25 06:00},
	date = {2012 Oct},
	dcom = {20130111},
	dep = {20120923},
	doi = {10.1038/ncb2581},
	edat = {2012/09/25 06:00},
	gr = {U54 CA143874/CA/NCI NIH HHS/United States; U54CA143874/CA/NCI NIH HHS/United States},
	issn = {1476-4679 (Electronic); 1465-7392 (Print); 1465-7392 (Linking)},
	jid = {100890575},
	journal = {Nat Cell Biol},
	jt = {Nature cell biology},
	language = {eng},
	lid = {10.1038/ncb2581 {$[$}doi{$]$}},
	lr = {20220318},
	mh = {Animals; Calcium-Binding Proteins; Cell Lineage; Cells, Cultured; Gene Knock-In Techniques; Intercellular Signaling Peptides and Proteins/analysis/genetics/*metabolism; Intestinal Mucosa/*metabolism; Mice; Organoids/metabolism; Receptors, Notch/antagonists \& inhibitors/metabolism; Stem Cells/*metabolism; Wnt3A Protein/pharmacology},
	mhda = {2013/01/12 06:00},
	mid = {NIHMS506945},
	month = {Oct},
	number = {10},
	own = {NLM},
	pages = {1099--1104},
	phst = {2012/05/23 00:00 {$[$}received{$]$}; 2012/08/16 00:00 {$[$}accepted{$]$}; 2012/09/25 06:00 {$[$}entrez{$]$}; 2012/09/25 06:00 {$[$}pubmed{$]$}; 2013/01/12 06:00 {$[$}medline{$]$}; 2013/10/03 00:00 {$[$}pmc-release{$]$}},
	pl = {England},
	pmc = {PMC3789123},
	pmcr = {2013/10/03},
	pmid = {23000963},
	pst = {ppublish},
	pt = {Journal Article; Research Support, N.I.H., Extramural; Research Support, Non-U.S. Gov't},
	rn = {0 (Calcium-Binding Proteins); 0 (Dlk1 protein, mouse); 0 (Intercellular Signaling Peptides and Proteins); 0 (Receptors, Notch); 0 (Wnt3A Protein); 0 (Wnt3a protein, mouse)},
	sb = {IM},
	status = {MEDLINE},
	title = {{Dll1+ secretory progenitor cells revert to stem cells upon crypt damage}},
	volume = {14},
	year = {2012}}

@article{Briggs:2018hg,
	author = {Briggs, James A. and Weinreb, Caleb and Wagner, Daniel E. and Megason, Sean and Peshkin, Leonid and Kirschner, Marc W. and Klein, Allon M.},
	doi = {10.1126/science.aar5780},
	issn = {0036-8075},
	journal = {Science},
	number = {6392},
	pages = {eaar5780},
	pmcid = {PMC6038144},
	pmid = {29700227},
	title = {{The dynamics of gene expression in vertebrate embryogenesis at single-cell resolution}},
	volume = {360},
	year = {2018}}

@article{Burns:1970tm,
	author = {Burns, F J and Tannock, I F},
	journal = {Cell Tissue Kinet},
	number = {4},
	pages = {321 -- 334},
	title = {{On the existence of a G0-phase in the cell cycle}},
	volume = {3},
	year = {1970}}

@article{Rao:2002aa,
	address = {Department of Bioengineering, University of California, Berkeley, California 94720, USA. c_rao@lbl.gov},
	author = {Rao, Christopher V and Wolf, Denise M and Arkin, Adam P},
	crdt = {2002/11/15 04:00},
	date = {2002 Nov 14},
	dcom = {20021212},
	doi = {10.1038/nature01258},
	edat = {2002/11/15 04:00},
	ein = {Nature. 2003 Jan 9;421(6919):190.},
	issn = {0028-0836 (Print); 0028-0836 (Linking)},
	jid = {0410462},
	journal = {Nature},
	jt = {Nature},
	language = {eng},
	lr = {20091119},
	mh = {Animals; *Cell Physiological Phenomena; Feedback, Physiological; Isomerism; Kinetics; *Models, Biological; Probability; Stochastic Processes},
	mhda = {2002/12/13 04:00},
	month = {Nov},
	number = {6912},
	own = {NLM},
	pages = {231--237},
	phst = {2002/11/15 04:00 {$[$}pubmed{$]$}; 2002/12/13 04:00 {$[$}medline{$]$}; 2002/11/15 04:00 {$[$}entrez{$]$}},
	pii = {nature01258},
	pl = {England},
	pmid = {12432408},
	pst = {ppublish},
	pt = {Journal Article; Review},
	rf = {86},
	sb = {IM},
	status = {MEDLINE},
	title = {{Control, exploitation and tolerance of intracellular noise}},
	volume = {420},
	year = {2002}}

@article{Elowitz:2002aa,
	author = {Elowitz, Michael B and Levine, Arnold J and Siggia, Eric D and Swain, Peter S},
	doi = {10.1126/science.1070919},
	journal = {Science},
	language = {English},
	number = {5584},
	pages = {1183 -- 1186},
	title = {{Stochastic gene expression in a single cell}},
	volume = {297},
	year = {2002}}

@article{Zhou:2016gd,
	author = {Zhou, Peijie and Li, Tiejun},
	doi = {10.1063/1.4943096},
	issn = {0021-9606},
	journal = {J Chem Phys},
	pages = {094109},
	pmid = {26957159},
	title = {{Construction of the landscape for multi-stable systems: Potential landscape, quasi-potential, A-type integral and beyond}},
	volume = {144},
	year = {2016}}

@article{Graf:2009aa,
	address = {Center for Genomic Regulation and ICREA, 08003 Barcelona, Spain. thomas.graf@crg.es},
	author = {Graf, Thomas and Enver, Tariq},
	crdt = {2009/12/04 06:00},
	date = {2009 Dec 3},
	dcom = {20091208},
	doi = {10.1038/nature08533},
	edat = {2009/12/04 06:00},
	gr = {MC{\_}U137973817/MRC{\_}/Medical Research Council/United Kingdom},
	issn = {1476-4687 (Electronic); 0028-0836 (Linking)},
	jid = {0410462},
	journal = {Nature},
	jt = {Nature},
	language = {eng},
	lid = {10.1038/nature08533 {$[$}doi{$]$}},
	lr = {20250529},
	mh = {Animals; Cell Differentiation; Cell Lineage/*physiology; Cellular Reprogramming/*genetics; *Gene Expression Regulation, Developmental; Gene Regulatory Networks/physiology; Humans; Pluripotent Stem Cells/cytology/*metabolism; Transcription Factors/*metabolism},
	mhda = {2009/12/16 06:00},
	month = {Dec},
	number = {7273},
	own = {NLM},
	pages = {587--594},
	phst = {2009/12/04 06:00 {$[$}entrez{$]$}; 2009/12/04 06:00 {$[$}pubmed{$]$}; 2009/12/16 06:00 {$[$}medline{$]$}},
	pii = {nature08533},
	pl = {England},
	pmid = {19956253},
	pst = {ppublish},
	pt = {Journal Article; Research Support, Non-U.S. Gov't; Review},
	rf = {98},
	rn = {0 (Transcription Factors)},
	sb = {IM},
	status = {MEDLINE},
	title = {{Forcing cells to change lineages}},
	volume = {462},
	year = {2009}}

@article{Wagner:2016aa,
	address = {Department of Electrical Engineering and Computer Science and the Center for Computational Biology, University of California, Berkeley, California, USA.; Howard Hughes Medical Institute, Department of Biology, Massachusetts Institute of Technology, Cambridge, Massachusetts, USA.; Broad Institute of MIT and Harvard, Cambridge, Massachusetts, USA.; Department of Electrical Engineering and Computer Science and the Center for Computational Biology, University of California, Berkeley, California, USA.; Ragon Institute of Massachusetts General Hospital, Massachusetts Institute of Technology, and Harvard University, Boston, Massachusetts, USA.},
	author = {Wagner, Allon and Regev, Aviv and Yosef, Nir},
	cois = {COMPETING FINANCIAL INTERESTS A.R. is a member of the Scientific Advisory Board for Thermo Fisher Scientific and Syros Pharmaceuticals and a consultant for Driver Group. A.W. and N.Y. declare no competing financial interests.},
	crdt = {2016/11/09 06:00},
	date = {2016 Nov 8},
	dcom = {20170605},
	doi = {10.1038/nbt.3711},
	edat = {2016/11/09 06:00},
	gr = {U24 CA180922/CA/NCI NIH HHS/United States; P30 CA014051/CA/NCI NIH HHS/United States; U01 MH105979/MH/NIMH NIH HHS/United States; U01 MH105960/MH/NIMH NIH HHS/United States; HHMI/Howard Hughes Medical Institute/United States; P50 HG006193/HG/NHGRI NIH HHS/United States; U24 AI118672/AI/NIAID NIH HHS/United States; RM1 HG006193/HG/NHGRI NIH HHS/United States},
	issn = {1546-1696 (Electronic); 1087-0156 (Print); 1087-0156 (Linking)},
	jid = {9604648},
	journal = {Nat Biotechnol},
	jt = {Nature biotechnology},
	language = {eng},
	lid = {10.1038/nbt.3711 {$[$}doi{$]$}},
	lr = {20250530},
	mh = {Animals; Cell Physiological Phenomena/*physiology; Computer Simulation; Gene Expression Profiling/*methods; Gene Expression Regulation/*physiology; High-Throughput Nucleotide Sequencing/*methods; Humans; Models, Biological; Proteome/*metabolism; Signal Transduction/physiology; Transcriptome/*physiology},
	mhda = {2017/06/06 06:00},
	mid = {NIHMS857730},
	month = {Nov},
	number = {11},
	own = {NLM},
	pages = {1145--1160},
	phst = {2016/11/09 06:00 {$[$}entrez{$]$}; 2016/11/09 06:00 {$[$}pubmed{$]$}; 2017/06/06 06:00 {$[$}medline{$]$}; 2017/11/08 00:00 {$[$}pmc-release{$]$}},
	pii = {nbt.3711},
	pl = {United States},
	pmc = {PMC5465644},
	pmcr = {2017/11/08},
	pmid = {27824854},
	pst = {ppublish},
	pt = {Journal Article; Review},
	rn = {0 (Proteome)},
	sb = {IM},
	status = {MEDLINE},
	title = {{Revealing the vectors of cellular identity with single-cell genomics}},
	volume = {34},
	year = {2016}}

@article{Davidson:2010aa,
	address = {Division of Biology 156-29, California Institute of Technology, Pasadena, California 91125, USA. Davidson@caltech.edu},
	author = {Davidson, Eric H},
	crdt = {2010/12/18 06:00},
	date = {2010 Dec 16},
	dcom = {20110210},
	doi = {10.1038/nature09645},
	edat = {2010/12/18 06:00},
	gr = {P01 HD037105/HD/NICHD NIH HHS/United States; R01 GM061005/GM/NIGMS NIH HHS/United States; HD-37105/HD/NICHD NIH HHS/United States; GM-61005/GM/NIGMS NIH HHS/United States},
	issn = {1476-4687 (Electronic); 0028-0836 (Print); 0028-0836 (Linking)},
	jid = {0410462},
	journal = {Nature},
	jt = {Nature},
	language = {eng},
	lid = {10.1038/nature09645 {$[$}doi{$]$}},
	lr = {20241207},
	mh = {Animal Structures/embryology/metabolism; Animals; Evolution, Molecular; Gene Expression Regulation, Developmental/*genetics; Gene Regulatory Networks/*genetics; Models, Genetic; Pluripotent Stem Cells/cytology/metabolism; Structure-Activity Relationship},
	mhda = {2011/02/11 06:00},
	mid = {NIHMS565406},
	month = {Dec},
	number = {7326},
	own = {NLM},
	pages = {911--920},
	phst = {2010/12/18 06:00 {$[$}entrez{$]$}; 2010/12/18 06:00 {$[$}pubmed{$]$}; 2011/02/11 06:00 {$[$}medline{$]$}; 2014/03/27 00:00 {$[$}pmc-release{$]$}},
	pii = {nature09645},
	pl = {England},
	pmc = {PMC3967874},
	pmcr = {2014/03/27},
	pmid = {21164479},
	pst = {ppublish},
	pt = {Journal Article; Research Support, N.I.H., Extramural; Research Support, Non-U.S. Gov't; Review},
	sb = {IM},
	status = {MEDLINE},
	title = {{Emerging properties of animal gene regulatory networks}},
	volume = {468},
	year = {2010}}

@article{Allis:2016aa,
	address = {Laboratory of Chromatin Biology and Epigenetics, The Rockefeller University, 1230 York Avenue, New York 10065, New York, USA.; Max Planck Institute of Immunobiology and Epigenetics, St{\"u}beweg 51, Freiburg D-79108, Germany.},
	author = {Allis, C David and Jenuwein, Thomas},
	crdt = {2016/06/28 06:00},
	date = {2016 Aug},
	dcom = {20170512},
	dep = {20160627},
	doi = {10.1038/nrg.2016.59},
	edat = {2016/06/28 06:00},
	issn = {1471-0064 (Electronic); 1471-0056 (Linking)},
	jid = {100962779},
	journal = {Nat Rev Genet},
	jt = {Nature reviews. Genetics},
	language = {eng},
	lid = {10.1038/nrg.2016.59 {$[$}doi{$]$}},
	lr = {20220408},
	mh = {Animals; Chromatin/*genetics; *DNA Methylation; Epigenesis, Genetic/*genetics; Humans},
	mhda = {2017/05/13 06:00},
	month = {Aug},
	number = {8},
	own = {NLM},
	pages = {487--500},
	phst = {2016/06/28 06:00 {$[$}entrez{$]$}; 2016/06/28 06:00 {$[$}pubmed{$]$}; 2017/05/13 06:00 {$[$}medline{$]$}},
	pii = {nrg.2016.59},
	pl = {England},
	pmid = {27346641},
	pst = {ppublish},
	pt = {Journal Article; Review},
	rn = {0 (Chromatin)},
	sb = {IM},
	status = {MEDLINE},
	title = {{The molecular hallmarks of epigenetic control}},
	volume = {17},
	year = {2016}}

@article{Jaenisch:2003aa,
	address = {Whitehead Institute for Biomedical Research and Department of Biology, Massachusetts Institute of Technology, 9 Cambridge Center, Cambridge, MA 02142, USA.},
	author = {Jaenisch, Rudolf and Bird, Adrian},
	crdt = {2003/03/01 04:00},
	date = {2003 Mar},
	dcom = {20030325},
	doi = {10.1038/ng1089},
	edat = {2003/03/01 04:00},
	issn = {1061-4036 (Print); 1061-4036 (Linking)},
	jid = {9216904},
	journal = {Nat Genet},
	jt = {Nature genetics},
	language = {eng},
	lr = {20240109},
	mh = {Aging/genetics; Animals; Cloning, Organism; DNA Methylation; Diet; Dosage Compensation, Genetic; *Gene Expression Regulation, Developmental; Genetic Diseases, Inborn/genetics/therapy; *Genome; Genomic Imprinting; Humans; Mice; Models, Genetic; Mutation; Neoplasms/genetics; Phenotype; Signal Transduction},
	mhda = {2003/03/26 05:00},
	month = {Mar},
	own = {NLM},
	pages = {245--254},
	phst = {2003/03/01 04:00 {$[$}pubmed{$]$}; 2003/03/26 05:00 {$[$}medline{$]$}; 2003/03/01 04:00 {$[$}entrez{$]$}},
	pii = {ng1089},
	pl = {United States},
	pmid = {12610534},
	pst = {ppublish},
	pt = {Journal Article; Review},
	rf = {186},
	sb = {IM},
	status = {MEDLINE},
	title = {{Epigenetic regulation of gene expression: how the genome integrates intrinsic and environmental signals}},
	volume = {33 Suppl},
	year = {2003}}

@article{Vittadello:2025aa,
	author = {Vittadello, Sean T. and Diaz, L{\'e}o and Liu, Yujing and Zanca, Adriana and Stumpf, Michael P. H.},
	date = {2025/09/24},
	doi = {10.1007/s00285-025-02287-w},
	id = {Vittadello2025},
	isbn = {1432-1416},
	journal = {J Math Biol},
	number = {5},
	pages = {48},
	title = {{Towards a mathematical framework for modelling cell fate dynamics}},
	url = {https://doi.org/10.1007/s00285-025-02287-w},
	volume = {91},
	year = {2025}}

@article{Wang02012015,
	author = {Jin Wang},
	doi = {10.1080/00018732.2015.1037068},
	journal = {Adv Phys},
	number = {1},
	pages = {1--137},
	publisher = {Taylor \& Francis},
	title = {{Landscape and flux theory of non-equilibrium dynamical systems with application to biology}},
	url = {https://doi.org/10.1080/00018732.2015.1037068},
	volume = {64},
	year = {2015}}

@article{Lei:2014aa,
	address = {Zhou Pei-Yuan Center for Applied Mathematics, Ministry of Education Key Laboratory of Bioinformatics, Tsinghua University, Beijing 100084, China.},
	author = {Lei, Jinzhi and Levin, Simon A and Nie, Qing},
	cin = {Proc Natl Acad Sci U S A. 2014 Mar 11;111(10):3653-4. doi: 10.1073/pnas.1401030111. PMID: 24591612},
	cois = {The authors declare no conflict of interest.},
	crdt = {2014/02/07 06:00},
	date = {2014 Mar 11},
	dcom = {20140508},
	dep = {20140205},
	doi = {10.1073/pnas.1324267111},
	edat = {2014/02/07 06:00},
	gr = {P50 GM076516/GM/NIGMS NIH HHS/United States; R01 GM067247/GM/NIGMS NIH HHS/United States; R01GM67247/GM/NIGMS NIH HHS/United States; P50GM76516/GM/NIGMS NIH HHS/United States},
	issn = {1091-6490 (Electronic); 0027-8424 (Print); 0027-8424 (Linking)},
	jid = {7505876},
	journal = {Proc Natl Acad Sci USA},
	jt = {Proceedings of the National Academy of Sciences of the United States of America},
	language = {eng},
	lid = {10.1073/pnas.1324267111 {$[$}doi{$]$}},
	lr = {20211021},
	mh = {Adult; Adult Stem Cells/*cytology; Apoptosis/physiology; Cell Differentiation/physiology; Cell Proliferation; Epigenesis, Genetic/*physiology; Gene Expression Regulation/*physiology; Humans; *Models, Biological; Regeneration/*physiology; Systems Biology},
	mhda = {2014/05/09 06:00},
	month = {Mar},
	number = {10},
	oto = {NOTNLM},
	own = {NLM},
	pages = {E880-E887},
	phst = {2014/02/07 06:00 {$[$}entrez{$]$}; 2014/02/07 06:00 {$[$}pubmed{$]$}; 2014/05/09 06:00 {$[$}medline{$]$}; 2014/08/05 00:00 {$[$}pmc-release{$]$}},
	pii = {1324267111; 201324267},
	pl = {United States},
	pmc = {PMC3956188},
	pmcr = {2014/08/05},
	pmid = {24501127},
	pst = {ppublish},
	pt = {Journal Article; Research Support, N.I.H., Extramural; Research Support, Non-U.S. Gov't; Research Support, U.S. Gov't, Non-P.H.S.},
	sb = {IM},
	status = {MEDLINE},
	title = {{Mathematical model of adult stem cell regeneration with cross-talk between genetic and epigenetic regulation}},
	volume = {111},
	year = {2014}}

@article{Anreiter:2017aa,
	address = {Department of Ecology and Evolutionary Biology, University of Toronto, Toronto, ON, Canada M5S 3B2.; Child and Brain Development Program, Canadian Institute for Advanced Research, Toronto, ON, Canada M5G 1M1.; Department of Physiology and Pharmacology, Schulich School of Medicine and Dentistry, Western University, London, ON, Canada N6A 5C1.; Department of Biology, Faculty of Science, Western University, London, ON, Canada N6G 2M1.; Department of Ecology and Evolutionary Biology, University of Toronto, Toronto, ON, Canada M5S 3B2; marla.sokolowski@utoronto.ca.; Child and Brain Development Program, Canadian Institute for Advanced Research, Toronto, ON, Canada M5G 1M1.},
	auid = {ORCID: 0000-0002-9563-0579},
	author = {Anreiter, Ina and Kramer, Jamie M and Sokolowski, Marla B},
	cin = {Proc Natl Acad Sci U S A. 2017 Nov 21;114(47):12365-12367. doi: 10.1073/pnas.1717376114. PMID: 29114044},
	cois = {The authors declare no conflict of interest.},
	crdt = {2017/10/29 06:00},
	date = {2017 Nov 21},
	dcom = {20180626},
	dep = {20171016},
	doi = {10.1073/pnas.1710770114},
	edat = {2017/10/29 06:00},
	ein = {Proc Natl Acad Sci U S A. 2019 Apr 23;116(17):8630-8631. doi: 10.1073/pnas.1905223116. PMID: 30988192},
	gr = {P40 OD018537/OD/NIH HHS/United States},
	issn = {1091-6490 (Electronic); 0027-8424 (Print); 0027-8424 (Linking)},
	jid = {7505876},
	journal = {Proc Natl Acad Sci USA},
	jt = {Proceedings of the National Academy of Sciences of the United States of America},
	language = {eng},
	lid = {10.1073/pnas.1710770114 {$[$}doi{$]$}},
	lr = {20190614},
	mh = {Alleles; Animals; Base Sequence; Cyclic GMP-Dependent Protein Kinases/*genetics/metabolism; Drosophila Proteins/*genetics/metabolism; Drosophila melanogaster/*genetics/metabolism; *Epigenesis, Genetic; Exploratory Behavior; Gene Deletion; Genetic Variation; Histone-Lysine N-Methyltransferase/deficiency/*genetics; Histones/*genetics/metabolism; Methylation; Phenotype; Promoter Regions, Genetic},
	mhda = {2018/06/27 06:00},
	month = {Nov},
	number = {47},
	oto = {NOTNLM},
	own = {NLM},
	pages = {12518--12523},
	phst = {2017/10/29 06:00 {$[$}pubmed{$]$}; 2018/06/27 06:00 {$[$}medline{$]$}; 2017/10/29 06:00 {$[$}entrez{$]$}; 2018/05/21 00:00 {$[$}pmc-release{$]$}},
	pii = {1710770114; 201710770},
	pl = {United States},
	pmc = {PMC5703298},
	pmcr = {2018/05/21},
	pmid = {29078350},
	pst = {ppublish},
	pt = {Journal Article; Research Support, N.I.H., Extramural; Research Support, Non-U.S. Gov't},
	rn = {0 (Drosophila Proteins); 0 (Histones); EC 2.1.1.43 (Histone-Lysine N-Methyltransferase); EC 2.7.11.12 (Cyclic GMP-Dependent Protein Kinases); EC 2.7.11.12 (for protein, Drosophila)},
	sb = {IM},
	si = {GENBANK/CP023329; GENBANK/CP023334; GENBANK/CP023335; GENBANK/CP023340},
	status = {MEDLINE},
	title = {{Epigenetic mechanisms modulate differences in Drosophila foraging behavior}},
	volume = {114},
	year = {2017}}

@article{Probst:2009aa,
	address = {Laboratory of Nuclear Dynamics and Genome Plasticity, UMR218 Centre National de la Recherche Scientifique/Institut Curie, 26, rue d'Ulm, 75231 Paris Cedex 05, France.},
	author = {Probst, Aline V and Dunleavy, Elaine and Almouzni, Genevi{\`e}ve},
	crdt = {2009/02/24 09:00},
	date = {2009 Mar},
	dcom = {20090306},
	doi = {10.1038/nrm2640},
	edat = {2009/02/24 09:00},
	issn = {1471-0080 (Electronic); 1471-0072 (Linking)},
	jid = {100962782},
	journal = {Nat Rev Mol Cell Biol},
	jt = {Nature reviews. Molecular cell biology},
	language = {eng},
	lid = {10.1038/nrm2640 {$[$}doi{$]$}},
	lr = {20211020},
	mh = {Animals; Cell Cycle/*genetics; Chromatin/genetics; DNA Methylation; DNA Replication/genetics; *Epigenesis, Genetic; Genetic Markers/genetics; Histones/genetics/metabolism; Humans; Models, Genetic},
	mhda = {2009/03/07 09:00},
	month = {Mar},
	number = {3},
	own = {NLM},
	pages = {192--206},
	phst = {2009/02/24 09:00 {$[$}entrez{$]$}; 2009/02/24 09:00 {$[$}pubmed{$]$}; 2009/03/07 09:00 {$[$}medline{$]$}},
	pii = {nrm2640},
	pl = {England},
	pmid = {19234478},
	pst = {ppublish},
	pt = {Journal Article; Research Support, Non-U.S. Gov't; Review},
	rf = {183},
	rn = {0 (Chromatin); 0 (Genetic Markers); 0 (Histones)},
	sb = {IM},
	status = {MEDLINE},
	title = {{Epigenetic inheritance during the cell cycle}},
	volume = {10},
	year = {2009}}

@article{Klein:2011aa,
	address = {Department of Systems Biology, Harvard Medical School, 200 Longwood Avenue, Boston, MA 02115, USA. allon_klein@hms.harvard.edu},
	author = {Klein, Allon M and Simons, Benjamin D},
	crdt = {2011/07/14 06:00},
	date = {2011 Aug},
	dcom = {20110919},
	doi = {10.1242/dev.060103},
	edat = {2011/07/14 06:00},
	issn = {1477-9129 (Electronic); 0950-1991 (Linking)},
	jid = {8701744},
	journal = {Development},
	jt = {Development (Cambridge, England)},
	language = {eng},
	lid = {10.1242/dev.060103 {$[$}doi{$]$}},
	lr = {20220408},
	mh = {Animals; Cell Differentiation/physiology; Cell Division/*physiology; *Cell Lineage; Germ Cells/cytology/physiology; Homeostasis; Models, Biological; Stem Cells/*cytology/*physiology; Stochastic Processes},
	mhda = {2011/09/20 06:00},
	month = {Aug},
	number = {15},
	own = {NLM},
	pages = {3103--3111},
	phst = {2011/07/14 06:00 {$[$}entrez{$]$}; 2011/07/14 06:00 {$[$}pubmed{$]$}; 2011/09/20 06:00 {$[$}medline{$]$}},
	pii = {138/15/3103},
	pl = {England},
	pmid = {21750026},
	pst = {ppublish},
	pt = {Journal Article; Research Support, Non-U.S. Gov't},
	sb = {IM},
	status = {MEDLINE},
	title = {{Universal patterns of stem cell fate in cycling adult tissues}},
	volume = {138},
	year = {2011}}

@article{Lander:2009aa,
	address = {Department of Developmental and Cell Biology, University of California, Irvine, Irvine, California, USA. adlander@uci.edu},
	author = {Lander, Arthur D and Gokoffski, Kimberly K and Wan, Frederic Y M and Nie, Qing and Calof, Anne L},
	cin = {PLoS Biol. 2009 Jan 20;7(1):e21. doi: 10.1371/journal.pbio.1000021. PMID: 19166272},
	cois = {Competing interests. The authors have declared that no competing interests exist.},
	crdt = {2009/01/27 09:00},
	date = {2009 Jan 20},
	dcom = {20090219},
	doi = {10.1371/journal.pbio.1000015},
	edat = {2009/01/27 09:00},
	gr = {T32 GM008620/GM/NIGMS NIH HHS/United States; R01-DC03583/DC/NIDCD NIH HHS/United States; R01 GM067247/GM/NIGMS NIH HHS/United States; R01-GM067247/GM/NIGMS NIH HHS/United States; R01 GM075309/GM/NIGMS NIH HHS/United States; GM08620/GM/NIGMS NIH HHS/United States; P50 GM076516/GM/NIGMS NIH HHS/United States; P30 CA062203/CA/NCI NIH HHS/United States; R01-GM075309/GM/NIGMS NIH HHS/United States; P50GM076516/GM/NIGMS NIH HHS/United States; R01 DC003583/DC/NIDCD NIH HHS/United States},
	issn = {1545-7885 (Electronic); 1544-9173 (Print); 1544-9173 (Linking)},
	jid = {101183755},
	journal = {PLoS Biol},
	jt = {PLoS biology},
	language = {eng},
	lid = {10.1371/journal.pbio.1000015 {$[$}doi{$]$}; e1000015},
	lr = {20250529},
	mh = {Activins/metabolism; Animals; Cell Differentiation/*physiology; Cell Lineage/*physiology; Cell Movement/physiology; *Cell Proliferation; Chalones/*metabolism; Feedback, Physiological; Growth Differentiation Factors/*metabolism; Mice; Models, Biological; Olfactory Mucosa/cytology/*physiology; Olfactory Receptor Neurons/cytology; *Regeneration; Signal Transduction/physiology; Stem Cells/*physiology},
	mhda = {2009/02/20 09:00},
	month = {Jan},
	number = {1},
	own = {NLM},
	pages = {e15},
	phst = {2008/08/11 00:00 {$[$}received{$]$}; 2008/12/06 00:00 {$[$}accepted{$]$}; 2009/01/27 09:00 {$[$}entrez{$]$}; 2009/01/27 09:00 {$[$}pubmed{$]$}; 2009/02/20 09:00 {$[$}medline{$]$}; 2009/01/20 00:00 {$[$}pmc-release{$]$}},
	pii = {08-PLBI-RA-3368; 08-PLBI-RA-3368R2},
	pl = {United States},
	pmc = {PMC2628408},
	pmcr = {2009/01/20},
	pmid = {19166268},
	pst = {ppublish},
	pt = {Journal Article; Research Support, N.I.H., Extramural},
	rn = {0 (Chalones); 0 (Growth Differentiation Factors); 104625-48-1 (Activins)},
	sb = {IM},
	status = {MEDLINE},
	title = {{Cell lineages and the logic of proliferative control}},
	volume = {7},
	year = {2009}}

@article{AO200963,
	author = {Ping Ao},
	doi = {https://doi.org/10.1016/S1673-8527(08)60093-4},
	issn = {1673-8527},
	journal = {J Genet Genomics},
	number = {2},
	pages = {63-73},
	title = {{Global view of bionetwork dynamics: adaptive landscape}},
	url = {https://www.sciencedirect.com/science/article/pii/S1673852708600934},
	volume = {36},
	year = {2009}}

@article{Huang:2012aa,
	address = {Institute for Systems Biology, Seattle, WA, USA. sui.huang{\char64}systemsbiology.org},
	author = {Huang, Sui},
	cin = {Bioessays. 2012 Feb;34(2):87. doi: 10.1002/bies.201290000. PMID: 22234900; Bioessays. 2012 Aug;34(8):711-2. doi: 10.1002/bies.201200038. PMID: 22488123},
	copyright = {Copyright {\copyright}2012 WILEY Periodicals, Inc.},
	crdt = {2011/11/22 06:00},
	date = {2012 Feb},
	dcom = {20120509},
	dep = {20111118},
	doi = {10.1002/bies.201100031},
	edat = {2011/11/22 06:00},
	gr = {Canadian Institutes of Health Research/Canada},
	issn = {1521-1878 (Electronic); 0265-9247 (Linking)},
	jid = {8510851},
	journal = {Bioessays},
	jt = {BioEssays : news and reviews in molecular, cellular and developmental biology},
	language = {eng},
	lid = {10.1002/bies.201100031 {$[$}doi{$]$}},
	lr = {20120911},
	mh = {Animals; Biological Evolution; *Epigenesis, Genetic; *Gene Expression Regulation; Gene Regulatory Networks/*physiology; Genotype; Humans; *Models, Genetic; Phenotype; *Selection, Genetic},
	mhda = {2012/05/10 06:00},
	month = {Feb},
	number = {2},
	own = {NLM},
	pages = {149--157},
	phst = {2011/11/22 06:00 {$[$}entrez{$]$}; 2011/11/22 06:00 {$[$}pubmed{$]$}; 2012/05/10 06:00 {$[$}medline{$]$}},
	pl = {United States},
	pmid = {22102361},
	pst = {ppublish},
	pt = {Journal Article; Research Support, Non-U.S. Gov't},
	sb = {IM},
	status = {MEDLINE},
	title = {{The molecular and mathematical basis of Waddington's epigenetic landscape: a framework for post-Darwinian biology?}},
	volume = {34},
	year = {2012}}

@book{Waddington:1940,
	address = {Cambridge},
	author = {Conrad Hal Waddington},
	publisher = {Cambridge Biological Studies},
	title = {Organisers and Genes},
	year = {1940}}

@article{shi2022energy,
	author = {J. Shi and K. Aihara and T. Li and L. Chen},
	journal = {Natl Sci Rev},
	number = {8},
	pages = {nwac116},
	title = {{Energy landscape decomposition for cell differentiation with proliferation effect}},
	volume = {9},
	year = {2022}}

@article{huang2007bifurcation,
	author = {S. Huang and Y. Guo and G. May and T. Enver},
	journal = {Dev Biol},
	number = {2},
	pages = {695--713},
	title = {Bifurcation dynamics in lineage-commitment in bipotent progenitor cells},
	volume = {305},
	year = {2007}}

@article{weinreb2018,
	author = {C. Weinreb and S. Wolock and B. K. Tusi and M. Socolovsky and A. M. Klein},
	journal = {Proc Natl Acad Sci USA},
	number = {10},
	pages = {E2467--E2476},
	title = {{Fundamental limits on dynamic inference from single-cell snapshots}},
	volume = {115},
	year = {2018}}

@article{ferrell2012,
	author = {J. E. Ferrell},
	journal = {Current Biology},
	number = {11},
	pages = {R458--R466},
	title = {{Bistability, bifurcations, and Waddington's epigenetic landscape}},
	volume = {22},
	year = {2012}}

@book{Waddington1957,
	address = {London},
	author = {C. H. Waddington},
	publisher = {Allen \& Unwin},
	title = {The Strategy of the Genes},
	year = {1957}}

@article{Shakiba2022Waddington,
	author = {N. Shakiba and C. Li and J. Garcia-Ojalvo and K.-H. Cho and K. Patil and A. Walczak and Y.-Y. Liu and S. Kuehn and Q. Nie and A. Klein and G. Deco and M. Kringelbach and S. Iyer-Biswas},
	doi = {10.1016/j.cels.2021.12.003},
	journal = {Cell Systems},
	number = {1},
	pages = {4--9},
	title = {{How can Waddington-like landscapes facilitate insights beyond developmental biology?}},
	volume = {13},
	year = {2022}}

@article{wang2011landscape,
	author = {J. Wang and K. Zhang and L. Xu and E. Wang},
	journal = {Proc Natl Acad Sci USA},
	number = {20},
	pages = {8257--8262},
	title = {{Quantifying the {Waddington} landscape and biological paths for development and differentiation}},
	volume = {108},
	year = {2011}}

\end{document}